%% file: CNSE_arxiv.tex
\documentclass[twocolumn,aps,prx,showpacs,amsmath,superscriptaddress,longbibliography,notitlepage]{revtex4-2}
\usepackage{amssymb}
\usepackage{mathrsfs}
\usepackage{graphicx}
\usepackage{float}
\usepackage[caption=false]{subfig}
\usepackage[pdftex,colorlinks,linkcolor={red!75!black},citecolor={red!75!black},urlcolor={blue!75!black}]{hyperref}
\usepackage{verbatim}
\usepackage{epstopdf}
\usepackage[normalem]{ulem}
\usepackage{xcolor}
\usepackage{bm}
\usepackage{soul}
\usepackage{ulem}

\newcommand{\clr}{\color{red!75!black}}

\newcommand{\Rnum}[1]{\uppercase\expandafter{\romannumeral #1\relax}}

\def\CH{\textcolor{orange}}

\usepackage{enumitem}
\usepackage{comment}

\begin{document}

\title{Many-body critical non-Hermitian skin effect}
\author{Yi Qin}
\affiliation{Science, Mathematics and Technology, Singapore University of Technology and Design, Singapore 487372, Singapore}
\author{Yee Sin Ang}
\affiliation{Science, Mathematics and Technology, Singapore University of Technology and Design, Singapore 487372, Singapore}
\author{Ching Hua Lee}\email{phylch@nus.edu.sg}
\affiliation{Department of Physics, National University of Singapore, Singapore 117542}
\author{Linhu Li}\email{lilinhu@quantumsc.cn}
\affiliation{Quantum Science Center of Guangdong-Hong Kong-Macao Greater Bay Area (Guangdong), Shenzhen, China}

\begin{abstract}
Criticality in non-Hermitian systems unveils unique phase transitions and scaling behaviors beyond Hermitian paradigms, offering new insights into the interplay between gain/loss, non-reciprocity, and complex energy spectra. In this paper, we uncover a new class of many-body critical non-Hermitian skin effect (CSE) originating from the interplay between multiple non-Hermitian pumping channels and Hubbard interactions. In particular, criticality in the real-to-complex transitions can selectively emerge within the subspace of bound states or scattering states, as well as their interacting admixtures. These mechanisms possess no single-particle analog and can be diagnosed through a specially defined correlation function. As more particles are involved, higher-order CSEs naturally arise, with greatly enhanced effective coupling strengths and hence greater experimental accessibility. Our results reveal an enriched landscape of non-Hermitian critical phenomena in interacting many-body systems, and pave the way for investigating unconventional non-Hermitian criticality in the context of various interaction-induced particle clustering configurations.
\end{abstract}

\maketitle
\emph{{\clr Introduction}.---}
Non-Hermitian systems are far more sensitive than their Hermitian counterparts,
possessing massive localization of eigenstates at open boundaries that greatly diverge from those under periodic boundary conditions (PBCs), known as the non-Hermitian skin effect (NHSE)~\cite{LeeTony2016PRL, Martinez2018PRB,KunstPRL2018,Kawabata2018PRB,Liu2019PRL,Yao2018PRL,Lee2019,okuma2023non,zhang2022review,lin2023top}.
 In such systems, an anomalous critical skin effect (CSE)~\cite{Li2020NC, Liu2021, rafi2022CSE, rafi2025critical,liu2024non,yang2024percolation,xu2025exciton,cai2024non,wei2025generalized} arises when different NHSE~\cite{Yao2018, McDonald2018, Lee2019, Kawabata2019, Zhang2020, Okuma2020, li2022non, KunstPRL2018, Budich2020, Li2021, Roccati2021, yang2022designing,Mu2022,arouca2020unconventional, lei2024activating,tai2023zoology, qin2023universal, qin2023kinked, Borgnia2020, Deng2022PRB, roccati2023hermitian, meng2024exceptional,Poli2015, Helbig2020, Hofmann2020, xiaoLei2020, Sebastian2020, Palacios2021, ZhangXiujuan2021,xue2024topologically,wu2025hybrid, longhi2025er, glio2025non, li2025phase, shen2025non,guo2025skin} channels are weakly coupled to each other, such that even an infinitesimal coupling with different NHSE channels can further drastically change the spectral structure and eigenstate distribution in a size-dependent manner~\cite{Li2020NC}.
Recently, investigations of non-Hermitian physics have been extended to many-body systems,
with strong correlation between particles leading to rich variations of NHSE beyond single-particle scenarios,
including new forms of NHSE existing exclusively with multiple particles~\cite{Faugno2022PRL} or the spin degree of freedom~\cite{yoshida2024non},
and distinct NHSE channels in subspaces with different particle configurations~\cite{QinYi2024PRL,wang2025non, lee2021many,shen2022non,li2023non,kawabata2022many,shen2024enhanced,gliozzi2024many,yang2024non,glio2025non,li2024Dis,hu2025many}. 

As such, many-body interactions are expected to generate even more exotic critical non-Hermitian phenomena. However, CSE has never been investigated in many-body systems,
where particle interactions and the structure of the many-body Hilbert space may greatly affect the critical behaviors~\cite{lee2015geometric,mukherjee2021minimal,moudgalya2022hilbert,moudgalya2022quantum,brighi2023hilbert,adler2024observation,FromholzPRB2020, GergsPRB2016,rachel2018IN,WangPRB2015}, and even enhance the sensitivity of a system near critical points~\cite{ding2022EN}. 


In this paper, we present the first systematic investigation of many-body CSE from the perspective of multi-particle spectra and the organization of many-body eigenstates, revealing universal features that transcend specific model details.
Specifically, we report new classes of many-body CSE, where nontrivial correlations arise from the interplay between scattering and bound-states, leading to novel scaling behavior. 
Various types of many-body CSE emerge from the overlapping of energy clusters composed of bound states, scattering states, or their mixtures, 
as controllable by tuning the Hubbard interaction and sublattice-dependent onsite potentials.

By analyzing real-space correlations, we map out phase diagrams of distinct many-body CSE regimes, revealing how interactions fundamentally reshape critical behavior beyond single-particle physics. In particular, new higher-order CSEs are observed when more particles are involved, with stronger effective coupling strengths, which makes for feasible for experimental realization. Our results uncover new interaction-induced mechanisms for CSEs, as verified in multi-particle and fermionic systems (see Supplementary Materials~\cite{SuppMat}), and establish a new paradigm for non-Hermitian many-body criticality.

\noindent\emph{{\clr Paradigmatic model}.---} 
In general, many-body CSE emerges from mixing between subspaces with different non-Hermitian localization. 
To provide a clear picture,
we consider a minimal 1D non-Hermitian bosonic ladder with Bose-Hubbard interaction,
described by~\cite{QinYi2024PRL,kim2024collective}
\begin{align}
\hat{H}= & -J\sum_{x=1}^{L-1}\sum_{\sigma=A,B} \left(e^{\alpha_\sigma} \hat{a}_{x, \sigma}^{\dagger} \hat{a}_{x+1, \sigma}+e^{-\alpha_\sigma} \hat{a}_{x+1, \sigma}^{\dagger} \hat{a}_{x, \sigma}\right) \notag \\
& +J_p \sum_{x=1}^L\left(\hat{a}_{x, A}^{\dagger} \hat{a}_{x, B}+\hat{a}_{x, B}^{\dagger} \hat{a}_{x, A}\right) +\mu \sum_{x=1}^L\left(\hat{n}_{x, A}-\hat{n}_{x, B}\right)\notag \\
& +\frac{U}{2} \sum_{x=1}^{L}\sum_{\sigma=A,B} \hat{n}_{x, \sigma}\left(\hat{n}_{x, \sigma}-1\right),
\label{eq:Hamiltonian}
\end{align}
where $\sigma = A, B$ represents the two sublattices, $Je^{\pm \alpha_{\sigma}}$ ($\alpha_{A} =-\alpha_{B}= \alpha$) represents the non-reciprocal hopping amplitudes for the $\sigma$-sublattice, and $J_p$ denotes the hopping amplitude between the two sublattices.
Without loss of generality, we will fixed the non-reciprocal hopping at $Je^{\alpha}=1$ and $Je^{-\alpha}=0.5$ in the following discussion.
$\mu$ ($-\mu$) represents the onsite chemical potential for the \textit{A} (\textit{B})-sublattice, $U$ is the onsite interaction strength, $L$ is the system size, and $\hat{n}_{x,\sigma} = \hat{a}_{x,\sigma}^\dagger \hat{a}_{x,\sigma}$ denotes the number operator for bosons on the $\sigma$ sublattice of the $x$th unit cell.

\begin{figure}[!htbp]
\centering
\includegraphics[width=1\linewidth]{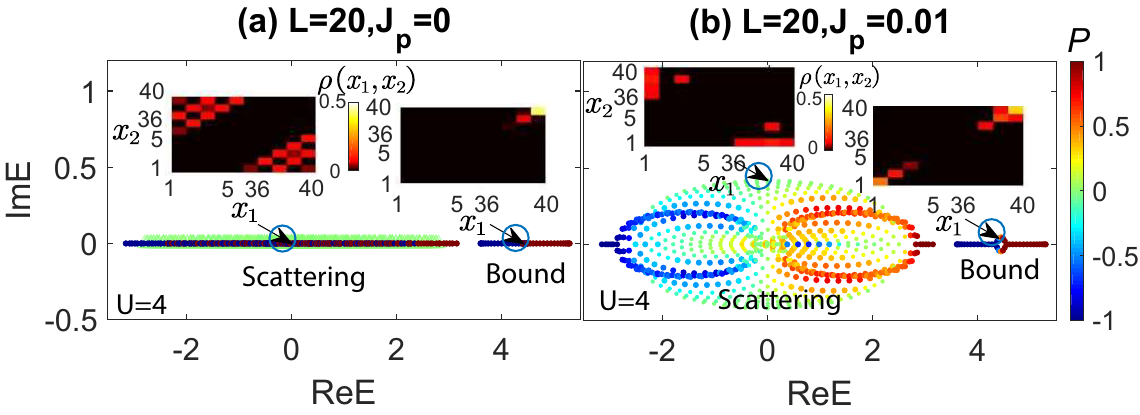}
\includegraphics[width=0.95\linewidth]{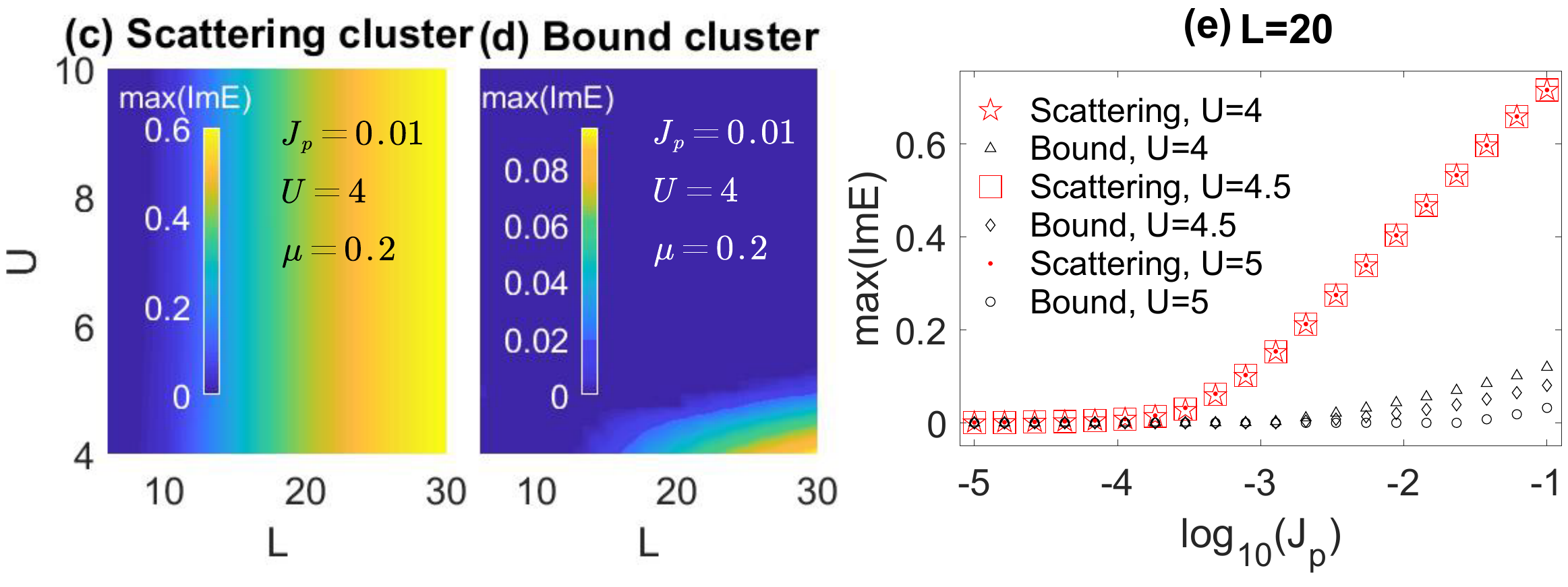}
\caption{\label{fig:FigCB} 
Distinct CSEs for scattering and bound states.
(a), (b)  Energy spectra for $N=2$ particles with inter-sublattice hopping $J_p=0$ and $0.01$, respectively. 
Eigenenergies $E$ are colored by the sublattice polarization $P$ of their eigenstates. Complex $E$ emerge at a small but nonzero $J_p$, manifesting the CSE. 
Insets show the 2-particle density $\rho(x_1,x_2)$, with sublattices \( A \) ($x_{1,2}\in[1,L]$) and \( B \) ($x_{1,2}\in[L+1,2L]$) for the eigenstates indicated by the arrows (highlighted with circles).
For scattering states, particle distribution in one sublattice greatly affects that in the other; for bound states, $J_p$ 
forces the edge localization to transition from uni-directional to bipolar across both sublattices.
(c) and (d) shows max(Im$E$) for the scattering and bound clusters for $J_p=0.01$, respectively. 
(e) max(Im$E$) for the scattering (bound) cluster versus $J_p$, as indicated by red (black) markers.
In (c) to (e), the interaction $U$ affects only max(Im$E$) of bound states, which take nonzero values at much bigger $L$ or $J_p$ compared with those of scattering states. 
In all panels, $\mu=0.2$, $U=4$, and $L=20$ unless otherwise specified. 
}
\end{figure} 

\

In the decoupled limit with $J_p=0$, the system reduces to two decoupled Hatano-Nelson models~\cite{HN1996PRL,HN1997PRB}, each exhibiting NHSE and real spectrum under open boundary conditions.
However, at the single-particle level, a weak coupling $J_p$ between the two sublattices can notably induce a 
real-complex transition of eigeneneriges, with eigenstates exhibiting scale-free behavior with increasing systems size. 
This is know as the CSE, where the threshold of $J_p$ to induce the transition tends to zero in the thermodynamic limit~\cite{Li2020NC,qin2023universal}.

\noindent\emph{{\clr Selective many-body CSE for non-overlapping scattering states and bound states}.---}
A strong interaction $U$ causes the appearance of energetically separated clusters known as scattering and bound states. Below, we show how they can each exhibit distinct critical scaling behavior. As a warm-up, we demonstrate this for $N=2$ particles in Fig.~\ref{fig:FigCB},
whose eigenenergies are separated into two groups centered at $E\approx 0$ and near $E\approx U$, representing the scattering and bound states, respectively.
In Figs. \ref{fig:FigCB}(a) and (b), they are colored according to their sublattice polarization 
$P=(N_A-N_B)/(N_A+N_B),N_\sigma=\sum_x\langle \hat{n}_{x,\sigma}\rangle.$
In Fig. \ref{fig:FigCB}(a) with $J_p=0$, all eigenstates have real eigenenergies with $P=1,0,-1$, as the two sublattices are decoupled. In particular, $P=0$ corresponds to a scattering state with one particle on each sublattice.
Upon turning on a weak inter-sublattice coupling $J_p$ [Fig. \ref{fig:FigCB}(b)], complex eigenenergies surprisingly appear for both scattering states and bound states, even though the non-Hermitian hoppings remain totally \emph{unchanged}. This arises because of the emergent correlation between different $P$ sectors, such that $P$ now takes on continuous values; 
to elucidate that, we examine the two-particle density
\begin{equation}
\rho(x_1,x_2)=\langle  \hat{b}^\dagger_{x_1}\hat{b}_{x_1}\hat{b}^\dagger_{x_2}\hat{b}_{x_2} \rangle,
\end{equation}
with $\hat{b}_{x_{1,2}}=\hat{a}_{x_{1,2},A}$ for $x_{1,2}\in[1,L]$ or $\hat{a}_{x_{1,2}-L,B}$ for  $x_{1,2}\in[L+1,2L]$,
as plotted in the insets. 
Evidently, a weak $J_p$ can abruptly induce strong correlation between the two particles -- for instance, in the left inset in (b) for a scattering state, $\rho(x_1,x_2)$ almost vanishes unless one particle is at $x=1$. In the right inset, the bound-particle pair simultaneously localize at the $x=1$ and $x=L$ ends, markedly different from the $J_p=0$ case in (a) where it is localized only at one end.
Such abrupt transitions in the energy spectra and eigenstate profiles driven by very small $J_p$ constitute the hallmark of many-body CSE.

A crucial observation is that strong onsite interactions $U$ can suppress the CSE (and hence the appearance of complex $E$) by binding the particles together. In Figs.~\ref{fig:FigCB}(c) and (d), we plot max(Im$E$) for scattering and bound states at fixed weak inter-sublattice coupling $J_p=0.01$, as a function of $U$ and the system size $L$.
While the scattering states (c) always acquire complex eigenenergies when $L\geq 10$, the bound states (d) only exhibit imaginary eigenenergies for small $U$. 
In Fig.~\ref{fig:FigCB}(e) where max(Im$E$) is shown against varying $J_p$, scattering states (red) exhibit unchanged propensity for the CSE as $U$ is varied. However, for bound states (black), the $U$ interactions evidently delay the onset of the CSE, with larger $U$ leading to much higher threshold for $J_p$ couplings to lead to complex energies.


\begin{figure}[!htbp]
\centering
\includegraphics[width=0.95\linewidth]{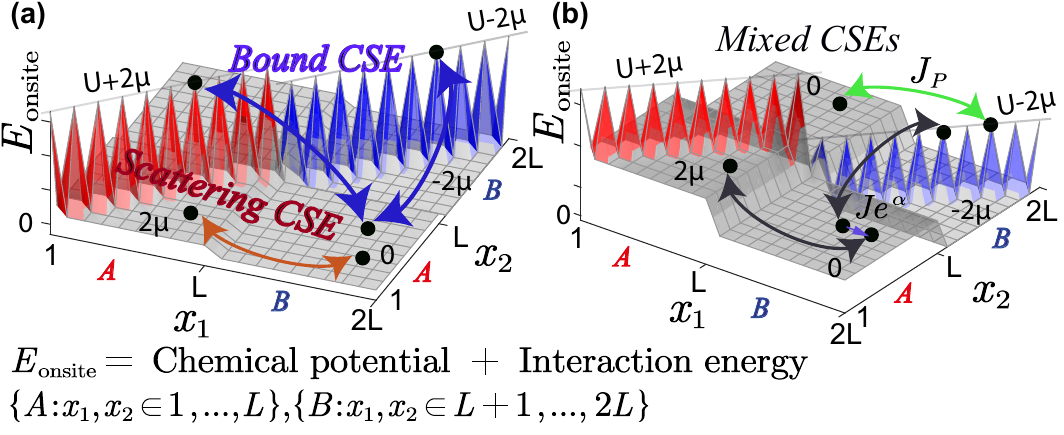}
\caption{\label{fig:sketch} 
Schematics of the mechanisms behind many-body CSE. Each panel shows the characteristic energy $E_{\rm onsite}$ for two particles located at positions $x_1$ and $x_2$. Red (blue) spikes of height $U$ correspond to energetically higher bound states at sublattice $A$ ($B$). Gray indicates scattering states. Colored arrows indicate different kinds of CSEs.
(a) Scattering (bound) CSE emerges from first (second)-order inter-sublattice $J_p$ hopping processes for $U>>\mu\approx 0$.
(b) Two distinct types of mixed CSEs also emerge from first and second-order $J_p$ hopping processes, arising from correlation between bound and scattering states when they have approximately equal energies.
}
\end{figure} 
To understand the weaker CSE with bound states, we perform a perturbative treatment in the subspace of bound particles, with $\mu=0$ chosen for the sack of simplicity.
Up to the second-order perturbation,
the effective Hamiltonian projected in this subspace is given by
\begin{equation}
\begin{array}{ll}
&{\hat H_{{\rm{eff}}}} = E_0 + {\widehat P_{{\rm{int}}}}{\hat H_{{\rm{hop}}}}{\widehat P_{{\rm{int}}}}  \\
&+ {\widehat P_{{\rm{int}}}}{\hat H_{{\rm{hop}}}}{\left( {E_0 - {{\hat H}_{{\rm{int}}}}} \right)^{ - 1}}{\hat H_{{\rm{hop}}}}{\widehat P_{{\rm{int}}}} + O\left( {\hat H_{{\rm{hop}}}^3} \right),
\end{array}
\end{equation}
with ${\widehat P_{{\mathop{\rm int}} }} = \sum\nolimits_{x = 1}^{L}\sum_{\sigma=A,B} {| {{\beta _{x,\sigma}}} \rangle \rangle} \langle {\langle {{\beta _{x,\sigma}}} } |$ the projector onto the Fock basis  $|\beta _{x,\sigma} \rangle \rangle$ with both particles occupying sublattice $\sigma$ at position $x$, 
and $E_0=U$ the unperburbed eigenenergy. 
Since hopping processes only move one particle at a time, the first-order term  ${\widehat P_{{\rm{int}}}}{\hat H_{{\rm{hop}}}}{\widehat P_{{\rm{int}}}}$ is zero. 
Expanding the second-order term, we obtain
\begin{small}
\begin{eqnarray}
&&H_{\mathrm{eff}} = U + \sum_{\sigma=A,B}\left[\sum_{x=1}^L  \left( \frac{\sqrt{2}J_{p}^{2}}{U} + \frac{2\sqrt{2}J^2}{U} \right) |\beta_{x,\sigma}\rangle \rangle \langle \langle \beta_{x,\sigma}| \right. \nonumber \\
&&\left.+ \frac{\sqrt{2}J^2}{U} \sum_{x=1}^{L-1}\left(e^{2\alpha_\sigma} |\beta_{x,\sigma}\rangle \rangle \langle \langle \beta_{x+1,\sigma}| + e^{-2\alpha_\sigma} |\beta_{x+1,\sigma}\rangle \rangle \langle \langle \beta_{x,\sigma}| \right)\right] \nonumber \\
&&+ \frac{2\sqrt{2}J_{p}^{2}}{U} \sum_{x=1}^L \left[ |\beta_{x,B}\rangle \rangle \langle \langle \beta_{x,A}| + |\beta_{x,A}\rangle \rangle \langle \langle \beta_{x,B}| \right]
\end{eqnarray}
\end{small}with $\alpha_A=-\alpha_B=\alpha$.
This Hamiltonian represents a single-particle ladder model with two sublattices spanned by $|\beta_{x,A}\rangle \rangle$ and $ |\beta_{x,B}\rangle \rangle$, possessing opposite non-reciprocal hopping $\sqrt{2}J^2e^{\pm \alpha}/U$ and inter-sublattice hopping $2\sqrt{2}J_p^2/U$.
Since CSE occurs when $J_p\ll J\approx 1$, the second-order inter-sublattice hopping amplitude becomes much weaker than the original one, 
thus a larger $J_p$ is required to induce a critical transition for bound states. A schematic illustrating the first and second-order processes of scattering and bound CSEs, respectively, is given in Fig.~\ref{fig:sketch}(a).
Note that these processes play a significant role only when the associated particle configurations have approximately the same characteristic energies $E_{\rm onsite}$ [so that eigenenergies with different $P$ overlap in Fig. \ref{fig:FigCB}(a)], which requires $\mu$ to be small in the current case.

In more generic many-body scenarios, interactions between different energetically separated sectors can be similarly decomposed into effective non-Hermitian hoppings of different strengths, which may then compete or collude to give rise to CSE behavior with no non-interacting analogs. 

\begin{figure*}[!htbp]
\centering
\includegraphics[width=1\linewidth]{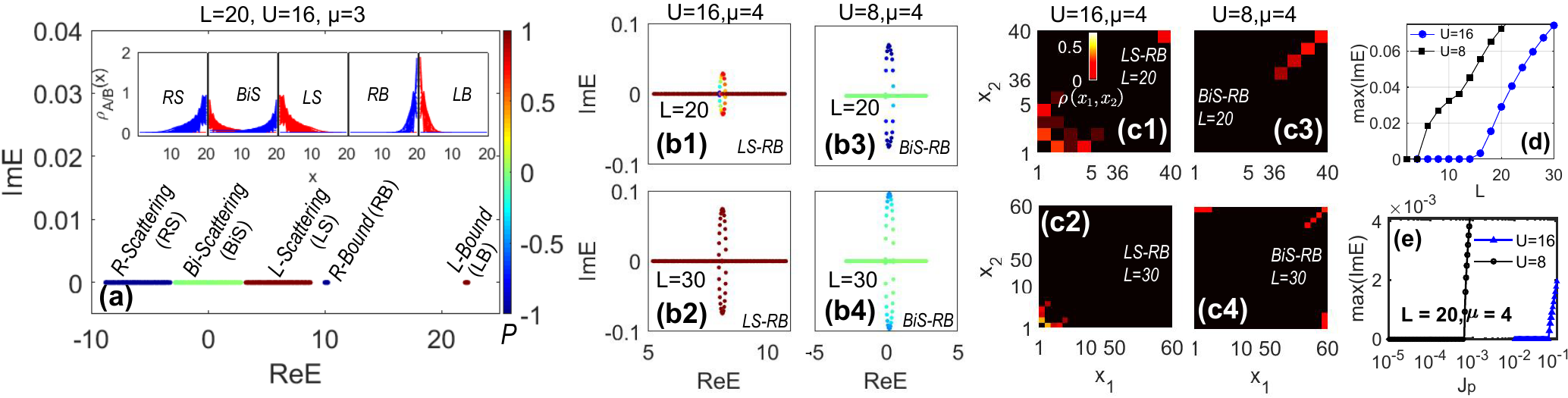}
\caption{\label{fig:Fig_CBmixed} 
New CSE channels within mixed scattering-bound states with $N=2$ particles and $J_p=0.01$.
(a) The five cluster types before the onset of CSE: R-Scattering (RS), Bi-Scattering (BiS), L-Scattering (LS), R-Bound (RB), and L-Bound (LB) clusters, colored by sublattice polarization $P$. Insets show their corresponding characteristic spatial profiles (red and blue represent sublattices A and B).
(b) Complex spectra for the mixed clusters for parameters that support CSEs.
(b1) and (b2) show the joint LS-RB cluster, and (b3) and (b4) show the joint BiS-RB cluster.
In each case, only eigenstates with ${\rm Im}E\neq0$ exhibit $L$-dependent $P$.
(c1) to (c4) the distribution $\rho(x_1,x_2)$ of eigenstates with the maximum ${\rm Im}E$ in (b1) to (b4), respectively.
When increasing the size $L$, a transition from bipolar to left (right to bipolar) localization is observed for the RS-LB (BiS-RB) cluster. Note that the bipolar localization in (c4) 
features nonzero density with $(x_{1,2},x_{2,1})\approx (1,2L)$, indicating a long-range correlation between different edges of the two sublattices.
(d) Maximum ${\rm Im}E$ of joint LS-RB (blue) and BiS-RB (black) clusters, respectively, versus the system size $L$.
(e) The same quantities as in (d), but versus the inter-sublattice hopping $J_p$.
The critical transition occurs at smaller $L$ or $J_p$ for the BiS-LB cluster, indicating its higher sensitivity. 
}
\end{figure*}

\noindent\emph{{\clr Emergent competitive many-body CSE for mixed scattering-bound states}.---}
Beyond the above discussed scenarios, new forms of competitive critical behaviors can arise when scattering and bound states become energetically comparable and mix. To elucidate them systematically, we first showcase the catalog of 
five scattering and bound clusters separated in energy in a certain regime of $\mu$ and $U$, such that CSE is absent and all 
eigenenergies are real [Fig.~\ref{fig:Fig_CBmixed}(a)].
These clusters exhibit different  localizations [right(R), left(L) or bipolar(Bi)] and characters [scattering(S) or bound(B)], and are labeled 
accordingly as RS, BiS, LS, RB, and LB in Fig.~\ref{fig:Fig_CBmixed}(a),
with red/blue representing A/B sublattice occupation.


To induce new mixed CSE channels between scattering and bound states, we further tune the onsite energy $\mu$ and interaction $U$ such that the desired clusters overlap in Re$E$.
In Fig.~\ref{fig:Fig_CBmixed}(b1) and (b2),
the left-localized scattering (LS) states and the right-localized bound (RB) states merge in their real energy (see Supplemental Materials~
\cite{SuppMat}), and complex eigenenergies appear for this joint LS-RB cluster, indicating the occurrence of a critical transition due to the correlation between singlons and doublons.
Notably, for our chosen parameters, this joint cluster exhibits bipolar localization when $L=20$, and only left localization when $L=30$, as shown in Fig.~\ref{fig:Fig_CBmixed} (b1) and (b2).
This is because the Hilbert subspace of scattering states scales as $L^2$, 
while that of bound states scales as $L$. Thus, for a sufficiently large system, the joint LS-RB cluster behavior is expected to be dominated by that of scattering states, which possess left localization in our case [see Fig.~\ref{fig:Fig_CBmixed}(a)].
The transition from bipolar to left localization can be more clearly seen from the density distribution, as shown in Fig. \ref{fig:Fig_CBmixed}(c1) and (c2).

Another example of mixed CSE is shown in Fig.~\ref{fig:Fig_CBmixed}(b3) and (b4), where the interaction is tuned to have bipolar scattering (BiS) states and right localized bound (RB) states joining together (see Supplemental Materials~
\cite{SuppMat}). 
Similar to the previous case, the emergence of complex eigenenergies marks the critical transition, and the distribution of critical states also varies with the system's size.
Explicitly, critical states distribute mostly on right edge of sublattice $B$ (with $P\approx -1$) as the R-bound cluster when $L=20$, as further demonstrated by the density distribution $\rho(x_1,x_2)$ in Fig. \ref{fig:Fig_CBmixed}(c3).
In contrast, for a larger $L=30$, bipolar distribution on both sublattices ($P\approx 0)$ as the Bi-scattering cluster appears for the critical states,
reflecting by the large density around $(x_{1,2},x_{2,1})\approx(1,2L)$ in Fig.~\ref{fig:Fig_CBmixed}(c4).
Finally, we note the tantalizing third possibility of mixture between clusters RS and RB is forbidden, as it requires $U=0$, which also eliminates any robust bound state.

In Fig. \ref{fig:Fig_CBmixed}(d) and (e), we illustrate max(Im$E$) as a function of $L$ and $J_p$, respectively, for the two types of mixed CSEs discussed above. Evidently, the BiS-RB cluster (black) is more sensitive and requires a smaller $J_p$ or $L$ to induce the critical transition. 
This is because before merging together, LS and RB clusters have the two particles both occupying one of sublattices A or B, and the BiS cluster has one particle occupying each sublattice.
Thus, as shown in Fig.~\ref{fig:sketch}(b), particle exchange between BiS and RB clusters is a first-order process of the inter-sublattice hopping, in contrast to the second-order one between LS and RB clusters,
making the former more sensitive to parameters that induce the exchange.

\begin{figure}[htbp]
\centering
\includegraphics[width=1\linewidth]{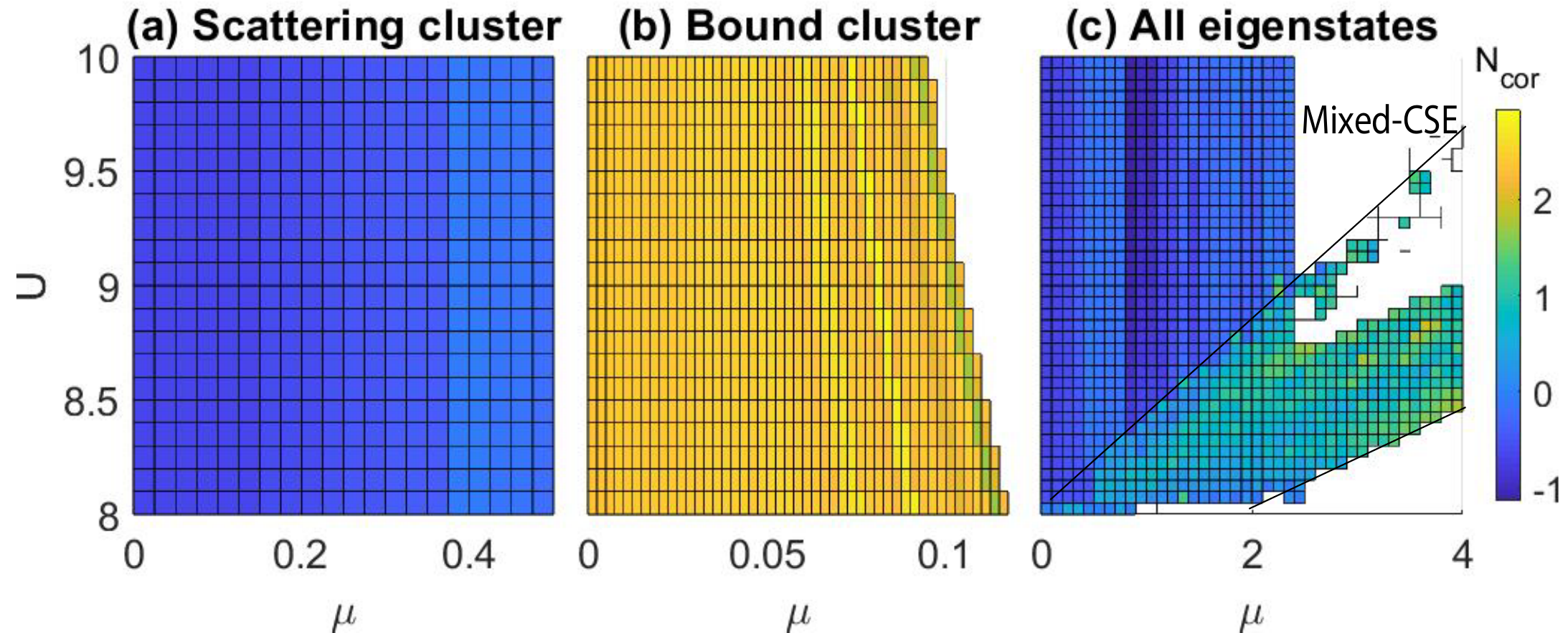}
\caption{\label{fig:Cor} Diagnostic of
many-body CESs
from the correlation $N_{\rm cor}$.
(a) to (c) $N_{\rm cor}$ for the state with the maximum imaginary energy in scattering clusters, bound clusters, and for all eigenstates, respectively.
In (a) and (b), a smaller parameter range of the potential $\mu$ is chosen, so that the scattering and bound clusters are well separated in energy.
The mixed CSE is characterized by $N_{\rm cor}$ taking intermediate values between that of scattering and bound CSEs, in the parameter region between the two dark lines in (c).
The system's size is set to $L=15$ and the inter-sublattice coupling is $J_p=0.01$. 
}
\end{figure}
\CH{\emph{{\clr Characterizing CSEs through correlations}.---}}
To quantitatively characterize and distinguish various types of many-body CSEs, we compute the 
correlation quantity
\begin{equation}
\mathcal{N}_{\rm cor}  =\sum_{x_1, x_2} \Gamma_{x_1, x_1} \Gamma_{x_2, x_2}-\Gamma_{x_1, x_2}^2,
\end{equation}
\noindent with 
$\Gamma_{x_1, x_2}  =\langle \hat{b}_{x_1}^{\dagger} \hat{b}_{x_2}^{\dagger} \hat{b}_{x_2} \hat{b}_{x_1}\rangle$
calculated for the max(Im$E$) eigenstate.
By definition, scattering states with two particles separated has negative $\mathcal{N}_{\rm cor}$,
while bound states have positive $\mathcal{N}_{\rm cor}<4(1-1/L)$~(see Supplemental Materials~
\cite{SuppMat}). 
Consistently, we observe $\mathcal{N}_{\rm cor}<0$ (blue) and $\mathcal{N}_{\rm cor}\approx 2.8$ (yellow) for 
scattering and bound clusters, respectively, as shown in Fig.~\ref{fig:Cor}(a) and (b).
On the other hand, mixed CSE states originate from the mixture of scattering and bound states, and thus have $\mathcal{N}_{\rm cor}$ taking values in between these $\mathcal{N}_{\rm cor}$. Indeed, as shown in Fig.~\ref{fig:Cor}(c), $0<\mathcal{N}_{\rm cor}<2$ (green) in the parameter region with mixed CSE. Note that in each case, 
max(Im$E$)$=0$ when the CSE vanishes, and we leave these parameter regions blank. 
\begin{figure}[htbp]
\centering
\includegraphics[width=1\linewidth]{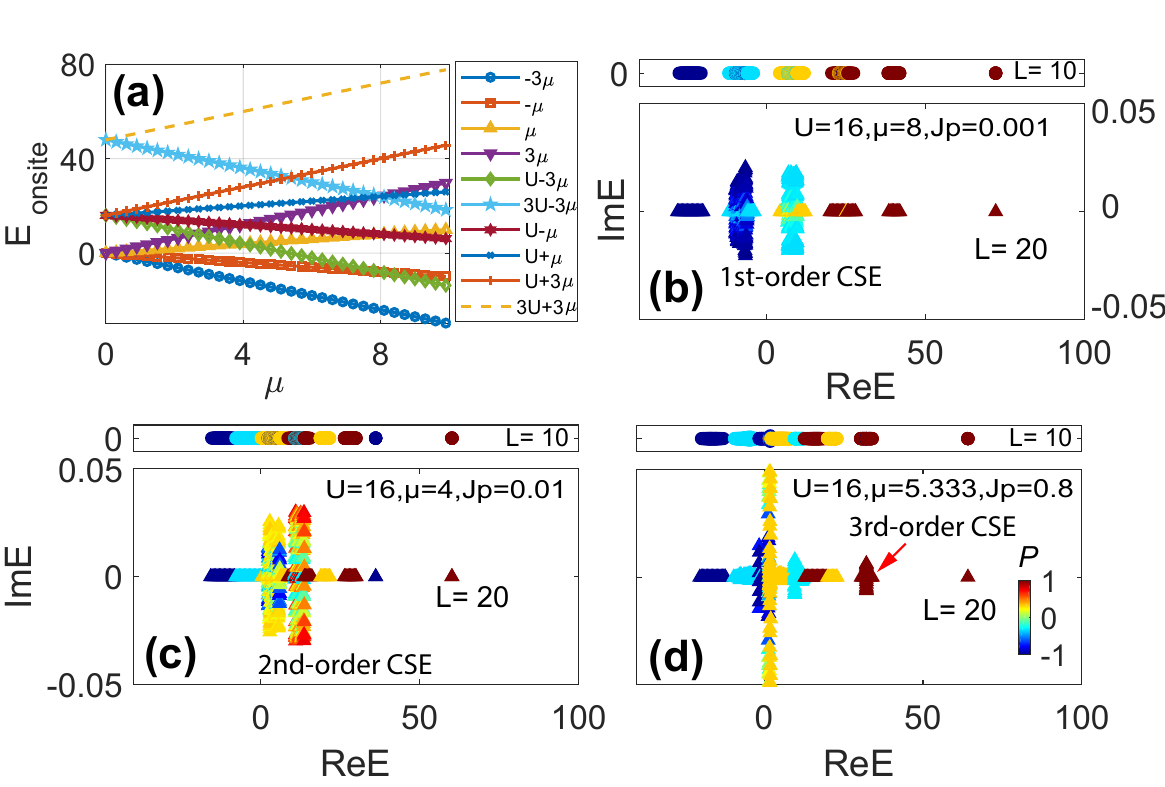}
\caption{\label{fig:N3} Many-body CSE for $N = 3$.
(a) Onsite energy $E_{\rm onsite}$ includes only the chemical potential and interaction energy. 
Crossings between different $E_{\rm onsite}$ curves enable multiple mechanisms that lead to CSEs.
(b) to (d) first, second, and third orders of CSEs, respectively, characterized by the real-complex transitions of eigenenergies when increasing
the system size $L$.
Parameters are indicated in the figure. A higher order CSE requires a stronger $J_p$ to occur.
}
\end{figure}

\noindent\emph{{\clr Many-body CSE beyond two particles}.---} 
The many-body regime beyond the two-particle case exhibits high order CSEs, which reflect the intricate interplay between non-Hermitian pumping and interaction-induced scattering and bound states. To further investigate the many-body CSE, we present several examples involving three particles in Fig.~\ref{fig:N3}. We plot the characteristic energy $E_{\rm onsite}$ as a function of the chemical potential $\mu$, with interaction fixed at $U=16$. 
Parameter regimes of different types of CSEs can thus be identified by the crossing of different $E_{\rm onsite}$,
as shown in Fig.~\ref{fig:N3}(a). 
In Fig.~\ref{fig:N3}(b) to (d), we display different orders of CSEs with size-dependent eigenenergies. The order $n$ of CSEs corresponds to particles collectively jumping between sublattices, with an amplitude proportional to $J_p^n$. In this case, \( J_p \) can be of the same order as the other hopping parameters. For example, the scattering and BiS-RB CSEs are first-order, while the Bound and LS-RB CSEs are second-order, as shown in Fig.~\ref{fig:sketch}.
We find that as the order increases, a stronger coupling strength $J_p$ is required to induced real-complex transition of eigenenergies (e.g, $J_p\approx0.8$ for the third order one),~yet the threshold of $J_p$ decreases as increasing the system size $L$~\cite{SuppMat}.


\noindent\emph{{\clr Discussion}.---}
We have investigated new mechanisms behind the many-body non-Hermitian critical skin effect (CSEs), focusing on the crucial role of correlations between different clusters in various types of many-body CSEs. Most generally, CSE occurs whenever there exists subsystems with different skin decay lengths  that are spatially separated, but energetically connected. Crucially, our findings reveal a qualitative departure from the non-interacting scenario, where the CSE typically manifests uniformly across all states. In contrast, interactions can selectively induce, delay, or reshape the onset of CSEs, depending on how they restructure energy clusters and their couplings.  In addition, since the interaction $U$ acts to energetically separate or mix different scattering and bound clusters, many-body CSEs can also be induced with attractive interactions $U<0$~\cite{SuppMat},  further attesting to the generality of many-body CSEs.

Our work provides a framework for investigating CSEs in the many-body context, particularly with the statistic properties of particles taking into consideration i.e. in bosonic, fermionic (example is given in the supplemental materials \cite{SuppMat} with nearest neighbor interactions), or even anyonic systems~\cite{Liu2018PRL,Zhang2023CP,Joyce2023arxiv,qin2025CP}.
  With larger particle numbers, the CSE would also effectively act in higher-dimensional lattices that are likely imbued with non-trivial higher-order topology. The combination of the sensitivity of the CSE and the robustness of topological modes can inspire many-body quantum sensing applications~\cite{koch2022quantum,sarkar2022free,MukhopadhyayPRL2024,montenegro2024R}. Finally, 
we point out that the CSE may be realized in ultracold atomic lattices where non-Hermitian skin couplings can be implemented with laser-induced atom loss~\cite{liang2022dynamic,zhao2025two},
as well as in electrical~\cite{sahin2025topolectrical,yang2024circuit,zheng2022exploring,Zhang2023CP,zhang2023electrical,shang2024observation,zhang2025observation,zou2024experimental,stegmaier2024topological,nagulu2022chip} or quantum circuit~\cite{koh2024realization,shen2025observation,koh2025interacting,shen2025robust,smith2022crossing,zhang2025observation2} lattices where 
transitions between different Fock states can be mapped onto a network with high connectivity.

\section*{Acknowledgements}
We acknowledge support from the Ministry of Education, Singapore (MOE) Tier II grant (Award ID: MOE-T2EP50222-0003), Tier I grant (WBS no: A-8002656-00-00), and National Natural Science Foundation of China (Grant No.12474159).

\input{CNSE_v16.bbl}
\newpage

\begin{widetext}

	\setcounter{equation}{0} \setcounter{figure}{0} \setcounter{table}{0} %
	\renewcommand{\theequation}{S\arabic{equation}} \renewcommand{\thefigure}{S%
		\arabic{figure}} \renewcommand{\bibnumfmt}[1]{[S#1]} 
	\renewcommand{\citenumfont}[1]{S#1}
	
\begin{center}
\textbf{\large Supplementary Materials}
\end{center}

\appendix   
\setcounter{table}{0}   
\setcounter{figure}{0}
\renewcommand{\thetable}{A\arabic{table}}
\renewcommand{\thefigure}{A\arabic{figure}}

\maketitle

\section{Perturbation treatment for the bound cluster states}
In this section, we derive the effective model for the bound states critical skin effect. The model Hamiltonian can be written as~\cite{SQinYi2024PRL,Skim2024collective} 
\begin{equation}
\begin{aligned}
\hat{H} = & -\sum_{j=1,\sigma=A,B}^{L-1} \left[ J_{L}^{\sigma}\hat{a}_{j,\sigma}^{\dagger}\hat{a}_{j+1,\sigma} + J_{R}^{\sigma}\hat{a}_{j+1,\sigma}^{\dagger}\hat{a}_{j,\sigma} \right] \\
& + J_p\sum_{j=1}^L \left( \hat{a}_{j,A}^{\dagger}\hat{a}_{j,B} + \hat{a}_{j,B}^{\dagger}\hat{a}_{j,A} \right) \\
& + \frac{U}{2} \sum_{j=1,\sigma=A,B}^L \hat{n}_{j,\sigma}(\hat{n}_{j,\sigma} - 1) \\
& + \mu \sum_{j=1}^L \left( \hat{n}_{j,A} - \hat{n}_{j,B} \right).
\end{aligned}
\end{equation}

In the paired subspace, we take the unperturbed term as  
\begin{equation}
\hat{H}_0 = \frac{U}{2} \sum_{j=1,\sigma=A,B}^L \hat{n}_{j,\sigma}(\hat{n}_{j,\sigma}-1) 
+ \mu \sum_{j=1}^L \left( \hat{n}_{j,A} - \hat{n}_{j,B} \right).
\end{equation}

The perturbation term is given by  
\begin{equation}
\begin{aligned}
\hat{H}_{\mathrm{hop}} = & -\sum_{j=1,\sigma=A,B}^{L-1} \left[ 
J_{L}^{\sigma} \hat{a}_{j,\sigma}^{\dagger} \hat{a}_{j+1,\sigma} 
+ J_{R}^{\sigma} \hat{a}_{j+1,\sigma}^{\dagger} \hat{a}_{j,\sigma} \right] \\
& + J_p \sum_{j=1}^L \left( 
\hat{a}_{j,A}^{\dagger} \hat{a}_{j,B} 
+ \hat{a}_{j,B}^{\dagger} \hat{a}_{j,A} \right).
\end{aligned}
\end{equation}
Then, we project the Hamiltonian into the paired subspace $|a_{l,\sigma}^{2}\rangle =\frac{1}{\sqrt{2}}\left( \hat{a}^{\dagger} \right) _{l,\sigma}^{2}|\mathrm{vac}\rangle \equiv |\left. \beta _{l,\sigma} \right> \rangle ,\left( l=1,2,...,L;\sigma =A,B \right) ,$ 
with $\left| {{\rm{vac}}} \right\rangle $ the vacuum state. This basis constitutes a sub-Hilbert space of the Fock space, whose basis can be denoted as $\left| {{l_1}{\sigma _1},{l_2}{\sigma _2}} \right\rangle ,{\rm{ 1}} \le {l_1} \le {l_2} \le L,{\sigma _{1,2}} = A,B.$ , with $L$ the number of unit cell), and  $({l_1},{l_2})$ the positions of the bosons.
\begin{equation}
\hat{H}_{\text {eff }}=E_0+\hat{P}_{\text {int }} \hat{H}_{\text {hop }} \hat{P}_{\text {int }}+\hat{P}_{\text {int }} \hat{H}_{\text {hop }}\left(E_0-\hat{H}_{\text {int }}\right)^{-1} \hat{H}_{\text {hop }} \hat{P}_{\text {int }}+O\left(\hat{H}_{\mathrm{hop}}^3\right),
\end{equation}
$
\hat{P}_{\mathrm{int}}=\sum_{l=1,\sigma =A,B}^L{\left| \beta _{l,\sigma} \right. \rangle \rangle \left. \langle \left. \langle \beta _{l,\sigma} \right| \right.}
$
 the projector onto the doubly occupied sub-Hilbert space and $E=U-2 \delta_{\sigma, A} \mu+2 \delta_{\sigma, B} \mu$. The firstorder perturbation $\widehat{P}_{\text {int }} \hat{H}_{\text {hop }} \hat{P}_{\text {int }}$ is zero.


$$
\begin{aligned}
& \left.\langle\langle\beta_{j,\sigma}\right| \hat{H}_{\mathrm{hop}}\left(E_0-\hat{H}_0\right)^{-1} \hat{H}_{\mathrm{hop}} \mid \beta_{i,\sigma}\rangle\rangle \\
& =\frac{1}{2}\langle\mathrm{vac}| \hat{a}_{j,\sigma}^2 \hat{H}_{\mathrm{hop}}\left(E_0-\hat{H}_0\right)^{-1} \hat{H}_{\mathrm{hop}}\left(\hat{a}_{i,\sigma}^{\dagger}\right)^2|\mathrm{vac}\rangle . \\
& {\left[-\sum_{l=1}^{L-1}\left(J_L^A \hat{a}_{l, A}^{\dagger} \hat{a}_{l+1, A}+J_R^A \hat{a}_{l+1, A}^{\dagger} \hat{a}_{l, A}\right)-\sum_{l=1}^{L-1}\left(J_L^B \hat{a}_{l, B}^{\dagger} \hat{a}_{l+1, B}+J_R^B \hat{a}_{l+1, B}^{\dagger} \hat{a}_{l, B}\right)\right.} \\
& \left.+J_p \sum_{l=1}^L\left(\hat{a}_{l, A}^{\dagger} \hat{a}_{l, B}+\hat{a}_{l, B}^{\dagger} \hat{a}_{l, A}\right)\right] \frac{\left(\hat{a}_{i, \sigma}^{\dagger}\right)^2}{\sqrt{2}}|\mathrm{vac}\rangle \\
& =-\delta_{\sigma, A}\left(J_L^A \hat{a}_{i-1, A}^{\dagger} \hat{a}_{i, A}^{\dagger}+J_R^A \hat{a}_{i+1, A}^{\dagger} \hat{a}_{i, A}^{\dagger}\right)|\mathrm{vac}\rangle \\
& -\delta_{\sigma, B}\left(J_L^B \hat{a}_{i-1, B}^{\dagger} \hat{a}_{i, B}^{\dagger}+J_R^B \hat{a}_{i+1, B}^{\dagger} \hat{a}_{i, B}^{\dagger}\right)|\mathrm{vac}\rangle \\
& +\delta_{\sigma, A} J_p \hat{a}_{i, B}^{\dagger} \hat{a}_{i, A}|\mathrm{vac}\rangle+\delta_{\sigma, B} J_p \hat{a}_{i, A}^{\dagger} \hat{a}_{i, B}|\mathrm{vac}\rangle
\end{aligned}
$$
Thus, we can divide the calculations into four sectors.
\begin{equation}
\begin{aligned}
& \left(E_0-\hat{H}_0\right)^{-1} \hat{H}_{\mathrm{hop}}\left(\hat{a}_{i, \sigma}^{\dagger}\right)^2|\mathrm{vac}\rangle \\
& =-\delta_{A, \sigma}\left(E_0-\hat{H}_0\right)^{-1}\left(J_L^A \hat{a}_{i-1, A}^{\dagger} \hat{a}_{i, A}^{\dagger}+J_R^A \hat{a}_{i+1, A}^{\dagger} \hat{a}_{i, A}^{\dagger}\right)|\mathrm{vac}\rangle \\
& -\delta_{B, \sigma}\left(E_0-\hat{H}_0\right)^{-1}\left(J_L^B \hat{a}_{i-1, B}^{\dagger} \hat{a}_{i, B}^{\dagger}+J_R^B \hat{a}_{i+1, B}^{\dagger} \hat{a}_{i, B}^{\dagger}\right)|\mathrm{vac}\rangle \\
& +\delta_{A, \sigma}\left(E_0-\hat{H}_0\right)^{-1} J_p \hat{a}_{i, B}^{\dagger} \hat{a}_{i, A}|\mathrm{vac}\rangle \\
& +\delta_{B, \sigma}\left(E_0-\hat{H}_0\right)^{-1} J_p \hat{a}_{i, A}^{\dagger} \hat{a}_{i, B}|\mathrm{vac}\rangle \\
& =-\delta_{A, \sigma} \frac{1}{U+2 \mu-2 \mu}\left(J_L^A \hat{a}_{i-1, A}^{\dagger} \hat{a}_{i, A}^{\dagger}+J_R^A \hat{a}_{i+1, A}^{\dagger} \hat{a}_{i, A}^{\dagger}\right)|\mathrm{vac}\rangle \\
& -\delta_{B, \sigma} \frac{1}{U-2 \mu-(-2 \mu)}\left(J_L^B \hat{a}_{i-1, B}^{\dagger} \hat{a}_{i, B}^{\dagger}+J_R^B \hat{a}_{i+1, B}^{\dagger} \hat{a}_{i, B}^{\dagger}\right)|\mathrm{vac}\rangle \\
& +\delta_{A, \sigma} \frac{1}{U+2 \mu} J_p \hat{a}_{i, B}^{\dagger} \hat{a}_{i, A}|\mathrm{vac}\rangle \\
& +\delta_{B, \sigma} \frac{1}{U-2 \mu} J_p \hat{a}_{i, A}^{\dagger} \hat{a}_{i, B}|\mathrm{vac}\rangle \\
& =-\delta_{A, \sigma} \frac{1}{U}\left(J_L^A \hat{a}_{i-1, A}^{\dagger} \hat{a}_{i, A}^{\dagger}+J_R^A \hat{a}_{i+1, A}^{\dagger} \hat{a}_{i, A}^{\dagger}\right)|\mathrm{vac}\rangle \\
& -\delta_{B, \sigma} \frac{1}{U}\left(J_L^B \hat{a}_{i-1, B}^{\dagger} \hat{a}_{i, B}^{\dagger}+J_R^B \hat{a}_{i+1, B}^{\dagger} \hat{a}_{i, B}^{\dagger}\right)|\mathrm{vac}\rangle \\
& +\delta_{A, \sigma} \frac{1}{U+2 \mu} J_p \hat{a}_{i, B}^{\dagger} \hat{a}_{i, A}|\mathrm{vac}\rangle \\
& +\delta_{B, \sigma} \frac{1}{U-2 \mu} J_p \hat{a}_{i, A}^{\dagger} \hat{a}_{i, B}|\mathrm{vac}\rangle
\end{aligned}
\end{equation}
Further derivation leads to 
\begin{equation*}
\begin{aligned}
&\begin{aligned}
& \hat{H}_{\text {hopp }}\left(E_0-\hat{H}_0\right)^{-1} \hat{H}_{\text {bop }}\left(\hat{a}_{i, \sigma}^{\dagger}\right)^2|\mathrm{vac}\rangle \\ 
& = \left[-\sum_{i=1}^{L-1}\left(J_{L}^A \hat{a}_{i-1, A}^{\dagger} \hat{a}_{i, A} + J_{R}^{A} \hat{a}_{i+1, A}^{\dagger} \hat{a}_{i, A}\right)\right. \\ 
& - \sum_{l=1}^{L-1}\left(J_L^B \hat{a}_{i-1, B}^{\dagger} \hat{a}_{i, B}+ J_R^B \hat{a}_{i+1, B}^{\dagger} \hat{a}_{i, B}\right) \\ 
& \left.+ J_p \sum_{i=1}^L\left(\hat{a}_{i, B}^{\dagger} \hat{a}_{i, A} + \hat{a}_{i, A}^{\dagger} \hat{a}_{i, B}\right)\right] \\
& \times\left[-\delta_{\sigma, A} \frac{1}{U}\left(J_L^A \hat{a}_{i-1, A}^{\dagger} \hat{a}_{i, A}^{\dagger} + J_R^A \hat{a}_{i+1, A}^{\dagger} \hat{a}_{i, A}^{\dagger}\right)|\mathrm{vac}\rangle\right. \\
& - \delta_{\sigma, B} \frac{1}{U}\left(J_L^B \hat{a}_{i-1, B}^{\dagger} \hat{a}_{i, B}^{\dagger} + J_R^B \hat{a}_{i+1, B}^{\dagger} \hat{a}_{i, B}^{\dagger}\right)|\mathrm{vac}\rangle \\ 
& \left.\left.+ \delta_{\sigma, A} \frac{1}{U+2 \mu} J_p \hat{a}_{i, B}^{\dagger} \hat{a}_{i, A}^{\dagger}|\mathrm{vac}\rangle + \delta_{\sigma, B} \frac{1}{U-2 \mu} J_p \hat{a}_{i, A}^{\dagger} \hat{a}_{i, B}^{\dagger}|\mathrm{vac}\rangle\right]\right]
\end{aligned}
\end{aligned}
\end{equation*}

\begin{equation*}
\begin{aligned}
& = \delta_{\sigma, A} \frac{1}{U}\left[J_L^A J_L^A \hat{a}_{i-2, A}^{\dagger} \hat{a}_{i, A}^{\dagger} + \sqrt{2} J_L^A J_L^A \hat{a}_{i-1, A}^{\dagger} \hat{a}_{i-1, A}^{\dagger} \right. \\ 
& + 2J_L^A J_R^A \hat{a}_{i-1, A}^{\dagger} \hat{a}_{i+1, A}^{\dagger} + 2\sqrt{2} J_L^A J_R^A \hat{a}_{i, A}^{\dagger} \hat{a}_{i, A}^{\dagger} + \sqrt{2} J_R^A J_R^A \hat{a}_{i+1, A}^{\dagger} \hat{a}_{i+1, A}^{\dagger} \\ 
& \left.+ J_R^A J_R^A \hat{a}_{i+2, A}^{\dagger} \hat{a}_{i, A}^{\dagger}\right]|\mathrm{vac}\rangle \\
& + \delta_{\sigma, B} \frac{1}{U}\left[J_L^B J_L^B \hat{a}_{i-2, B}^{\dagger} \hat{a}_{i, B}^{\dagger} + \sqrt{2} J_L^B J_L^B \hat{a}_{i-1, B}^{\dagger} \hat{a}_{i-1, B}^{\dagger} \right. \\ 
& + 2J_L^B J_R^B \hat{a}_{i-1, B}^{\dagger} \hat{a}_{i+1, B}^{\dagger} + 2\sqrt{2} J_L^B J_R^B \hat{a}_{i, B}^{\dagger} \hat{a}_{i, B}^{\dagger} + \sqrt{2} J_R^B J_R^B \hat{a}_{i+1, B}^{\dagger} \hat{a}_{i+1, B}^{\dagger} \\ 
& \left.+ J_R^B J_R^B \hat{a}_{i+2, B}^{\dagger} \hat{a}_{i, B}^{\dagger}\right]|\mathrm{vac}\rangle \\ 
& + \delta_{\sigma, A} \frac{\sqrt{2}}{U+2 \mu} J_p J_p \hat{a}_{i, A}^{\dagger} \hat{a}_{i, A}^{\dagger}|\mathrm{vac}\rangle + \delta_{\sigma, B} \frac{\sqrt{2}}{U-2 \mu} J_p J_p \hat{a}_{i, B}^{\dagger} \hat{a}_{i, B}^{\dagger}|\mathrm{vac}\rangle\\
& + \delta_{\sigma, A} \frac{\sqrt{2}}{U-2 \mu} J_p J_p \hat{a}_{i, A}^{\dagger} \hat{a}_{i, A}^{\dagger}|\mathrm{vac}\rangle + \delta_{\sigma, B} \frac{\sqrt{2}}{U+2 \mu} J_p J_p \hat{a}_{i, B}^{\dagger} \hat{a}_{i, B}^{\dagger}|\mathrm{vac}\rangle
\end{aligned}
\end{equation*}

Finally, we obtain the results. 

\begin{equation}
\begin{aligned}
& \frac{1}{2}\langle\operatorname{vac}| \hat{a}_{j, \sigma^{\prime}}^2 \hat{H}_{\mathrm{hop}}\left(E_0-\hat{H}_0\right)^{-1} \hat{H}_{\mathrm{hop}}\left(\hat{a}_{i, \sigma}^{\dagger}\right)^2|\mathrm{vac}\rangle \\
& =\delta_{A, \sigma} \delta_{A, \sigma^{\prime}} \frac{\sqrt{2}}{U}\left[J_L^A J_L^A \delta_{j, i-1}+2 J_L^A J_R^A \delta_{j, i}+J_R^A J_R^A \delta_{j, i+1}\right] \\
& +\delta_{B, \sigma} \delta_{B, \sigma^{\prime}} \frac{\sqrt{2}}{U}\left[J_L^B J_L^B \delta_{j, i-1}+2 J_L^B J_R^B \delta_{j, i}+J_R^B J_R^B \delta_{j, i+1}\right] \\
& +\delta_{A, \sigma} \delta_{B, \sigma^{\prime}}  J_p J_p\left[\frac{\sqrt{2}}{U+2 \mu}\delta_{j, i}+\frac{\sqrt{2}}{U-2 \mu}\delta_{j, i}\right] \\
& +\delta_{B, \sigma} \delta_{A, \sigma^{\prime}}  J_p J_p\left[\frac{\sqrt{2}}{U-2 \mu}\delta_{j, i}+\delta_{j, i}\frac{\sqrt{2}}{U+2 \mu}\right]
\end{aligned}
\end{equation}
Thus, the effective Hamiltonian can be written as 
\begin{equation}
\begin{aligned}
H_{\text{eff}} = & \, U + \sum_{l=1}^L \left[ 
\left( \frac{\sqrt{2} J_p^2}{U} + \frac{2 \sqrt{2} J_L^A J_R^A}{U} \right) 
\left|\beta_{l, A}\right\rangle\!\rangle 
\left\langle\!\langle \beta_{l, A}\right| \right] \\
& + \sum_{l=1}^L \left[ 
\left( \frac{\sqrt{2} J_p^2}{U} + \frac{2 \sqrt{2} J_L^B J_R^B}{U} \right) 
\left|\beta_{l, B}\right\rangle\!\rangle 
\left\langle\!\langle \beta_{l, B}\right| \right] \\
& + \frac{\sqrt{2}}{U} \sum_{l=1}^{L-1} \left[ 
(J_L^A)^2 \left|\beta_{l, A}\right\rangle\!\rangle 
\left\langle\!\langle \beta_{l+1, A}\right| + 
(J_R^A)^2 \left|\beta_{l+1, A}\right\rangle\!\rangle 
\left\langle\!\langle \beta_{l, A}\right| \right] \\
& + \frac{\sqrt{2}}{U} \sum_{l=1}^{L-1} \left[ 
(J_L^B)^2 \left|\beta_{l, B}\right\rangle\!\rangle 
\left\langle\!\langle \beta_{l+1, B}\right| + 
(J_R^B)^2 \left|\beta_{l+1, B}\right\rangle\!\rangle 
\left\langle\!\langle \beta_{l, B}\right| \right] \\
& + \sum_{l=1,\pm}^L \left[ 
\frac{\sqrt{2} J_p^2}{U\pm+ 2 \mu} \left|\beta_{l, B}\right\rangle\!\rangle 
\left\langle\!\langle \beta_{l, A}\right| + 
\frac{\sqrt{2} J_p^2}{U \pm 2 \mu} \left|\beta_{l, A}\right\rangle\!\rangle 
\left\langle\!\langle \beta_{l, B}\right| \right].
\end{aligned}
\end{equation}
We find that the effective Hamiltonian of bound pair behaves like a single particle and the critical skin effect is naturally occurs. For simplicity, we will set $\mu =0$ in the maintext and we have 
%
\begin{equation}
\begin{aligned}
H_{\text {eff }} &= U+\sum_{l=1}^L\left[\left(\frac{\sqrt{2} J_p^2}{U}+\frac{2 \sqrt{2} J_L^A J_R^A}{U}\right)\left\lvert\beta_{l, A}\right\rangle\!\left\langle\beta_{l, A}\right\rvert\right] \\
&\quad + \sum_{l=1}^L\left[\left(\frac{\sqrt{2} J_p^2}{U}+\frac{2 \sqrt{2} J_L^B J_R^B}{U}\right)\left\lvert\beta_{l, B}\right\rangle\!\left\langle\beta_{l, B}\right\rvert\right] \\
&\quad + \frac{\sqrt{2}}{U} \sum_{l=1}^{L-1}\left[(J_L^A)^2\left\lvert\beta_{l, A}\right\rangle\!\left\langle\beta_{l+1, A}\right\rvert + (J_R^A)^2\left\lvert\beta_{l+1, A}\right\rangle\!\left\langle\beta_{l, A}\right\rvert\right] \\
&\quad + \frac{\sqrt{2}}{U} \sum_{l=1}^{L-1}\left[(J_L^B)^2\left\lvert\beta_{l, B}\right\rangle\!\left\langle\beta_{l+1, B}\right\rvert + (J_R^B)^2\left\lvert\beta_{l+1, B}\right\rangle\!\left\langle\beta_{l, B}\right\rvert\right] \\
&\quad + \frac{2\sqrt{2} J_p^2}{U} \sum_{l=1}^L\left[\left\lvert\beta_{l, B}\right\rangle\!\left\langle\beta_{l, A}\right\rvert + \left\lvert\beta_{l, A}\right\rangle\!\left\langle\beta_{l, B}\right\rvert\right].
\end{aligned}
\end{equation}
Note that in the main text we have $J_L^{A/B}=Je^{\pm\alpha},J_R^{A/B}=Je^{\mp\alpha}$, thus the effective Hamiltonian written as 

\begin{align}
H_{\mathrm{eff}} &= U + \sum_{l=1}^L \left[ \left( \frac{\sqrt{2}J_{p}^{2}}{U} + \frac{2\sqrt{2}J^2}{U} \right) |\beta_{l,A}\rangle \rangle \langle \langle \beta_{l,A}| \right] \nonumber \\
& + \sum_{l=1}^L \left[ \left( \frac{\sqrt{2}J_{p}^{2}}{U} + \frac{2\sqrt{2}J^2}{U} \right) |\beta_{l,B}\rangle \rangle \langle \langle \beta_{l,B}| \right] \nonumber \\
&+ \frac{\sqrt{2}J^2}{U} \sum_{l=1}^{L-1} \left[ e^{2\alpha} |\beta_{l,A}\rangle \rangle \langle \langle \beta_{l+1,A}| + e^{-2\alpha} |\beta_{l+1,A}\rangle \rangle \langle \langle \beta_{l,A}| \right] \nonumber \\
&+ \frac{\sqrt{2}J^2}{U} \sum_{l=1}^{L-1} \left[ e^{-2\alpha} |\beta_{l,B}\rangle \rangle \langle \langle \beta_{l+1,B}| + e^{2\alpha} |\beta_{l+1,B}\rangle \rangle \langle \langle \beta_{l,B}| \right] \nonumber \\
&+ \frac{2\sqrt{2}J_{p}^{2}}{U} \sum_{l=1}^L \left[ |\beta_{l,B}\rangle \rangle \langle \langle \beta_{l,A}| + |\beta_{l,A}\rangle \rangle \langle \langle \beta_{l,B}| \right].
\end{align}

\section{The critical beheaviour for the scattering and bound cluster states}

In the main text, we primarily present the results for a fixed chemical potential in both the scattering and bound CSE cases. However, the critical behavior can indeed be tuned by the onsite interaction \( U \) and the chemical potential \( \mu \), as shown in Fig.~\ref{fig:Umu_phase} (a) and (b). For the scattering cluster, \(\max(\text{Im}E)\) remains unchanged with increasing interaction \( U \) but tends to become real as the chemical potential \( \mu \) increases. In contrast, for the bound state cluster, \(\max(\text{Im}E)\) decreases with increasing interaction strength \( U \) or chemical potential \( \mu \). In both cases, no CSE occurs if the chemical potential \( \mu \) is sufficiently large to separate the overlapping clusters.

\begin{figure}[htbp]
\centering
\includegraphics[width=0.75\linewidth]{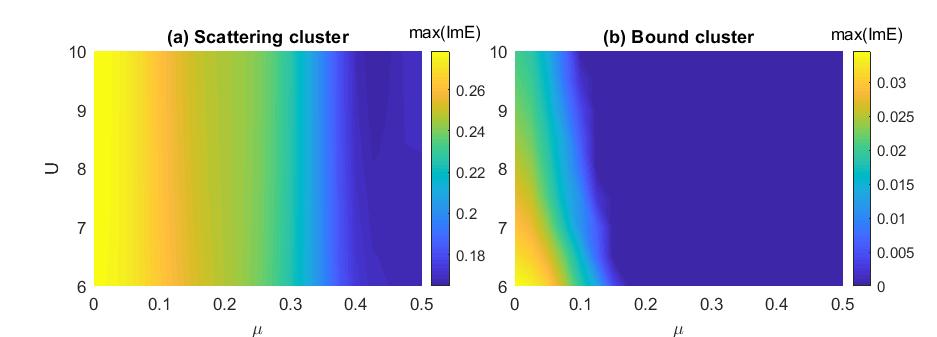}
\caption{\label{fig:Umu_phase} The maximum imaginary energy states with different $U$ and $\mu$. (a) and (b) maximum imaginary energies for the scattering and bound cluster of $L=15,J_p=0.01$ respectively.  Other parameters are the same as in the main text ($Je^{\alpha}=1$ and $Je^{-\alpha}=0.5$).. }
\end{figure}

\section{Density distributions in real space and more examples for mixed CSE }
To better highlighting our discovery, we have focused only on the joint cluster when discussing the mixed CSE in the main text.
In this section, we provide the full information of density distributions and eigenenergies for each cluster at both \( U = 16, \mu = 4 \) and \( U = 8, \mu = 4 \) (the same parameters as in Fig. 3 in the main text), as shown in Fig.~\ref{fig:U8U16L20L30}. 
The transitions that emerge with increasing system size can be clearly see from the density distribution for the LS-RB cluster [Fig. \ref{fig:U8U16L20L30}(a3) and (b3)], but not for the BiS-RB cluster Fig. \ref{fig:U8U16L20L30}(c2) and (d2)].
Additionally, we demonstrate the occurrence of a mixed CSE under negative interaction \( U = -16, \mu = 4 \), as shown in Fig.~\ref{fig:Uf16L30L20}. In this case, the density distribution of the RS-LB cluster evolves from bipolar-localized to right-localized as the system size increases, indicating the emergence of a mixed CSE.
\begin{figure*}[!htbp]
\centering
\includegraphics[width=0.85\linewidth]{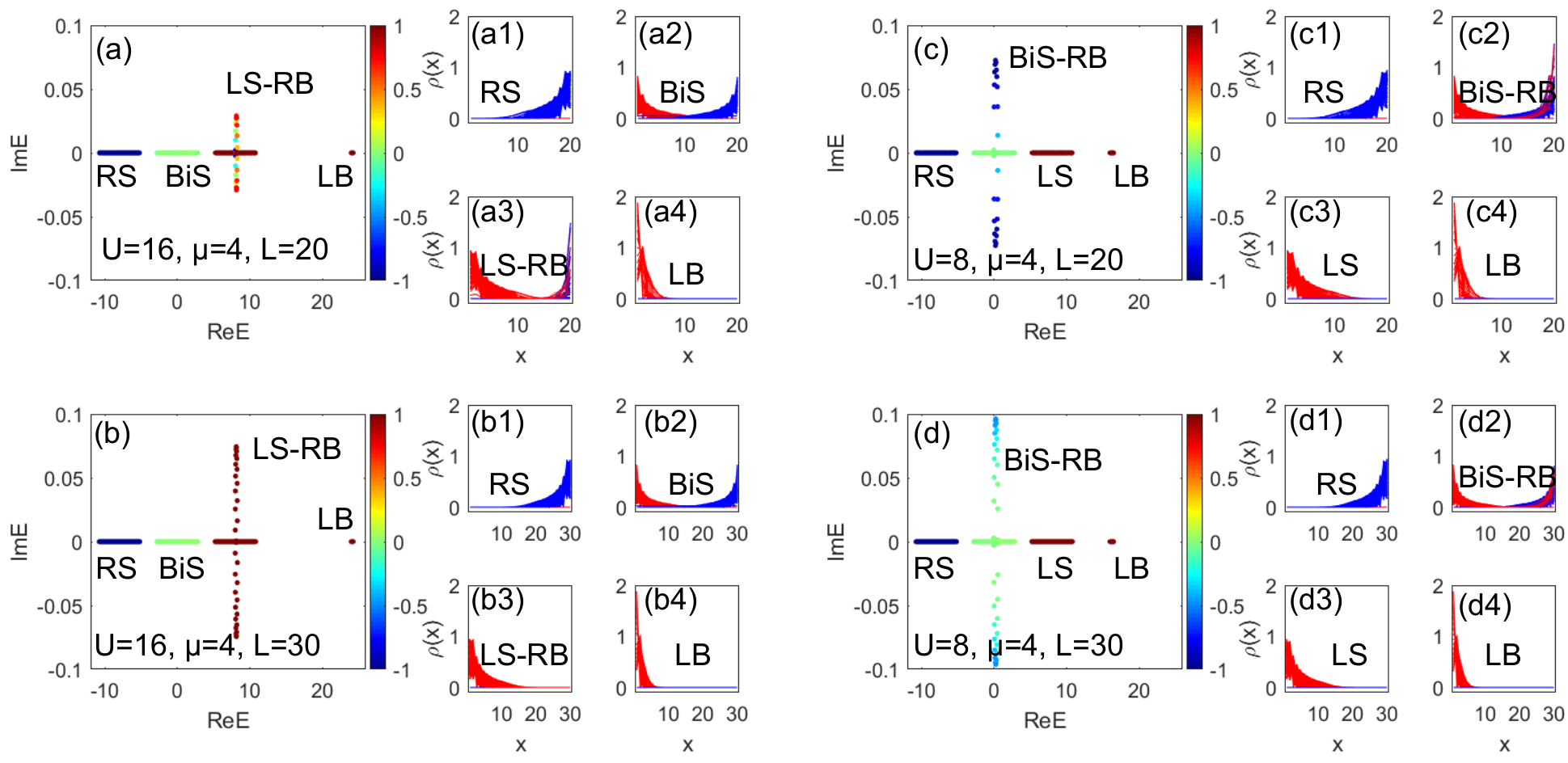}
\caption{\label{fig:U8U16L20L30} The energy spectrum and density distributions versus system size $L$ with $U=16, \mu=4$ and $U=8, \mu=4$ for $N=2$. (a) The energy spectrum for $L=20$ with  $U=16, \mu=4$. (a1-a4) The densities of the energy clusters ordered by the real energy in (a) indicated by RS, BiS, LS-RB, and LB. (b) the same as (a) but for $L=30$.  (c) The energy spectrum for $L=20$ with  $U=8, \mu=4$. (c1-c4) The densities of the energy clusters ordered by the real energy in (c) indicated by RS, BiS-RB, LS, and LB. (d) the same as (c) but for $L=30$. In all density plots, the red (blue) curve represents the density on the A (B) sublattice. $J_p=0.01$ and other parameters are the same as in the main text ($Je^{\alpha}=1$ and $Je^{-\alpha}=0.5$).}
\end{figure*}

Then, we present a more intricate example of the mixed skin effect, involving the mixing of three scattering clusters, where one of the bound-state clusters is embedded within a scattering continuum. The corresponding energy spectrum and density distributions are shown in Fig.~\ref{fig:FigMixedCB_C}. The mixing between the bound-state cluster and the scattering cluster occurs when ${\rm Re} E$ approximatly falls between $2$ and $3$.
The transition from real to complex eigenvalues is computed and illustrated in Fig.~\ref{fig:FigMixedCB_C}(d).


\begin{figure*}[!htbp]
\centering
\includegraphics[width=1\linewidth]{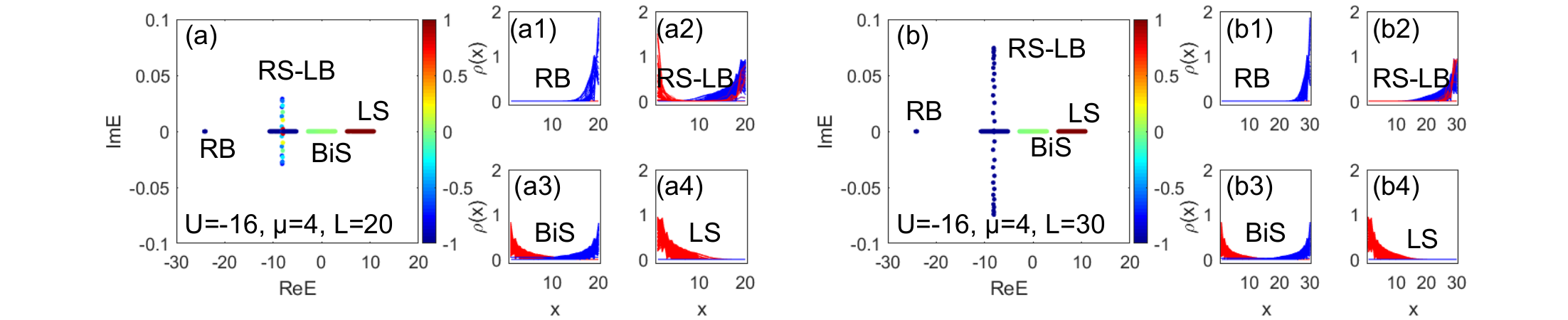}
\caption{\label{fig:Uf16L30L20} The energy spectrum and density distributions versus system size $L$ with $U=-16, \mu=4$ for $N=2$. (a) The energy spectrum for $L=20$. (a1-a4) The densities of the energy clusters ordered by the real energy in (a) indicated by RS, BiS, LS-RB, and LB. (b) the same as (a) but for $L=30$. In all density plots, the red (blue) curve represents the density on the A (B) sublattice.  $J_p=0.01$ and other parameters are the same as in the main text ($Je^{\alpha}=1$ and $Je^{-\alpha}=0.5$).  }
\end{figure*} 


\begin{figure}[htbp]
\centering
\includegraphics[width=0.6\linewidth]{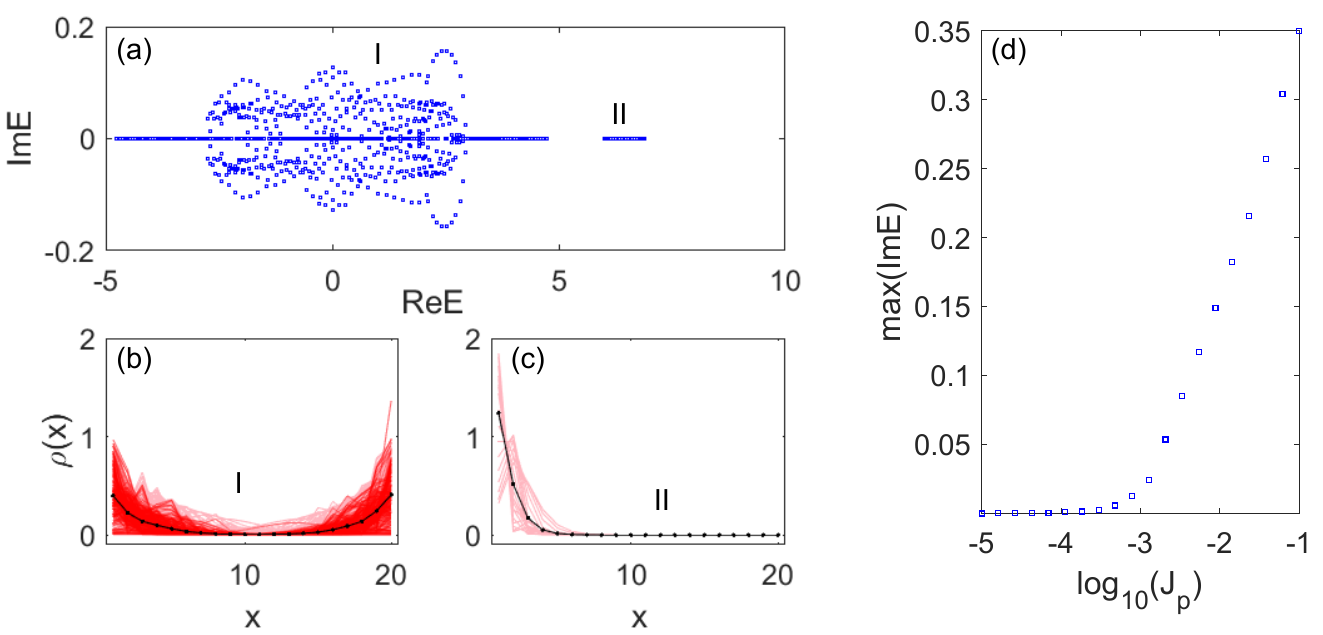}
\caption{\label{fig:FigMixedCB_C} The energy spectrum and density distributions of $L=20$ with $U=4, \mu=1,J_p=0.01$ for $N=2$. (a) The energy spectrum for $L=20$. 
Cluster I consists all the three scattering clusters and one bound cluster in Fig. 3(a) in the main text.
(b-c) The densities of the energy clusters ordered by the real energy in (a). The pink lines are density of each states and the black dot for the average density of each cluster. 
(d) The maximum imaginary energy for the mixed energy cluster I versus coupling strength $J_p$. Other parameters are the same as in the main text ($Je^{\alpha}=1$ and $Je^{-\alpha}=0.5$).  }
\end{figure}


\section{Entropy for the mixed CSE with $U=16,\mu=4$  }
In the main text, 
we have demonstrated the different density distributions at $L=20$ and $L=30$ for the joint cluster that exhibits mixed CSE.
To further investigate the transition with increasing system size,
we calculate the A-B chain entropy $S_A$ and the left-right entropy $S_{\rm left}$ of the largest imaginary energy state of the joint LS-RB cluster ([see Fig. \ref{fig:U8U16L20L30}(a) and (b), and Fig. 3(b) in the main text] with the increasing the system size $L$, as shown in Fig.~\ref{fig:Fig_CBmixedUP}. These quantities are defined as~\cite{Orito2022PRB}
\begin{align}
S_{A}=-{\rm Tr} \rho_{A}\ln \rho_{A}
\end{align}
with $\rho_{A}={\rm Tr}_{B} [\rho]={\rm Tr}_{B}[| \psi\rangle \langle \psi|]$ and 
\begin{align}
S_{\rm left}=-{\rm Tr} \rho_{\rm left}\ln \rho_{\rm left}
\end{align}
with $\rho_{\rm left}={\rm Tr}_{\rm right} [\rho]={\rm Tr}_{\rm right}[| \psi\rangle \langle \psi|]$. Here ``left" (``right") indicates the left (right) half of lattice, and $| \psi\rangle$ represents the eigenstate with maximum imaginary energy. 
We find that $S_A$ and $S_{\rm left}$ approximately take the same value across different system sizes.
Specifically, they increase and reach a maximum value then decay to zero with the increasing of the system size. The transition from nonzero to zero entropy corresponds to that the bipolar-like state occupying both chains evolves to a left-localized state mainly occupying chain A. This transition is further verified by the density ratio on chain A ($\rho_A/N$) and on the left half of the system ($\rho_{\rm left}/N$), which are also shown in Fig.~\ref{fig:Fig_CBmixedUP}. 


\begin{figure*}[!htbp]
\centering
\includegraphics[width=0.6\linewidth]{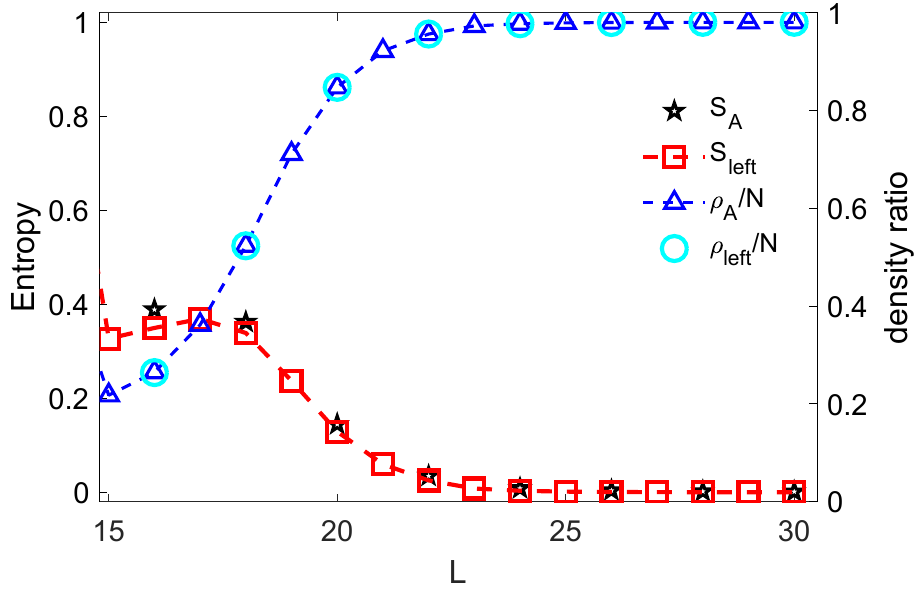}
\caption{\label{fig:Fig_CBmixedUP} 
Entropies and density ratios of the maximum state for the joint LS-RB cluster ([see Fig. \ref{fig:U8U16L20L30}(a) and (b), and Fig. 3(b) in the main text], defined between different chains or different halves of the system. $S_A$ (black star) is the entropy for the A chain by tracing out the B chain. $S_{\rm left}$ (red square) is the entropy for the left-half of the lattice. $\rho_A/N$ (blue triangular) is the ratio of the particle density on chain A to the total density.
$\rho_{\rm left}/N$ (blue triangular) is the ratio of the particle density on the left-half of the lattice to the total particle number $N$.
the A chain density to the total density. $\rho_{\rm left}/N$ (cyan circle) is the ratio of the left-half lattice to $N$. Other parameters are the same as in the main text ($Je^{\alpha}=1$ and $Je^{-\alpha}=0.5$). }
\end{figure*} 

\section{The third order of CSE for $N=3$}
In our study, the emergence of CSE is mainly characterized by the real-complex transition of eigenenergies, which occurs at different values of $J_p$ depending on the orders of CSE, as shown in Fig. 5 of the main text.
In particular, the 3rd-order CSE occurs at a relatively large interchain coupling comparable with other hopping parameters ($J_p=0.8$), which seems to diverge from the critical regime for the single-particle CSE with a vanishing $J_p$.
However, we note that in CSEs, the treshold of $J_p$ to induce the transition is expected to decrease as the system size increases.
In Fig.~\ref{fig:N3Jpc}, we demonstrate the maximum imaginary energy of the states with 3rd-order CSE for several different system sizes $L$.
It is seen that eigenenergies acuqire imaginary values at weaker $J_p$ for larger $L$, verifying the higher order criticality of the concerned states.

\begin{figure}[htbp]
\centering
\includegraphics[width=0.4\linewidth]{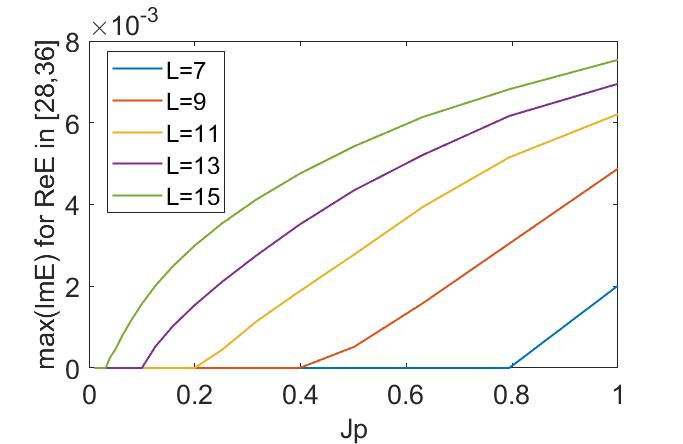}\caption{\label{fig:N3Jpc} 
Maximum imaginary energy of the states with 3rd-order CSE for $N=3$ at $U=16$ and $\mu=5.333$ [indicated in Fig. 5(d) of the main text].
The treshold of $J_p$ to induce nonzero imaginary energies is found to deacrease with increasing system size.
Other parameters are the same as in the main text ($Je^{\alpha}=1$ and $Je^{-\alpha}=0.5$).
}
\end{figure}

\section{Interaction induced correlation for bound state cluster}
In this section, we provide a derivation for the correlation of bound state cluster (see Eq. 5 in the main text). We obtain

\begin{equation}
\begin{aligned}
\Gamma_{xx} 
&= \langle \psi | \hat{b}_x^{\dagger} \hat{b}_x^{\dagger} \hat{b}_x \hat{b}_x | \psi \rangle \\
&= \left( \sum_{l'=1}^L a_{l'}^* \langle l'_2 | \right) 
   \left( \hat{n}_x^2 - \hat{n}_x \right)
   \left( \sum_{l=1}^L a_l | l_2 \rangle \right) \\
&= 2 \sum_{l'=1}^L \sum_{l=1}^L a_{l'}^* a_l \langle l'_2 | l_2 \rangle \delta_{l,x} \\
&= 2 |a_x|^2.
\end{aligned}
\end{equation}
Thus, the first term of $\mathcal{N}_{\rm cor} $ is 
\begin{equation}
\sum_{qr} \Gamma_{qq} \Gamma_{rr} 
= 4 \sum_{qr} |a_q|^2 |a_r|^2 = 4.
\end{equation}
The second term of  $\mathcal{N}_{\rm cor} $ is  
\begin{align}
\Gamma_{qr} 
&= \langle \psi | \hat{b}_q^{\dagger} \hat{b}_r^{\dagger} \hat{b}_r \hat{b}_q | \psi \rangle \notag \\
&= \left( \sum_{l'=1}^L a_{l'}^* \langle l'_2 | \right) \hat{n}_q \hat{n}_r 
   \left( \sum_{l=1}^L a_l | l_2 \rangle \right), \quad q \ne r \notag \\
&= \sum_{l'=1}^L \sum_{l=1}^L a_{l'}^* a_l \langle l'_2 | l_2 \rangle 
   \cdot 4 \delta_{l,q} \delta_{l,r} \notag \\
&= 4 \sum_{l=1}^L |a_l|^2 \delta_{l,q} \delta_{l,r} 
= 4 |a_q|^2 \delta_{r,q} = 0 \quad (q \ne r) \label{eq:Gqr_offdiag}
\end{align}
When $q=r$, we obtain 
$$
q=r:\sum_{qr}{\Gamma _{qr}^{2}}=4L\sum_q{|a_q|}^4. 
$$
Thus, we obtain 
\begin{align}
\sum_{q r} \Gamma_{qr}^2 = 4L\sum_q{|a_q|}^4.  \label{eq:Gqr_sum}
\end{align}
Finally, the correlation function becomes  $\mathcal{N}_{\rm cor}=4(1-L\sum_q{|a_q|}^4) $.
For an evenly distributed state
with $a_q=1/\sqrt{L}$, we have 
$$\mathcal{N}^{\rm even}_{\rm cor}=4(1-1/L),$$

Next we proof that this value is the upper bond of $\mathcal{N}_{\rm cor}$.
Consider the correlation function
\begin{equation}
\mathcal{N}_{\rm cor} = 4\left(1 - L \sum_{q=1}^L |a_q|^4\right),
\end{equation}
under the normalization condition
\begin{equation}
\sum_{q=1}^L |a_q|^2 = 1.
\end{equation}

Let $x_q = |a_q|^2 \geq 0$. Then the problem reduces to maximizing
\begin{equation}
\mathcal{N}_{\rm cor} = 4\left(1 - L \sum_{q=1}^L x_q^2\right)
\quad \text{subject to} \quad \sum_{q=1}^L x_q = 1.
\end{equation}

 Define the Lagrangian:
\begin{equation}
\mathcal{L}(x_1,\dots,x_L, \lambda) = 4\left(1 - L \sum_{q=1}^L x_q^2\right) - \lambda\left( \sum_{q=1}^L x_q - 1 \right).
\end{equation}

Taking derivatives:
\[
\frac{\partial \mathcal{L}}{\partial x_q} = -8L x_q - \lambda = 0 
\quad \Rightarrow \quad x_q = -\frac{\lambda}{8L}.
\]

Thus all $x_q$ are equal. Using the constraint $\sum_{q=1}^L x_q = 1$:
\[
x_q = \frac{1}{L}.
\]

Substitute back into $\mathcal{N}_{\rm cor}$:
\[
\sum_{q=1}^L x_q^2 = L \cdot \left(\frac{1}{L}\right)^2 = \frac{1}{L},
\]
\[
\Rightarrow \mathcal{N}_{\rm cor}^{\max} = 4\left(1 - \frac{1}{L} \right).
\]
Thus, we finally obtain 
\begin{equation}
\boxed{\mathcal{N}_{\rm cor}^{\max} = 4\left(1 - \frac{1}{L} \right)},
\end{equation}
attained when $|a_q|^2 = \frac{1}{L}$ for all $q$. 

\section{The critical skin effect for Fermions}
In the main text  we have provided examples and discussion of many-body CSE for bosons. In this section, we present results for many-body CSE with fermions. The fermoinic analog of our model can be written as 
\begin{equation}
\begin{aligned}
    \hat{H} = & -\sum_{j=1}^{L-1} \sum_{\sigma = A,B} J \left( e^{\alpha_\sigma} \hat{c}_{j,\sigma}^\dagger \hat{c}_{j+1,\sigma} + e^{-\alpha_\sigma} \hat{c}_{j+1,\sigma}^\dagger \hat{c}_{j,\sigma} \right) \\
    & + J_p \sum_{j=1}^L \left( \hat{c}_{j,A}^\dagger \hat{c}_{j,B} + \hat{c}_{j,B}^\dagger \hat{c}_{j,A} \right) \\
    & + U_{NN}\sum_{j=1}^{L-1} \sum_{\sigma = A,B} \hat{n}_{j,\sigma} \hat{n}_{j+1,\sigma} \\
    & + \mu \sum_{j=1}^L \left( \hat{n}_{j,A} - \hat{n}_{j,B} \right),
\end{aligned}
\end{equation}
where $\alpha_A = \alpha, \quad \alpha_B = -\alpha$.
The onsite-Hubbard interaction for bosons is replaced by a nearest-neighbor interaction $U_{NN}$ for fermions, while other parameters remain the same.
The main finding of our paper suggests that CSE can emerge when two parts are spatially separated but energetically connected (facilitating tunneling). 
Here, we present results for two fermions in Fig.~\ref{fig:FermionL}. In the fermionic system, the bound state consists of two nearest-neighbor fermions. The energy spectrum becomes complex with increasing system size, indicating the emergence of many-body CSE.

\begin{figure}[htbp]
\centering
\includegraphics[width=1\linewidth]{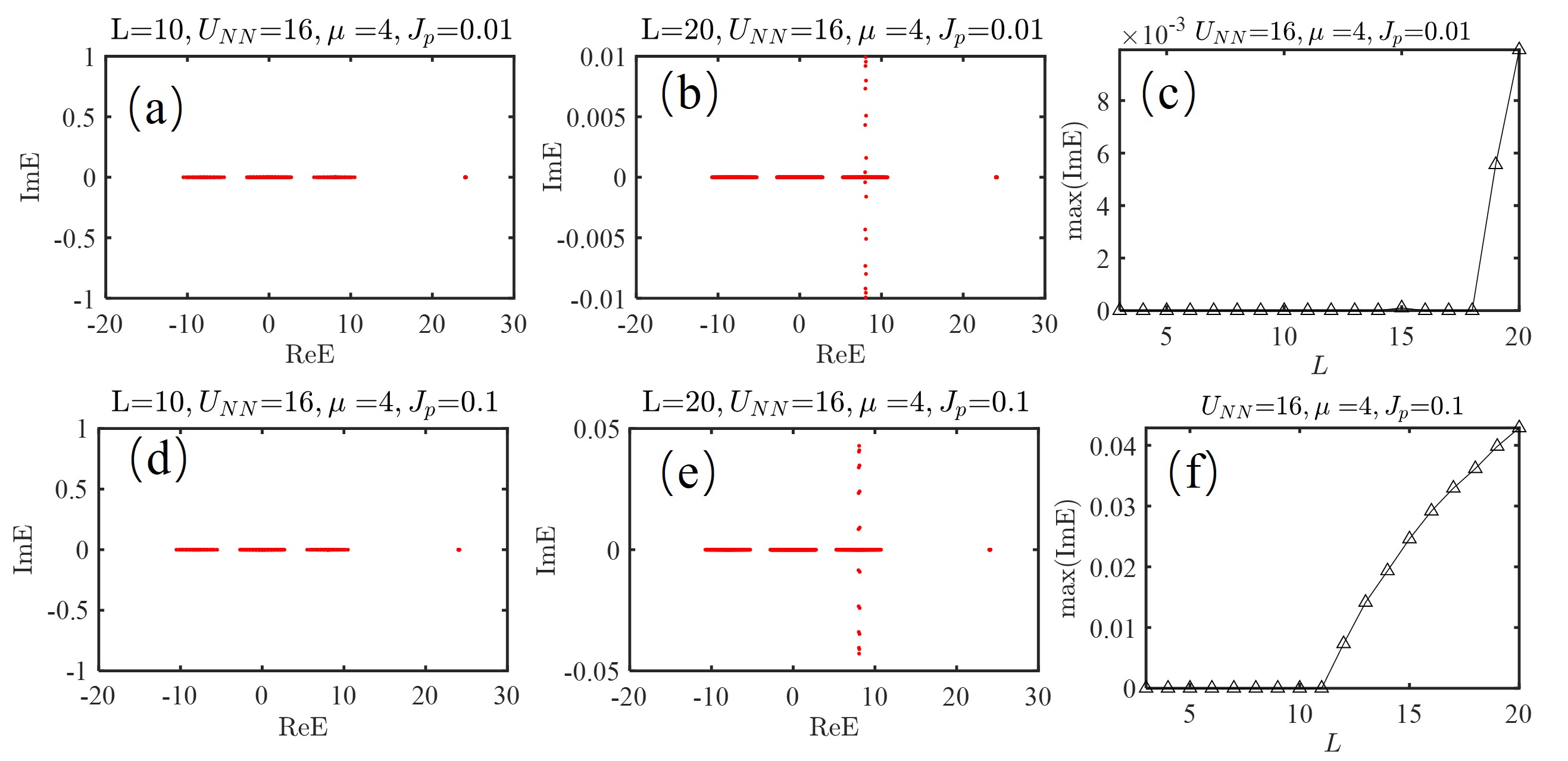}
\caption{\label{fig:FermionL} 
The many-body mixed CSE for fermions. The energy spectrum for (a) $L=10$ and (b) $L=20$, with $U_{NN}=16$, $\mu=4$, and $J_p=0.01$. (c) The maximum imaginary part of the energy versus $L$, where a sharp transition indicates the emergence of the critical skin effect. (d–f) Same as (a–c), but with a larger coupling strength $J_p=0.1$. 
As shown in (f), the real-complex transition of eigenenergies occurs at a smaller $L$ for stronger $J_p$.
Other parameters are the same as in the main text ($Je^{\alpha}=1$ and $Je^{-\alpha}=0.5$).
}
\end{figure}

\input{CNSE_sup_v1.bbl}
\end{widetext}

\end{document}

%% file: CNSE_sup_v1.bbl
%

%% file: CNSE_arxiv.bbl
\begin{thebibliography}{106}%
\makeatletter
\providecommand \@ifxundefined [1]{%
 \@ifx{#1\undefined}
}%
\providecommand \@ifnum [1]{%
 \ifnum #1\expandafter \@firstoftwo
 \else \expandafter \@secondoftwo
 \fi
}%
\providecommand \@ifx [1]{%
 \ifx #1\expandafter \@firstoftwo
 \else \expandafter \@secondoftwo
 \fi
}%
\providecommand \natexlab [1]{#1}%
\providecommand \enquote  [1]{``#1''}%
\providecommand \bibnamefont  [1]{#1}%
\providecommand \bibfnamefont [1]{#1}%
\providecommand \citenamefont [1]{#1}%
\providecommand \href@noop [0]{\@secondoftwo}%
\providecommand \href [0]{\begingroup \@sanitize@url \@href}%
\providecommand \@href[1]{\@@startlink{#1}\@@href}%
\providecommand \@@href[1]{\endgroup#1\@@endlink}%
\providecommand \@sanitize@url [0]{\catcode `\\12\catcode `\$12\catcode
  `\&12\catcode `\#12\catcode `\^12\catcode `\_12\catcode `\%12\relax}%
\providecommand \@@startlink[1]{}%
\providecommand \@@endlink[0]{}%
\providecommand \url  [0]{\begingroup\@sanitize@url \@url }%
\providecommand \@url [1]{\endgroup\@href {#1}{\urlprefix }}%
\providecommand \urlprefix  [0]{URL }%
\providecommand \Eprint [0]{\href }%
\providecommand \doibase [0]{https://doi.org/}%
\providecommand \selectlanguage [0]{\@gobble}%
\providecommand \bibinfo  [0]{\@secondoftwo}%
\providecommand \bibfield  [0]{\@secondoftwo}%
\providecommand \translation [1]{[#1]}%
\providecommand \BibitemOpen [0]{}%
\providecommand \bibitemStop [0]{}%
\providecommand \bibitemNoStop [0]{.\EOS\space}%
\providecommand \EOS [0]{\spacefactor3000\relax}%
\providecommand \BibitemShut  [1]{\csname bibitem#1\endcsname}%
\let\auto@bib@innerbib\@empty
\bibitem [{\citenamefont {Lee}(2016)}]{LeeTony2016PRL}%
  \BibitemOpen
  \bibfield  {author} {\bibinfo {author} {\bibfnamefont {T.~E.}\ \bibnamefont
  {Lee}},\ }\href {https://doi.org/10.1103/PhysRevLett.116.133903} {\bibfield
  {journal} {\bibinfo  {journal} {Phys. Rev. Lett.}\ }\textbf {\bibinfo
  {volume} {116}},\ \bibinfo {pages} {133903} (\bibinfo {year}
  {2016})}\BibitemShut {NoStop}%
\bibitem [{\citenamefont {Martinez~Alvarez}\ \emph {et~al.}(2018)\citenamefont
  {Martinez~Alvarez}, \citenamefont {Barrios~Vargas},\ and\ \citenamefont
  {Foa~Torres}}]{Martinez2018PRB}%
  \BibitemOpen
  \bibfield  {author} {\bibinfo {author} {\bibfnamefont {V.~M.}\ \bibnamefont
  {Martinez~Alvarez}}, \bibinfo {author} {\bibfnamefont {J.~E.}\ \bibnamefont
  {Barrios~Vargas}},\ and\ \bibinfo {author} {\bibfnamefont {L.~E.~F.}\
  \bibnamefont {Foa~Torres}},\ }\href
  {https://doi.org/10.1103/PhysRevB.97.121401} {\bibfield  {journal} {\bibinfo
  {journal} {Phys. Rev. B}\ }\textbf {\bibinfo {volume} {97}},\ \bibinfo
  {pages} {121401} (\bibinfo {year} {2018})}\BibitemShut {NoStop}%
\bibitem [{\citenamefont {Kunst}\ \emph {et~al.}(2018)\citenamefont {Kunst},
  \citenamefont {Edvardsson}, \citenamefont {Budich},\ and\ \citenamefont
  {Bergholtz}}]{KunstPRL2018}%
  \BibitemOpen
  \bibfield  {author} {\bibinfo {author} {\bibfnamefont {F.~K.}\ \bibnamefont
  {Kunst}}, \bibinfo {author} {\bibfnamefont {E.}~\bibnamefont {Edvardsson}},
  \bibinfo {author} {\bibfnamefont {J.~C.}\ \bibnamefont {Budich}},\ and\
  \bibinfo {author} {\bibfnamefont {E.~J.}\ \bibnamefont {Bergholtz}},\ }\href
  {https://doi.org/10.1103/PhysRevLett.121.026808} {\bibfield  {journal}
  {\bibinfo  {journal} {Phys. Rev. Lett.}\ }\textbf {\bibinfo {volume} {121}},\
  \bibinfo {pages} {026808} (\bibinfo {year} {2018})}\BibitemShut {NoStop}%
\bibitem [{\citenamefont {Kawabata}\ \emph {et~al.}(2018)\citenamefont
  {Kawabata}, \citenamefont {Shiozaki},\ and\ \citenamefont
  {Ueda}}]{Kawabata2018PRB}%
  \BibitemOpen
  \bibfield  {author} {\bibinfo {author} {\bibfnamefont {K.}~\bibnamefont
  {Kawabata}}, \bibinfo {author} {\bibfnamefont {K.}~\bibnamefont {Shiozaki}},\
  and\ \bibinfo {author} {\bibfnamefont {M.}~\bibnamefont {Ueda}},\ }\href
  {https://doi.org/10.1103/PhysRevB.98.165148} {\bibfield  {journal} {\bibinfo
  {journal} {Phys. Rev. B}\ }\textbf {\bibinfo {volume} {98}},\ \bibinfo
  {pages} {165148} (\bibinfo {year} {2018})}\BibitemShut {NoStop}%
\bibitem [{\citenamefont {Liu}\ \emph {et~al.}(2019)\citenamefont {Liu},
  \citenamefont {Zhang}, \citenamefont {Ai}, \citenamefont {Gong},
  \citenamefont {Kawabata}, \citenamefont {Ueda},\ and\ \citenamefont
  {Nori}}]{Liu2019PRL}%
  \BibitemOpen
  \bibfield  {author} {\bibinfo {author} {\bibfnamefont {T.}~\bibnamefont
  {Liu}}, \bibinfo {author} {\bibfnamefont {Y.-R.}\ \bibnamefont {Zhang}},
  \bibinfo {author} {\bibfnamefont {Q.}~\bibnamefont {Ai}}, \bibinfo {author}
  {\bibfnamefont {Z.}~\bibnamefont {Gong}}, \bibinfo {author} {\bibfnamefont
  {K.}~\bibnamefont {Kawabata}}, \bibinfo {author} {\bibfnamefont
  {M.}~\bibnamefont {Ueda}},\ and\ \bibinfo {author} {\bibfnamefont
  {F.}~\bibnamefont {Nori}},\ }\href
  {https://doi.org/10.1103/PhysRevLett.122.076801} {\bibfield  {journal}
  {\bibinfo  {journal} {Phys. Rev. Lett.}\ }\textbf {\bibinfo {volume} {122}},\
  \bibinfo {pages} {076801} (\bibinfo {year} {2019})}\BibitemShut {NoStop}%
\bibitem [{\citenamefont {Yao}\ \emph {et~al.}(2018)\citenamefont {Yao},
  \citenamefont {Song},\ and\ \citenamefont {Wang}}]{Yao2018PRL}%
  \BibitemOpen
  \bibfield  {author} {\bibinfo {author} {\bibfnamefont {S.}~\bibnamefont
  {Yao}}, \bibinfo {author} {\bibfnamefont {F.}~\bibnamefont {Song}},\ and\
  \bibinfo {author} {\bibfnamefont {Z.}~\bibnamefont {Wang}},\ }\href
  {https://doi.org/10.1103/PhysRevLett.121.136802} {\bibfield  {journal}
  {\bibinfo  {journal} {Phys. Rev. Lett.}\ }\textbf {\bibinfo {volume} {121}},\
  \bibinfo {pages} {136802} (\bibinfo {year} {2018})}\BibitemShut {NoStop}%
\bibitem [{\citenamefont {Lee}\ and\ \citenamefont {Thomale}(2019)}]{Lee2019}%
  \BibitemOpen
  \bibfield  {author} {\bibinfo {author} {\bibfnamefont {C.~H.}\ \bibnamefont
  {Lee}}\ and\ \bibinfo {author} {\bibfnamefont {R.}~\bibnamefont {Thomale}},\
  }\href {https://doi.org/10.1103/PhysRevB.99.201103} {\bibfield  {journal}
  {\bibinfo  {journal} {Phys. Rev. B}\ }\textbf {\bibinfo {volume} {99}},\
  \bibinfo {pages} {201103} (\bibinfo {year} {2019})}\BibitemShut {NoStop}%
\bibitem [{\citenamefont {Okuma}\ and\ \citenamefont
  {Sato}(2023)}]{okuma2023non}%
  \BibitemOpen
  \bibfield  {author} {\bibinfo {author} {\bibfnamefont {N.}~\bibnamefont
  {Okuma}}\ and\ \bibinfo {author} {\bibfnamefont {M.}~\bibnamefont {Sato}},\
  }\href@noop {} {\bibfield  {journal} {\bibinfo  {journal} {Annual Review of
  Condensed Matter Physics}\ }\textbf {\bibinfo {volume} {14}},\ \bibinfo
  {pages} {83} (\bibinfo {year} {2023})}\BibitemShut {NoStop}%
\bibitem [{\citenamefont {Zhang}\ \emph {et~al.}(2022)\citenamefont {Zhang},
  \citenamefont {Zhang}, \citenamefont {Lu},\ and\ \citenamefont
  {Chen}}]{zhang2022review}%
  \BibitemOpen
  \bibfield  {author} {\bibinfo {author} {\bibfnamefont {X.}~\bibnamefont
  {Zhang}}, \bibinfo {author} {\bibfnamefont {T.}~\bibnamefont {Zhang}},
  \bibinfo {author} {\bibfnamefont {M.-H.}\ \bibnamefont {Lu}},\ and\ \bibinfo
  {author} {\bibfnamefont {Y.-F.}\ \bibnamefont {Chen}},\ }\href@noop {}
  {\bibfield  {journal} {\bibinfo  {journal} {Advances in Physics: X}\ }\textbf
  {\bibinfo {volume} {7}},\ \bibinfo {pages} {2109431} (\bibinfo {year}
  {2022})}\BibitemShut {NoStop}%
\bibitem [{\citenamefont {Lin}\ \emph {et~al.}(2023)\citenamefont {Lin},
  \citenamefont {Tai}, \citenamefont {Li},\ and\ \citenamefont
  {Lee}}]{lin2023top}%
  \BibitemOpen
  \bibfield  {author} {\bibinfo {author} {\bibfnamefont {R.}~\bibnamefont
  {Lin}}, \bibinfo {author} {\bibfnamefont {T.}~\bibnamefont {Tai}}, \bibinfo
  {author} {\bibfnamefont {L.}~\bibnamefont {Li}},\ and\ \bibinfo {author}
  {\bibfnamefont {C.~H.}\ \bibnamefont {Lee}},\ }\href@noop {} {\bibfield
  {journal} {\bibinfo  {journal} {Frontiers of Physics}\ }\textbf {\bibinfo
  {volume} {18}},\ \bibinfo {pages} {53605} (\bibinfo {year}
  {2023})}\BibitemShut {NoStop}%
\bibitem [{\citenamefont {Li}\ \emph {et~al.}(2020)\citenamefont {Li},
  \citenamefont {Lee}, \citenamefont {Mu},\ and\ \citenamefont
  {Gong}}]{Li2020NC}%
  \BibitemOpen
  \bibfield  {author} {\bibinfo {author} {\bibfnamefont {L.}~\bibnamefont
  {Li}}, \bibinfo {author} {\bibfnamefont {C.~H.}\ \bibnamefont {Lee}},
  \bibinfo {author} {\bibfnamefont {S.}~\bibnamefont {Mu}},\ and\ \bibinfo
  {author} {\bibfnamefont {J.}~\bibnamefont {Gong}},\ }\bibfield  {journal}
  {\bibinfo  {journal} {Nature Communications}\ }\textbf {\bibinfo {volume}
  {11}},\ \href {https://doi.org/10.1038/s41467-020-18917-4}
  {10.1038/s41467-020-18917-4} (\bibinfo {year} {2020})\BibitemShut {NoStop}%
\bibitem [{\citenamefont {Liu}\ \emph {et~al.}(2021)\citenamefont {Liu},
  \citenamefont {Zeng}, \citenamefont {Li},\ and\ \citenamefont
  {Chen}}]{Liu2021}%
  \BibitemOpen
  \bibfield  {author} {\bibinfo {author} {\bibfnamefont {Y.}~\bibnamefont
  {Liu}}, \bibinfo {author} {\bibfnamefont {Y.}~\bibnamefont {Zeng}}, \bibinfo
  {author} {\bibfnamefont {L.}~\bibnamefont {Li}},\ and\ \bibinfo {author}
  {\bibfnamefont {S.}~\bibnamefont {Chen}},\ }\href
  {https://doi.org/10.1103/PhysRevB.104.085401} {\bibfield  {journal} {\bibinfo
   {journal} {Phys. Rev. B}\ }\textbf {\bibinfo {volume} {104}},\ \bibinfo
  {pages} {085401} (\bibinfo {year} {2021})}\BibitemShut {NoStop}%
\bibitem [{\citenamefont {Rafi-Ul-Islam}\ \emph {et~al.}(2022)\citenamefont
  {Rafi-Ul-Islam}, \citenamefont {Siu}, \citenamefont {Sahin}, \citenamefont
  {Lee},\ and\ \citenamefont {Jalil}}]{rafi2022CSE}%
  \BibitemOpen
  \bibfield  {author} {\bibinfo {author} {\bibfnamefont {S.}~\bibnamefont
  {Rafi-Ul-Islam}}, \bibinfo {author} {\bibfnamefont {Z.~B.}\ \bibnamefont
  {Siu}}, \bibinfo {author} {\bibfnamefont {H.}~\bibnamefont {Sahin}}, \bibinfo
  {author} {\bibfnamefont {C.~H.}\ \bibnamefont {Lee}},\ and\ \bibinfo {author}
  {\bibfnamefont {M.~B.}\ \bibnamefont {Jalil}},\ }\href@noop {} {\bibfield
  {journal} {\bibinfo  {journal} {Physical Review Research}\ }\textbf {\bibinfo
  {volume} {4}},\ \bibinfo {pages} {013243} (\bibinfo {year}
  {2022})}\BibitemShut {NoStop}%
\bibitem [{\citenamefont {Rafi-Ul-Islam}\ \emph {et~al.}(2025)\citenamefont
  {Rafi-Ul-Islam}, \citenamefont {Siu}, \citenamefont {Razo},\ and\
  \citenamefont {Jalil}}]{rafi2025critical}%
  \BibitemOpen
  \bibfield  {author} {\bibinfo {author} {\bibfnamefont {S.}~\bibnamefont
  {Rafi-Ul-Islam}}, \bibinfo {author} {\bibfnamefont {Z.~B.}\ \bibnamefont
  {Siu}}, \bibinfo {author} {\bibfnamefont {M.~S.~H.}\ \bibnamefont {Razo}},\
  and\ \bibinfo {author} {\bibfnamefont {M.~B.}\ \bibnamefont {Jalil}},\
  }\href@noop {} {\bibfield  {journal} {\bibinfo  {journal} {Physical Review
  B}\ }\textbf {\bibinfo {volume} {111}},\ \bibinfo {pages} {115415} (\bibinfo
  {year} {2025})}\BibitemShut {NoStop}%
\bibitem [{\citenamefont {Liu}\ \emph {et~al.}(2024)\citenamefont {Liu},
  \citenamefont {Jiang}, \citenamefont {Xue}, \citenamefont {Li}, \citenamefont
  {Gong}, \citenamefont {Liu},\ and\ \citenamefont {Lee}}]{liu2024non}%
  \BibitemOpen
  \bibfield  {author} {\bibinfo {author} {\bibfnamefont {S.}~\bibnamefont
  {Liu}}, \bibinfo {author} {\bibfnamefont {H.}~\bibnamefont {Jiang}}, \bibinfo
  {author} {\bibfnamefont {W.-T.}\ \bibnamefont {Xue}}, \bibinfo {author}
  {\bibfnamefont {Q.}~\bibnamefont {Li}}, \bibinfo {author} {\bibfnamefont
  {J.}~\bibnamefont {Gong}}, \bibinfo {author} {\bibfnamefont {X.}~\bibnamefont
  {Liu}},\ and\ \bibinfo {author} {\bibfnamefont {C.~H.}\ \bibnamefont {Lee}},\
  }\href@noop {} {\bibfield  {journal} {\bibinfo  {journal} {arXiv preprint
  arXiv:2408.02736}\ } (\bibinfo {year} {2024})}\BibitemShut {NoStop}%
\bibitem [{\citenamefont {Yang}\ and\ \citenamefont
  {Lee}(2024)}]{yang2024percolation}%
  \BibitemOpen
  \bibfield  {author} {\bibinfo {author} {\bibfnamefont {M.}~\bibnamefont
  {Yang}}\ and\ \bibinfo {author} {\bibfnamefont {C.~H.}\ \bibnamefont {Lee}},\
  }\href@noop {} {\bibfield  {journal} {\bibinfo  {journal} {Physical Review
  Letters}\ }\textbf {\bibinfo {volume} {133}},\ \bibinfo {pages} {136602}
  (\bibinfo {year} {2024})}\BibitemShut {NoStop}%
\bibitem [{\citenamefont {Xu}\ \emph {et~al.}(2025)\citenamefont {Xu},
  \citenamefont {Tian}, \citenamefont {An}, \citenamefont {Xiong},\ and\
  \citenamefont {Ghosh}}]{xu2025exciton}%
  \BibitemOpen
  \bibfield  {author} {\bibinfo {author} {\bibfnamefont {X.}~\bibnamefont
  {Xu}}, \bibinfo {author} {\bibfnamefont {L.}~\bibnamefont {Tian}}, \bibinfo
  {author} {\bibfnamefont {Z.}~\bibnamefont {An}}, \bibinfo {author}
  {\bibfnamefont {Q.}~\bibnamefont {Xiong}},\ and\ \bibinfo {author}
  {\bibfnamefont {S.}~\bibnamefont {Ghosh}},\ }\href@noop {} {\bibfield
  {journal} {\bibinfo  {journal} {Physical Review B}\ }\textbf {\bibinfo
  {volume} {111}},\ \bibinfo {pages} {L121301} (\bibinfo {year}
  {2025})}\BibitemShut {NoStop}%
\bibitem [{\citenamefont {Cai}\ \emph {et~al.}(2024)\citenamefont {Cai},
  \citenamefont {Liu},\ and\ \citenamefont {Yang}}]{cai2024non}%
  \BibitemOpen
  \bibfield  {author} {\bibinfo {author} {\bibfnamefont {Z.-F.}\ \bibnamefont
  {Cai}}, \bibinfo {author} {\bibfnamefont {T.}~\bibnamefont {Liu}},\ and\
  \bibinfo {author} {\bibfnamefont {Z.}~\bibnamefont {Yang}},\ }\href@noop {}
  {\bibfield  {journal} {\bibinfo  {journal} {Physical Review A}\ }\textbf
  {\bibinfo {volume} {109}},\ \bibinfo {pages} {063329} (\bibinfo {year}
  {2024})}\BibitemShut {NoStop}%
\bibitem [{\citenamefont {Wei}\ \emph {et~al.}(2025)\citenamefont {Wei},
  \citenamefont {Fan}, \citenamefont {Cao}, \citenamefont {Ma},\ and\
  \citenamefont {Kou}}]{wei2025generalized}%
  \BibitemOpen
  \bibfield  {author} {\bibinfo {author} {\bibfnamefont {Z.}~\bibnamefont
  {Wei}}, \bibinfo {author} {\bibfnamefont {J.-Y.}\ \bibnamefont {Fan}},
  \bibinfo {author} {\bibfnamefont {K.}~\bibnamefont {Cao}}, \bibinfo {author}
  {\bibfnamefont {X.-R.}\ \bibnamefont {Ma}},\ and\ \bibinfo {author}
  {\bibfnamefont {S.-P.}\ \bibnamefont {Kou}},\ }\href@noop {} {\bibfield
  {journal} {\bibinfo  {journal} {arXiv preprint arXiv:2505.10252}\ } (\bibinfo
  {year} {2025})}\BibitemShut {NoStop}%
\bibitem [{\citenamefont {Yao}\ and\ \citenamefont {Wang}(2018)}]{Yao2018}%
  \BibitemOpen
  \bibfield  {author} {\bibinfo {author} {\bibfnamefont {S.}~\bibnamefont
  {Yao}}\ and\ \bibinfo {author} {\bibfnamefont {Z.}~\bibnamefont {Wang}},\
  }\href {https://doi.org/10.1103/PhysRevLett.121.086803} {\bibfield  {journal}
  {\bibinfo  {journal} {Phys. Rev. Lett.}\ }\textbf {\bibinfo {volume} {121}},\
  \bibinfo {pages} {086803} (\bibinfo {year} {2018})}\BibitemShut {NoStop}%
\bibitem [{\citenamefont {McDonald}\ \emph {et~al.}(2018)\citenamefont
  {McDonald}, \citenamefont {Pereg-Barnea},\ and\ \citenamefont
  {Clerk}}]{McDonald2018}%
  \BibitemOpen
  \bibfield  {author} {\bibinfo {author} {\bibfnamefont {A.}~\bibnamefont
  {McDonald}}, \bibinfo {author} {\bibfnamefont {T.}~\bibnamefont
  {Pereg-Barnea}},\ and\ \bibinfo {author} {\bibfnamefont {A.~A.}\ \bibnamefont
  {Clerk}},\ }\href {https://doi.org/10.1103/PhysRevX.8.041031} {\bibfield
  {journal} {\bibinfo  {journal} {Phys. Rev. X}\ }\textbf {\bibinfo {volume}
  {8}},\ \bibinfo {pages} {041031} (\bibinfo {year} {2018})}\BibitemShut
  {NoStop}%
\bibitem [{\citenamefont {Kawabata}\ \emph {et~al.}(2019)\citenamefont
  {Kawabata}, \citenamefont {Shiozaki}, \citenamefont {Ueda},\ and\
  \citenamefont {Sato}}]{Kawabata2019}%
  \BibitemOpen
  \bibfield  {author} {\bibinfo {author} {\bibfnamefont {K.}~\bibnamefont
  {Kawabata}}, \bibinfo {author} {\bibfnamefont {K.}~\bibnamefont {Shiozaki}},
  \bibinfo {author} {\bibfnamefont {M.}~\bibnamefont {Ueda}},\ and\ \bibinfo
  {author} {\bibfnamefont {M.}~\bibnamefont {Sato}},\ }\href
  {https://doi.org/10.1103/PhysRevX.9.041015} {\bibfield  {journal} {\bibinfo
  {journal} {Phys. Rev. X}\ }\textbf {\bibinfo {volume} {9}},\ \bibinfo {pages}
  {041015} (\bibinfo {year} {2019})}\BibitemShut {NoStop}%
\bibitem [{\citenamefont {Zhang}\ \emph {et~al.}(2020)\citenamefont {Zhang},
  \citenamefont {Yang},\ and\ \citenamefont {Fang}}]{Zhang2020}%
  \BibitemOpen
  \bibfield  {author} {\bibinfo {author} {\bibfnamefont {K.}~\bibnamefont
  {Zhang}}, \bibinfo {author} {\bibfnamefont {Z.}~\bibnamefont {Yang}},\ and\
  \bibinfo {author} {\bibfnamefont {C.}~\bibnamefont {Fang}},\ }\href
  {https://doi.org/10.1103/PhysRevLett.125.126402} {\bibfield  {journal}
  {\bibinfo  {journal} {Phys. Rev. Lett.}\ }\textbf {\bibinfo {volume} {125}},\
  \bibinfo {pages} {126402} (\bibinfo {year} {2020})}\BibitemShut {NoStop}%
\bibitem [{\citenamefont {Okuma}\ \emph {et~al.}(2020)\citenamefont {Okuma},
  \citenamefont {Kawabata}, \citenamefont {Shiozaki},\ and\ \citenamefont
  {Sato}}]{Okuma2020}%
  \BibitemOpen
  \bibfield  {author} {\bibinfo {author} {\bibfnamefont {N.}~\bibnamefont
  {Okuma}}, \bibinfo {author} {\bibfnamefont {K.}~\bibnamefont {Kawabata}},
  \bibinfo {author} {\bibfnamefont {K.}~\bibnamefont {Shiozaki}},\ and\
  \bibinfo {author} {\bibfnamefont {M.}~\bibnamefont {Sato}},\ }\href
  {https://doi.org/10.1103/PhysRevLett.124.086801} {\bibfield  {journal}
  {\bibinfo  {journal} {Phys. Rev. Lett.}\ }\textbf {\bibinfo {volume} {124}},\
  \bibinfo {pages} {086801} (\bibinfo {year} {2020})}\BibitemShut {NoStop}%
\bibitem [{\citenamefont {Li}\ and\ \citenamefont {Lee}(2022)}]{li2022non}%
  \BibitemOpen
  \bibfield  {author} {\bibinfo {author} {\bibfnamefont {L.}~\bibnamefont
  {Li}}\ and\ \bibinfo {author} {\bibfnamefont {C.~H.}\ \bibnamefont {Lee}},\
  }\href@noop {} {\bibfield  {journal} {\bibinfo  {journal} {Science Bulletin}\
  }\textbf {\bibinfo {volume} {67}},\ \bibinfo {pages} {685} (\bibinfo {year}
  {2022})}\BibitemShut {NoStop}%
\bibitem [{\citenamefont {Budich}\ and\ \citenamefont
  {Bergholtz}(2020)}]{Budich2020}%
  \BibitemOpen
  \bibfield  {author} {\bibinfo {author} {\bibfnamefont {J.~C.}\ \bibnamefont
  {Budich}}\ and\ \bibinfo {author} {\bibfnamefont {E.~J.}\ \bibnamefont
  {Bergholtz}},\ }\href {https://doi.org/10.1103/PhysRevLett.125.180403}
  {\bibfield  {journal} {\bibinfo  {journal} {Phys. Rev. Lett.}\ }\textbf
  {\bibinfo {volume} {125}},\ \bibinfo {pages} {180403} (\bibinfo {year}
  {2020})}\BibitemShut {NoStop}%
\bibitem [{\citenamefont {Li}\ \emph {et~al.}(2021)\citenamefont {Li},
  \citenamefont {Lee},\ and\ \citenamefont {Gong}}]{Li2021}%
  \BibitemOpen
  \bibfield  {author} {\bibinfo {author} {\bibfnamefont {L.}~\bibnamefont
  {Li}}, \bibinfo {author} {\bibfnamefont {C.~H.}\ \bibnamefont {Lee}},\ and\
  \bibinfo {author} {\bibfnamefont {J.}~\bibnamefont {Gong}},\ }\href
  {https://doi.org/10.1038/s42005-021-00547-x} {\bibfield  {journal} {\bibinfo
  {journal} {Commun Phys}\ }\textbf {\bibinfo {volume} {4}},\ \bibinfo {pages}
  {42} (\bibinfo {year} {2021})}\BibitemShut {NoStop}%
\bibitem [{\citenamefont {Roccati}(2021)}]{Roccati2021}%
  \BibitemOpen
  \bibfield  {author} {\bibinfo {author} {\bibfnamefont {F.}~\bibnamefont
  {Roccati}},\ }\href {https://doi.org/10.1103/PhysRevA.104.022215} {\bibfield
  {journal} {\bibinfo  {journal} {Phys. Rev. A}\ }\textbf {\bibinfo {volume}
  {104}},\ \bibinfo {pages} {022215} (\bibinfo {year} {2021})}\BibitemShut
  {NoStop}%
\bibitem [{\citenamefont {Yang}\ \emph {et~al.}(2022)\citenamefont {Yang},
  \citenamefont {Tan}, \citenamefont {Tai}, \citenamefont {Koh}, \citenamefont
  {Li}, \citenamefont {Longhi},\ and\ \citenamefont {Lee}}]{yang2022designing}%
  \BibitemOpen
  \bibfield  {author} {\bibinfo {author} {\bibfnamefont {R.}~\bibnamefont
  {Yang}}, \bibinfo {author} {\bibfnamefont {J.~W.}\ \bibnamefont {Tan}},
  \bibinfo {author} {\bibfnamefont {T.}~\bibnamefont {Tai}}, \bibinfo {author}
  {\bibfnamefont {J.~M.}\ \bibnamefont {Koh}}, \bibinfo {author} {\bibfnamefont
  {L.}~\bibnamefont {Li}}, \bibinfo {author} {\bibfnamefont {S.}~\bibnamefont
  {Longhi}},\ and\ \bibinfo {author} {\bibfnamefont {C.~H.}\ \bibnamefont
  {Lee}},\ }\href@noop {} {\bibfield  {journal} {\bibinfo  {journal} {Science
  Bulletin}\ }\textbf {\bibinfo {volume} {67}},\ \bibinfo {pages} {1865}
  (\bibinfo {year} {2022})}\BibitemShut {NoStop}%
\bibitem [{\citenamefont {Mu}\ \emph {et~al.}(2022)\citenamefont {Mu},
  \citenamefont {Zhou}, \citenamefont {Li},\ and\ \citenamefont
  {Gong}}]{Mu2022}%
  \BibitemOpen
  \bibfield  {author} {\bibinfo {author} {\bibfnamefont {S.}~\bibnamefont
  {Mu}}, \bibinfo {author} {\bibfnamefont {L.}~\bibnamefont {Zhou}}, \bibinfo
  {author} {\bibfnamefont {L.}~\bibnamefont {Li}},\ and\ \bibinfo {author}
  {\bibfnamefont {J.}~\bibnamefont {Gong}},\ }\href
  {https://doi.org/10.1103/PhysRevB.105.205402} {\bibfield  {journal} {\bibinfo
   {journal} {Phys. Rev. B}\ }\textbf {\bibinfo {volume} {105}},\ \bibinfo
  {pages} {205402} (\bibinfo {year} {2022})}\BibitemShut {NoStop}%
\bibitem [{\citenamefont {Arouca}\ \emph {et~al.}(2020)\citenamefont {Arouca},
  \citenamefont {Lee},\ and\ \citenamefont
  {Morais~Smith}}]{arouca2020unconventional}%
  \BibitemOpen
  \bibfield  {author} {\bibinfo {author} {\bibfnamefont {R.}~\bibnamefont
  {Arouca}}, \bibinfo {author} {\bibfnamefont {C.}~\bibnamefont {Lee}},\ and\
  \bibinfo {author} {\bibfnamefont {C.}~\bibnamefont {Morais~Smith}},\
  }\href@noop {} {\bibfield  {journal} {\bibinfo  {journal} {Physical Review
  B}\ }\textbf {\bibinfo {volume} {102}},\ \bibinfo {pages} {245145} (\bibinfo
  {year} {2020})}\BibitemShut {NoStop}%
\bibitem [{\citenamefont {Lei}\ \emph {et~al.}(2024)\citenamefont {Lei},
  \citenamefont {Lee},\ and\ \citenamefont {Li}}]{lei2024activating}%
  \BibitemOpen
  \bibfield  {author} {\bibinfo {author} {\bibfnamefont {Z.}~\bibnamefont
  {Lei}}, \bibinfo {author} {\bibfnamefont {C.~H.}\ \bibnamefont {Lee}},\ and\
  \bibinfo {author} {\bibfnamefont {L.}~\bibnamefont {Li}},\ }\href@noop {}
  {\bibfield  {journal} {\bibinfo  {journal} {Communications Physics}\ }\textbf
  {\bibinfo {volume} {7}},\ \bibinfo {pages} {100} (\bibinfo {year}
  {2024})}\BibitemShut {NoStop}%
\bibitem [{\citenamefont {Tai}\ and\ \citenamefont
  {Lee}(2023)}]{tai2023zoology}%
  \BibitemOpen
  \bibfield  {author} {\bibinfo {author} {\bibfnamefont {T.}~\bibnamefont
  {Tai}}\ and\ \bibinfo {author} {\bibfnamefont {C.~H.}\ \bibnamefont {Lee}},\
  }\href {https://doi.org/10.1103/PhysRevB.107.L220301} {\bibfield  {journal}
  {\bibinfo  {journal} {Phys. Rev. B}\ }\textbf {\bibinfo {volume} {107}},\
  \bibinfo {pages} {L220301} (\bibinfo {year} {2023})}\BibitemShut {NoStop}%
\bibitem [{\citenamefont {Qin}\ \emph {et~al.}(2023)\citenamefont {Qin},
  \citenamefont {Ma}, \citenamefont {Shen},\ and\ \citenamefont
  {Lee}}]{qin2023universal}%
  \BibitemOpen
  \bibfield  {author} {\bibinfo {author} {\bibfnamefont {F.}~\bibnamefont
  {Qin}}, \bibinfo {author} {\bibfnamefont {Y.}~\bibnamefont {Ma}}, \bibinfo
  {author} {\bibfnamefont {R.}~\bibnamefont {Shen}},\ and\ \bibinfo {author}
  {\bibfnamefont {C.~H.}\ \bibnamefont {Lee}},\ }\href
  {https://doi.org/10.1103/PhysRevB.107.155430} {\bibfield  {journal} {\bibinfo
   {journal} {Phys. Rev. B}\ }\textbf {\bibinfo {volume} {107}},\ \bibinfo
  {pages} {155430} (\bibinfo {year} {2023})}\BibitemShut {NoStop}%
\bibitem [{\citenamefont {Qin}\ \emph {et~al.}()\citenamefont {Qin},
  \citenamefont {Shen}, \citenamefont {Li},\ and\ \citenamefont
  {Lee}}]{qin2023kinked}%
  \BibitemOpen
  \bibfield  {author} {\bibinfo {author} {\bibfnamefont {F.}~\bibnamefont
  {Qin}}, \bibinfo {author} {\bibfnamefont {R.}~\bibnamefont {Shen}}, \bibinfo
  {author} {\bibfnamefont {L.}~\bibnamefont {Li}},\ and\ \bibinfo {author}
  {\bibfnamefont {C.~H.}\ \bibnamefont {Lee}},\ }\href@noop {} {\ }\Eprint
  {https://arxiv.org/abs/2306.13139v1} {2306.13139v1} \BibitemShut {NoStop}%
\bibitem [{\citenamefont {Borgnia}\ \emph {et~al.}(2020)\citenamefont
  {Borgnia}, \citenamefont {Kruchkov},\ and\ \citenamefont
  {Slager}}]{Borgnia2020}%
  \BibitemOpen
  \bibfield  {author} {\bibinfo {author} {\bibfnamefont {D.~S.}\ \bibnamefont
  {Borgnia}}, \bibinfo {author} {\bibfnamefont {A.~J.}\ \bibnamefont
  {Kruchkov}},\ and\ \bibinfo {author} {\bibfnamefont {R.-J.}\ \bibnamefont
  {Slager}},\ }\href {https://doi.org/10.1103/PhysRevLett.124.056802}
  {\bibfield  {journal} {\bibinfo  {journal} {Phys. Rev. Lett.}\ }\textbf
  {\bibinfo {volume} {124}},\ \bibinfo {pages} {056802} (\bibinfo {year}
  {2020})}\BibitemShut {NoStop}%
\bibitem [{\citenamefont {Deng}\ and\ \citenamefont
  {Flebus}(2022)}]{Deng2022PRB}%
  \BibitemOpen
  \bibfield  {author} {\bibinfo {author} {\bibfnamefont {K.}~\bibnamefont
  {Deng}}\ and\ \bibinfo {author} {\bibfnamefont {B.}~\bibnamefont {Flebus}},\
  }\href {https://doi.org/10.1103/PhysRevB.105.L180406} {\bibfield  {journal}
  {\bibinfo  {journal} {Phys. Rev. B}\ }\textbf {\bibinfo {volume} {105}},\
  \bibinfo {pages} {L180406} (\bibinfo {year} {2022})}\BibitemShut {NoStop}%
\bibitem [{\citenamefont {Roccati}\ \emph {et~al.}(2023)\citenamefont
  {Roccati}, \citenamefont {Bello}, \citenamefont {Gong}, \citenamefont {Ueda},
  \citenamefont {Ciccarello}, \citenamefont {Chenu},\ and\ \citenamefont
  {Carollo}}]{roccati2023hermitian}%
  \BibitemOpen
  \bibfield  {author} {\bibinfo {author} {\bibfnamefont {F.}~\bibnamefont
  {Roccati}}, \bibinfo {author} {\bibfnamefont {M.}~\bibnamefont {Bello}},
  \bibinfo {author} {\bibfnamefont {Z.}~\bibnamefont {Gong}}, \bibinfo {author}
  {\bibfnamefont {M.}~\bibnamefont {Ueda}}, \bibinfo {author} {\bibfnamefont
  {F.}~\bibnamefont {Ciccarello}}, \bibinfo {author} {\bibfnamefont
  {A.}~\bibnamefont {Chenu}},\ and\ \bibinfo {author} {\bibfnamefont
  {A.}~\bibnamefont {Carollo}},\ }\href@noop {} {\bibfield  {journal} {\bibinfo
   {journal} {arXiv preprint arXiv:2303.00762}\ } (\bibinfo {year}
  {2023})}\BibitemShut {NoStop}%
\bibitem [{\citenamefont {Meng}\ \emph {et~al.}(2024)\citenamefont {Meng},
  \citenamefont {Ang},\ and\ \citenamefont {Lee}}]{meng2024exceptional}%
  \BibitemOpen
  \bibfield  {author} {\bibinfo {author} {\bibfnamefont {H.}~\bibnamefont
  {Meng}}, \bibinfo {author} {\bibfnamefont {Y.~S.}\ \bibnamefont {Ang}},\ and\
  \bibinfo {author} {\bibfnamefont {C.~H.}\ \bibnamefont {Lee}},\ }\href@noop
  {} {\bibfield  {journal} {\bibinfo  {journal} {Applied Physics Letters}\
  }\textbf {\bibinfo {volume} {124}} (\bibinfo {year} {2024})}\BibitemShut
  {NoStop}%
\bibitem [{\citenamefont {Poli}\ \emph {et~al.}(2015)\citenamefont {Poli},
  \citenamefont {Bellec}, \citenamefont {Kuhl}, \citenamefont {Mortessagne},\
  and\ \citenamefont {Schomerus}}]{Poli2015}%
  \BibitemOpen
  \bibfield  {author} {\bibinfo {author} {\bibfnamefont {C.}~\bibnamefont
  {Poli}}, \bibinfo {author} {\bibfnamefont {M.}~\bibnamefont {Bellec}},
  \bibinfo {author} {\bibfnamefont {U.}~\bibnamefont {Kuhl}}, \bibinfo {author}
  {\bibfnamefont {F.}~\bibnamefont {Mortessagne}},\ and\ \bibinfo {author}
  {\bibfnamefont {H.}~\bibnamefont {Schomerus}},\ }\href
  {https://doi.org/10.1038/ncomms7710} {\bibfield  {journal} {\bibinfo
  {journal} {Nat. Commun.}\ }\textbf {\bibinfo {volume} {6}},\ \bibinfo {pages}
  {6710} (\bibinfo {year} {2015})}\BibitemShut {NoStop}%
\bibitem [{\citenamefont {Helbig}\ \emph {et~al.}(2020)\citenamefont {Helbig},
  \citenamefont {Hofmann}, \citenamefont {Imhof}, \citenamefont {Abdelghany},
  \citenamefont {Kiessling}, \citenamefont {Molenkamp}, \citenamefont {Lee},
  \citenamefont {Szameit}, \citenamefont {Greiter},\ and\ \citenamefont
  {Thomale}}]{Helbig2020}%
  \BibitemOpen
  \bibfield  {author} {\bibinfo {author} {\bibfnamefont {T.}~\bibnamefont
  {Helbig}}, \bibinfo {author} {\bibfnamefont {T.}~\bibnamefont {Hofmann}},
  \bibinfo {author} {\bibfnamefont {S.}~\bibnamefont {Imhof}}, \bibinfo
  {author} {\bibfnamefont {M.}~\bibnamefont {Abdelghany}}, \bibinfo {author}
  {\bibfnamefont {T.}~\bibnamefont {Kiessling}}, \bibinfo {author}
  {\bibfnamefont {L.~W.}\ \bibnamefont {Molenkamp}}, \bibinfo {author}
  {\bibfnamefont {C.~H.}\ \bibnamefont {Lee}}, \bibinfo {author} {\bibfnamefont
  {A.}~\bibnamefont {Szameit}}, \bibinfo {author} {\bibfnamefont
  {M.}~\bibnamefont {Greiter}},\ and\ \bibinfo {author} {\bibfnamefont
  {R.}~\bibnamefont {Thomale}},\ }\href
  {https://doi.org/10.1038/s41567-020-0922-9} {\bibfield  {journal} {\bibinfo
  {journal} {Nat. Phys.}\ }\textbf {\bibinfo {volume} {16}},\ \bibinfo {pages}
  {747} (\bibinfo {year} {2020})}\BibitemShut {NoStop}%
\bibitem [{\citenamefont {Hofmann}\ \emph {et~al.}(2020)\citenamefont
  {Hofmann}, \citenamefont {Helbig}, \citenamefont {Schindler}, \citenamefont
  {Salgo}, \citenamefont {Brzezi\ifmmode~\acute{n}\else \'{n}\fi{}ska},
  \citenamefont {Greiter}, \citenamefont {Kiessling}, \citenamefont {Wolf},
  \citenamefont {Vollhardt}, \citenamefont {Kaba\ifmmode~\check{s}\else
  \v{s}\fi{}i}, \citenamefont {Lee}, \citenamefont {Bilu\ifmmode \check{s}\else
  \v{s}\fi{}i\ifmmode~\acute{c}\else \'{c}\fi{}}, \citenamefont {Thomale},\
  and\ \citenamefont {Neupert}}]{Hofmann2020}%
  \BibitemOpen
  \bibfield  {author} {\bibinfo {author} {\bibfnamefont {T.}~\bibnamefont
  {Hofmann}}, \bibinfo {author} {\bibfnamefont {T.}~\bibnamefont {Helbig}},
  \bibinfo {author} {\bibfnamefont {F.}~\bibnamefont {Schindler}}, \bibinfo
  {author} {\bibfnamefont {N.}~\bibnamefont {Salgo}}, \bibinfo {author}
  {\bibfnamefont {M.}~\bibnamefont {Brzezi\ifmmode~\acute{n}\else
  \'{n}\fi{}ska}}, \bibinfo {author} {\bibfnamefont {M.}~\bibnamefont
  {Greiter}}, \bibinfo {author} {\bibfnamefont {T.}~\bibnamefont {Kiessling}},
  \bibinfo {author} {\bibfnamefont {D.}~\bibnamefont {Wolf}}, \bibinfo {author}
  {\bibfnamefont {A.}~\bibnamefont {Vollhardt}}, \bibinfo {author}
  {\bibfnamefont {A.}~\bibnamefont {Kaba\ifmmode~\check{s}\else \v{s}\fi{}i}},
  \bibinfo {author} {\bibfnamefont {C.~H.}\ \bibnamefont {Lee}}, \bibinfo
  {author} {\bibfnamefont {A.}~\bibnamefont {Bilu\ifmmode \check{s}\else
  \v{s}\fi{}i\ifmmode~\acute{c}\else \'{c}\fi{}}}, \bibinfo {author}
  {\bibfnamefont {R.}~\bibnamefont {Thomale}},\ and\ \bibinfo {author}
  {\bibfnamefont {T.}~\bibnamefont {Neupert}},\ }\href
  {https://doi.org/10.1103/PhysRevResearch.2.023265} {\bibfield  {journal}
  {\bibinfo  {journal} {Phys. Rev. Res.}\ }\textbf {\bibinfo {volume} {2}},\
  \bibinfo {pages} {023265} (\bibinfo {year} {2020})}\BibitemShut {NoStop}%
\bibitem [{\citenamefont {Xiao}\ \emph {et~al.}(2020)\citenamefont {Xiao},
  \citenamefont {Deng}, \citenamefont {Wang}, \citenamefont {Zhu},
  \citenamefont {Wang}, \citenamefont {Yi},\ and\ \citenamefont
  {Xue}}]{xiaoLei2020}%
  \BibitemOpen
  \bibfield  {author} {\bibinfo {author} {\bibfnamefont {L.}~\bibnamefont
  {Xiao}}, \bibinfo {author} {\bibfnamefont {T.}~\bibnamefont {Deng}}, \bibinfo
  {author} {\bibfnamefont {K.}~\bibnamefont {Wang}}, \bibinfo {author}
  {\bibfnamefont {G.}~\bibnamefont {Zhu}}, \bibinfo {author} {\bibfnamefont
  {Z.}~\bibnamefont {Wang}}, \bibinfo {author} {\bibfnamefont {W.}~\bibnamefont
  {Yi}},\ and\ \bibinfo {author} {\bibfnamefont {P.}~\bibnamefont {Xue}},\
  }\href {https://doi.org/10.1038/s41567-020-0836-6} {\bibfield  {journal}
  {\bibinfo  {journal} {Nat. Phys.}\ }\textbf {\bibinfo {volume} {16}},\
  \bibinfo {pages} {761} (\bibinfo {year} {2020})}\BibitemShut {NoStop}%
\bibitem [{\citenamefont {Weidemann}\ \emph {et~al.}(2020)\citenamefont
  {Weidemann}, \citenamefont {Kremer}, \citenamefont {Helbig}, \citenamefont
  {Hofmann}, \citenamefont {Stegmaier}, \citenamefont {Greiter}, \citenamefont
  {Thomale},\ and\ \citenamefont {Szameit}}]{Sebastian2020}%
  \BibitemOpen
  \bibfield  {author} {\bibinfo {author} {\bibfnamefont {S.}~\bibnamefont
  {Weidemann}}, \bibinfo {author} {\bibfnamefont {M.}~\bibnamefont {Kremer}},
  \bibinfo {author} {\bibfnamefont {T.}~\bibnamefont {Helbig}}, \bibinfo
  {author} {\bibfnamefont {T.}~\bibnamefont {Hofmann}}, \bibinfo {author}
  {\bibfnamefont {A.}~\bibnamefont {Stegmaier}}, \bibinfo {author}
  {\bibfnamefont {M.}~\bibnamefont {Greiter}}, \bibinfo {author} {\bibfnamefont
  {R.}~\bibnamefont {Thomale}},\ and\ \bibinfo {author} {\bibfnamefont
  {A.}~\bibnamefont {Szameit}},\ }\href
  {https://doi.org/10.1126/science.aaz8727} {\bibfield  {journal} {\bibinfo
  {journal} {Science}\ }\textbf {\bibinfo {volume} {368}},\ \bibinfo {pages}
  {311} (\bibinfo {year} {2020})},\ \Eprint
  {https://arxiv.org/abs/https://www.science.org/doi/pdf/10.1126/science.aaz8727}
  {https://www.science.org/doi/pdf/10.1126/science.aaz8727} \BibitemShut
  {NoStop}%
\bibitem [{\citenamefont {Palacios}\ \emph {et~al.}(2021)\citenamefont
  {Palacios}, \citenamefont {Tchoumakov}, \citenamefont {Guix}, \citenamefont
  {Pagonabarraga}, \citenamefont {Sánchez},\ and\ \citenamefont
  {G.~Grushin}}]{Palacios2021}%
  \BibitemOpen
  \bibfield  {author} {\bibinfo {author} {\bibfnamefont {L.~S.}\ \bibnamefont
  {Palacios}}, \bibinfo {author} {\bibfnamefont {S.}~\bibnamefont
  {Tchoumakov}}, \bibinfo {author} {\bibfnamefont {M.}~\bibnamefont {Guix}},
  \bibinfo {author} {\bibfnamefont {I.}~\bibnamefont {Pagonabarraga}}, \bibinfo
  {author} {\bibfnamefont {S.}~\bibnamefont {Sánchez}},\ and\ \bibinfo
  {author} {\bibfnamefont {A.}~\bibnamefont {G.~Grushin}},\ }\href
  {https://doi.org/10.1038/s41467-021-24948-2} {\bibfield  {journal} {\bibinfo
  {journal} {Nat. Commun.}\ }\textbf {\bibinfo {volume} {12}},\ \bibinfo
  {pages} {4691} (\bibinfo {year} {2021})}\BibitemShut {NoStop}%
\bibitem [{\citenamefont {Zhang}\ \emph {et~al.}(2021)\citenamefont {Zhang},
  \citenamefont {Tian}, \citenamefont {Jiang}, \citenamefont {Lu},\ and\
  \citenamefont {Chen}}]{ZhangXiujuan2021}%
  \BibitemOpen
  \bibfield  {author} {\bibinfo {author} {\bibfnamefont {X.}~\bibnamefont
  {Zhang}}, \bibinfo {author} {\bibfnamefont {Y.}~\bibnamefont {Tian}},
  \bibinfo {author} {\bibfnamefont {J.-H.}\ \bibnamefont {Jiang}}, \bibinfo
  {author} {\bibfnamefont {M.-H.}\ \bibnamefont {Lu}},\ and\ \bibinfo {author}
  {\bibfnamefont {Y.-F.}\ \bibnamefont {Chen}},\ }\href
  {https://doi.org/10.1038/s41467-021-25716-y} {\bibfield  {journal} {\bibinfo
  {journal} {Nat. Commun.}\ }\textbf {\bibinfo {volume} {12}},\ \bibinfo
  {pages} {5377} (\bibinfo {year} {2021})}\BibitemShut {NoStop}%
\bibitem [{\citenamefont {Xue}\ and\ \citenamefont
  {Lee}(2024)}]{xue2024topologically}%
  \BibitemOpen
  \bibfield  {author} {\bibinfo {author} {\bibfnamefont {W.-T.}\ \bibnamefont
  {Xue}}\ and\ \bibinfo {author} {\bibfnamefont {C.~H.}\ \bibnamefont {Lee}},\
  }\href@noop {} {\bibfield  {journal} {\bibinfo  {journal} {arXiv preprint
  arXiv:2403.03259}\ } (\bibinfo {year} {2024})}\BibitemShut {NoStop}%
\bibitem [{\citenamefont {Wu}\ \emph {et~al.}(2025)\citenamefont {Wu},
  \citenamefont {Hu}, \citenamefont {He}, \citenamefont {Deng}, \citenamefont
  {Huang}, \citenamefont {Ke}, \citenamefont {Deng}, \citenamefont {Lu},\ and\
  \citenamefont {Liu}}]{wu2025hybrid}%
  \BibitemOpen
  \bibfield  {author} {\bibinfo {author} {\bibfnamefont {J.}~\bibnamefont
  {Wu}}, \bibinfo {author} {\bibfnamefont {Y.}~\bibnamefont {Hu}}, \bibinfo
  {author} {\bibfnamefont {Z.}~\bibnamefont {He}}, \bibinfo {author}
  {\bibfnamefont {K.}~\bibnamefont {Deng}}, \bibinfo {author} {\bibfnamefont
  {X.}~\bibnamefont {Huang}}, \bibinfo {author} {\bibfnamefont
  {M.}~\bibnamefont {Ke}}, \bibinfo {author} {\bibfnamefont {W.}~\bibnamefont
  {Deng}}, \bibinfo {author} {\bibfnamefont {J.}~\bibnamefont {Lu}},\ and\
  \bibinfo {author} {\bibfnamefont {Z.}~\bibnamefont {Liu}},\ }\href@noop {}
  {\bibfield  {journal} {\bibinfo  {journal} {Physical Review Letters}\
  }\textbf {\bibinfo {volume} {134}},\ \bibinfo {pages} {176601} (\bibinfo
  {year} {2025})}\BibitemShut {NoStop}%
\bibitem [{\citenamefont {Longhi}(2025)}]{longhi2025er}%
  \BibitemOpen
  \bibfield  {author} {\bibinfo {author} {\bibfnamefont {S.}~\bibnamefont
  {Longhi}},\ }\href@noop {} {\bibfield  {journal} {\bibinfo  {journal} {arXiv
  preprint arXiv:2504.14910}\ } (\bibinfo {year} {2025})}\BibitemShut {NoStop}%
\bibitem [{\citenamefont {Gliozzi}\ \emph {et~al.}(2025)\citenamefont
  {Gliozzi}, \citenamefont {Balducci}, \citenamefont {Hughes},\ and\
  \citenamefont {De~Tomasi}}]{glio2025non}%
  \BibitemOpen
  \bibfield  {author} {\bibinfo {author} {\bibfnamefont {J.}~\bibnamefont
  {Gliozzi}}, \bibinfo {author} {\bibfnamefont {F.}~\bibnamefont {Balducci}},
  \bibinfo {author} {\bibfnamefont {T.~L.}\ \bibnamefont {Hughes}},\ and\
  \bibinfo {author} {\bibfnamefont {G.}~\bibnamefont {De~Tomasi}},\ }\href@noop
  {} {\bibfield  {journal} {\bibinfo  {journal} {arXiv preprint
  arXiv:2504.10580}\ } (\bibinfo {year} {2025})}\BibitemShut {NoStop}%
\bibitem [{\citenamefont {Li}\ \emph {et~al.}(2025)\citenamefont {Li},
  \citenamefont {Jiang},\ and\ \citenamefont {Lee}}]{li2025phase}%
  \BibitemOpen
  \bibfield  {author} {\bibinfo {author} {\bibfnamefont {Q.}~\bibnamefont
  {Li}}, \bibinfo {author} {\bibfnamefont {H.}~\bibnamefont {Jiang}},\ and\
  \bibinfo {author} {\bibfnamefont {C.~H.}\ \bibnamefont {Lee}},\ }\href@noop
  {} {\bibfield  {journal} {\bibinfo  {journal} {arXiv preprint
  arXiv:2501.09785}\ } (\bibinfo {year} {2025})}\BibitemShut {NoStop}%
\bibitem [{\citenamefont {Shen}\ \emph
  {et~al.}(2025{\natexlab{a}})\citenamefont {Shen}, \citenamefont {Chan},\ and\
  \citenamefont {Lee}}]{shen2025non}%
  \BibitemOpen
  \bibfield  {author} {\bibinfo {author} {\bibfnamefont {R.}~\bibnamefont
  {Shen}}, \bibinfo {author} {\bibfnamefont {W.~J.}\ \bibnamefont {Chan}},\
  and\ \bibinfo {author} {\bibfnamefont {C.~H.}\ \bibnamefont {Lee}},\
  }\href@noop {} {\bibfield  {journal} {\bibinfo  {journal} {Physical Review
  B}\ }\textbf {\bibinfo {volume} {111}},\ \bibinfo {pages} {045420} (\bibinfo
  {year} {2025}{\natexlab{a}})}\BibitemShut {NoStop}%
\bibitem [{\citenamefont {Guo}\ \emph {et~al.}(2025)\citenamefont {Guo},
  \citenamefont {Yin}, \citenamefont {Zhang},\ and\ \citenamefont
  {Li}}]{guo2025skin}%
  \BibitemOpen
  \bibfield  {author} {\bibinfo {author} {\bibfnamefont {S.}~\bibnamefont
  {Guo}}, \bibinfo {author} {\bibfnamefont {S.}~\bibnamefont {Yin}}, \bibinfo
  {author} {\bibfnamefont {S.-X.}\ \bibnamefont {Zhang}},\ and\ \bibinfo
  {author} {\bibfnamefont {Z.-X.}\ \bibnamefont {Li}},\ }\href@noop {}
  {\bibfield  {journal} {\bibinfo  {journal} {arXiv preprint arXiv:2504.21631}\
  } (\bibinfo {year} {2025})}\BibitemShut {NoStop}%
\bibitem [{\citenamefont {Faugno}\ and\ \citenamefont
  {Ozawa}(2022)}]{Faugno2022PRL}%
  \BibitemOpen
  \bibfield  {author} {\bibinfo {author} {\bibfnamefont {W.~N.}\ \bibnamefont
  {Faugno}}\ and\ \bibinfo {author} {\bibfnamefont {T.}~\bibnamefont {Ozawa}},\
  }\href {https://doi.org/10.1103/PhysRevLett.129.180401} {\bibfield  {journal}
  {\bibinfo  {journal} {Phys. Rev. Lett.}\ }\textbf {\bibinfo {volume} {129}},\
  \bibinfo {pages} {180401} (\bibinfo {year} {2022})}\BibitemShut {NoStop}%
\bibitem [{\citenamefont {Yoshida}\ \emph {et~al.}(2024)\citenamefont
  {Yoshida}, \citenamefont {Zhang}, \citenamefont {Neupert},\ and\
  \citenamefont {Kawakami}}]{yoshida2024non}%
  \BibitemOpen
  \bibfield  {author} {\bibinfo {author} {\bibfnamefont {T.}~\bibnamefont
  {Yoshida}}, \bibinfo {author} {\bibfnamefont {S.-B.}\ \bibnamefont {Zhang}},
  \bibinfo {author} {\bibfnamefont {T.}~\bibnamefont {Neupert}},\ and\ \bibinfo
  {author} {\bibfnamefont {N.}~\bibnamefont {Kawakami}},\ }\href@noop {}
  {\bibfield  {journal} {\bibinfo  {journal} {Physical Review Letters}\
  }\textbf {\bibinfo {volume} {133}},\ \bibinfo {pages} {076502} (\bibinfo
  {year} {2024})}\BibitemShut {NoStop}%
\bibitem [{\citenamefont {Qin}\ and\ \citenamefont {Li}(2024)}]{QinYi2024PRL}%
  \BibitemOpen
  \bibfield  {author} {\bibinfo {author} {\bibfnamefont {Y.}~\bibnamefont
  {Qin}}\ and\ \bibinfo {author} {\bibfnamefont {L.}~\bibnamefont {Li}},\
  }\href {https://doi.org/10.1103/PhysRevLett.132.096501} {\bibfield  {journal}
  {\bibinfo  {journal} {Phys. Rev. Lett.}\ }\textbf {\bibinfo {volume} {132}},\
  \bibinfo {pages} {096501} (\bibinfo {year} {2024})}\BibitemShut {NoStop}%
\bibitem [{\citenamefont {Wang}\ and\ \citenamefont {Li}(2025)}]{wang2025non}%
  \BibitemOpen
  \bibfield  {author} {\bibinfo {author} {\bibfnamefont {Y.-A.}\ \bibnamefont
  {Wang}}\ and\ \bibinfo {author} {\bibfnamefont {L.}~\bibnamefont {Li}},\
  }\href@noop {} {\bibfield  {journal} {\bibinfo  {journal} {Chin. Phys.
  Lett.}\ } (\bibinfo {year} {2025})}\BibitemShut {NoStop}%
\bibitem [{\citenamefont {Lee}(2021)}]{lee2021many}%
  \BibitemOpen
  \bibfield  {author} {\bibinfo {author} {\bibfnamefont {C.~H.}\ \bibnamefont
  {Lee}},\ }\href {https://doi.org/10.1103/PhysRevB.104.195102} {\bibfield
  {journal} {\bibinfo  {journal} {Phys. Rev. B}\ }\textbf {\bibinfo {volume}
  {104}},\ \bibinfo {pages} {195102} (\bibinfo {year} {2021})}\BibitemShut
  {NoStop}%
\bibitem [{\citenamefont {Shen}\ and\ \citenamefont {Lee}(2022)}]{shen2022non}%
  \BibitemOpen
  \bibfield  {author} {\bibinfo {author} {\bibfnamefont {R.}~\bibnamefont
  {Shen}}\ and\ \bibinfo {author} {\bibfnamefont {C.~H.}\ \bibnamefont {Lee}},\
  }\href@noop {} {\bibfield  {journal} {\bibinfo  {journal} {Communications
  Physics}\ }\textbf {\bibinfo {volume} {5}},\ \bibinfo {pages} {238} (\bibinfo
  {year} {2022})}\BibitemShut {NoStop}%
\bibitem [{\citenamefont {Li}\ \emph {et~al.}(2023)\citenamefont {Li},
  \citenamefont {Yu},\ and\ \citenamefont {Zhong}}]{li2023non}%
  \BibitemOpen
  \bibfield  {author} {\bibinfo {author} {\bibfnamefont {H.-Z.}\ \bibnamefont
  {Li}}, \bibinfo {author} {\bibfnamefont {X.-J.}\ \bibnamefont {Yu}},\ and\
  \bibinfo {author} {\bibfnamefont {J.-X.}\ \bibnamefont {Zhong}},\ }\href@noop
  {} {\bibfield  {journal} {\bibinfo  {journal} {Physical Review A}\ }\textbf
  {\bibinfo {volume} {108}},\ \bibinfo {pages} {043301} (\bibinfo {year}
  {2023})}\BibitemShut {NoStop}%
\bibitem [{\citenamefont {Kawabata}\ \emph {et~al.}(2022)\citenamefont
  {Kawabata}, \citenamefont {Shiozaki},\ and\ \citenamefont
  {Ryu}}]{kawabata2022many}%
  \BibitemOpen
  \bibfield  {author} {\bibinfo {author} {\bibfnamefont {K.}~\bibnamefont
  {Kawabata}}, \bibinfo {author} {\bibfnamefont {K.}~\bibnamefont {Shiozaki}},\
  and\ \bibinfo {author} {\bibfnamefont {S.}~\bibnamefont {Ryu}},\ }\href@noop
  {} {\bibfield  {journal} {\bibinfo  {journal} {Physical Review B}\ }\textbf
  {\bibinfo {volume} {105}},\ \bibinfo {pages} {165137} (\bibinfo {year}
  {2022})}\BibitemShut {NoStop}%
\bibitem [{\citenamefont {Shen}\ \emph {et~al.}(2024)\citenamefont {Shen},
  \citenamefont {Qin}, \citenamefont {Desaules}, \citenamefont {Papi{\'c}},\
  and\ \citenamefont {Lee}}]{shen2024enhanced}%
  \BibitemOpen
  \bibfield  {author} {\bibinfo {author} {\bibfnamefont {R.}~\bibnamefont
  {Shen}}, \bibinfo {author} {\bibfnamefont {F.}~\bibnamefont {Qin}}, \bibinfo
  {author} {\bibfnamefont {J.-Y.}\ \bibnamefont {Desaules}}, \bibinfo {author}
  {\bibfnamefont {Z.}~\bibnamefont {Papi{\'c}}},\ and\ \bibinfo {author}
  {\bibfnamefont {C.~H.}\ \bibnamefont {Lee}},\ }\href@noop {} {\bibfield
  {journal} {\bibinfo  {journal} {Physical Review Letters}\ }\textbf {\bibinfo
  {volume} {133}},\ \bibinfo {pages} {216601} (\bibinfo {year}
  {2024})}\BibitemShut {NoStop}%
\bibitem [{\citenamefont {Gliozzi}\ \emph {et~al.}(2024)\citenamefont
  {Gliozzi}, \citenamefont {De~Tomasi},\ and\ \citenamefont
  {Hughes}}]{gliozzi2024many}%
  \BibitemOpen
  \bibfield  {author} {\bibinfo {author} {\bibfnamefont {J.}~\bibnamefont
  {Gliozzi}}, \bibinfo {author} {\bibfnamefont {G.}~\bibnamefont {De~Tomasi}},\
  and\ \bibinfo {author} {\bibfnamefont {T.~L.}\ \bibnamefont {Hughes}},\
  }\href@noop {} {\bibfield  {journal} {\bibinfo  {journal} {Physical review
  letters}\ }\textbf {\bibinfo {volume} {133}},\ \bibinfo {pages} {136503}
  (\bibinfo {year} {2024})}\BibitemShut {NoStop}%
\bibitem [{\citenamefont {Yang}\ \emph
  {et~al.}(2024{\natexlab{a}})\citenamefont {Yang}, \citenamefont {Yuan},\ and\
  \citenamefont {Lee}}]{yang2024non}%
  \BibitemOpen
  \bibfield  {author} {\bibinfo {author} {\bibfnamefont {M.}~\bibnamefont
  {Yang}}, \bibinfo {author} {\bibfnamefont {L.}~\bibnamefont {Yuan}},\ and\
  \bibinfo {author} {\bibfnamefont {C.~H.}\ \bibnamefont {Lee}},\ }\href@noop
  {} {\bibfield  {journal} {\bibinfo  {journal} {arXiv preprint
  arXiv:2410.01258}\ } (\bibinfo {year} {2024}{\natexlab{a}})}\BibitemShut
  {NoStop}%
\bibitem [{\citenamefont {Li}\ \emph {et~al.}(2024)\citenamefont {Li},
  \citenamefont {Cai}, \citenamefont {Liu},\ and\ \citenamefont
  {Nori}}]{li2024Dis}%
  \BibitemOpen
  \bibfield  {author} {\bibinfo {author} {\bibfnamefont {Y.}~\bibnamefont
  {Li}}, \bibinfo {author} {\bibfnamefont {Z.-F.}\ \bibnamefont {Cai}},
  \bibinfo {author} {\bibfnamefont {T.}~\bibnamefont {Liu}},\ and\ \bibinfo
  {author} {\bibfnamefont {F.}~\bibnamefont {Nori}},\ }\href@noop {} {\bibfield
   {journal} {\bibinfo  {journal} {arXiv preprint arXiv:2408.12451}\ }
  (\bibinfo {year} {2024})}\BibitemShut {NoStop}%
\bibitem [{\citenamefont {Hu}\ \emph {et~al.}(2025)\citenamefont {Hu},
  \citenamefont {Wang}, \citenamefont {Lian},\ and\ \citenamefont
  {Wang}}]{hu2025many}%
  \BibitemOpen
  \bibfield  {author} {\bibinfo {author} {\bibfnamefont {Y.-M.}\ \bibnamefont
  {Hu}}, \bibinfo {author} {\bibfnamefont {Z.}~\bibnamefont {Wang}}, \bibinfo
  {author} {\bibfnamefont {B.}~\bibnamefont {Lian}},\ and\ \bibinfo {author}
  {\bibfnamefont {Z.}~\bibnamefont {Wang}},\ }\href@noop {} {\bibfield
  {journal} {\bibinfo  {journal} {arXiv:2502.03534}\ } (\bibinfo {year}
  {2025})}\BibitemShut {NoStop}%
\bibitem [{\citenamefont {Lee}\ \emph {et~al.}(2015)\citenamefont {Lee},
  \citenamefont {Papi{\'c}},\ and\ \citenamefont {Thomale}}]{lee2015geometric}%
  \BibitemOpen
  \bibfield  {author} {\bibinfo {author} {\bibfnamefont {C.~H.}\ \bibnamefont
  {Lee}}, \bibinfo {author} {\bibfnamefont {Z.}~\bibnamefont {Papi{\'c}}},\
  and\ \bibinfo {author} {\bibfnamefont {R.}~\bibnamefont {Thomale}},\
  }\href@noop {} {\bibfield  {journal} {\bibinfo  {journal} {Physical Review
  X}\ }\textbf {\bibinfo {volume} {5}},\ \bibinfo {pages} {041003} (\bibinfo
  {year} {2015})}\BibitemShut {NoStop}%
\bibitem [{\citenamefont {Mukherjee}\ \emph {et~al.}(2021)\citenamefont
  {Mukherjee}, \citenamefont {Banerjee}, \citenamefont {Sengupta},\ and\
  \citenamefont {Sen}}]{mukherjee2021minimal}%
  \BibitemOpen
  \bibfield  {author} {\bibinfo {author} {\bibfnamefont {B.}~\bibnamefont
  {Mukherjee}}, \bibinfo {author} {\bibfnamefont {D.}~\bibnamefont {Banerjee}},
  \bibinfo {author} {\bibfnamefont {K.}~\bibnamefont {Sengupta}},\ and\
  \bibinfo {author} {\bibfnamefont {A.}~\bibnamefont {Sen}},\ }\href@noop {}
  {\bibfield  {journal} {\bibinfo  {journal} {Physical Review B}\ }\textbf
  {\bibinfo {volume} {104}},\ \bibinfo {pages} {155117} (\bibinfo {year}
  {2021})}\BibitemShut {NoStop}%
\bibitem [{\citenamefont {Moudgalya}\ and\ \citenamefont
  {Motrunich}(2022)}]{moudgalya2022hilbert}%
  \BibitemOpen
  \bibfield  {author} {\bibinfo {author} {\bibfnamefont {S.}~\bibnamefont
  {Moudgalya}}\ and\ \bibinfo {author} {\bibfnamefont {O.~I.}\ \bibnamefont
  {Motrunich}},\ }\href@noop {} {\bibfield  {journal} {\bibinfo  {journal}
  {Physical Review X}\ }\textbf {\bibinfo {volume} {12}},\ \bibinfo {pages}
  {011050} (\bibinfo {year} {2022})}\BibitemShut {NoStop}%
\bibitem [{\citenamefont {Moudgalya}\ \emph {et~al.}(2022)\citenamefont
  {Moudgalya}, \citenamefont {Bernevig},\ and\ \citenamefont
  {Regnault}}]{moudgalya2022quantum}%
  \BibitemOpen
  \bibfield  {author} {\bibinfo {author} {\bibfnamefont {S.}~\bibnamefont
  {Moudgalya}}, \bibinfo {author} {\bibfnamefont {B.~A.}\ \bibnamefont
  {Bernevig}},\ and\ \bibinfo {author} {\bibfnamefont {N.}~\bibnamefont
  {Regnault}},\ }\href@noop {} {\bibfield  {journal} {\bibinfo  {journal}
  {Reports on Progress in Physics}\ }\textbf {\bibinfo {volume} {85}},\
  \bibinfo {pages} {086501} (\bibinfo {year} {2022})}\BibitemShut {NoStop}%
\bibitem [{\citenamefont {Brighi}\ \emph {et~al.}(2023)\citenamefont {Brighi},
  \citenamefont {Ljubotina},\ and\ \citenamefont {Serbyn}}]{brighi2023hilbert}%
  \BibitemOpen
  \bibfield  {author} {\bibinfo {author} {\bibfnamefont {P.}~\bibnamefont
  {Brighi}}, \bibinfo {author} {\bibfnamefont {M.}~\bibnamefont {Ljubotina}},\
  and\ \bibinfo {author} {\bibfnamefont {M.}~\bibnamefont {Serbyn}},\
  }\href@noop {} {\bibfield  {journal} {\bibinfo  {journal} {SciPost Physics}\
  }\textbf {\bibinfo {volume} {15}},\ \bibinfo {pages} {093} (\bibinfo {year}
  {2023})}\BibitemShut {NoStop}%
\bibitem [{\citenamefont {Adler}\ \emph {et~al.}(2024)\citenamefont {Adler},
  \citenamefont {Wei}, \citenamefont {Will}, \citenamefont {Srakaew},
  \citenamefont {Agrawal}, \citenamefont {Weckesser}, \citenamefont {Moessner},
  \citenamefont {Pollmann}, \citenamefont {Bloch},\ and\ \citenamefont
  {Zeiher}}]{adler2024observation}%
  \BibitemOpen
  \bibfield  {author} {\bibinfo {author} {\bibfnamefont {D.}~\bibnamefont
  {Adler}}, \bibinfo {author} {\bibfnamefont {D.}~\bibnamefont {Wei}}, \bibinfo
  {author} {\bibfnamefont {M.}~\bibnamefont {Will}}, \bibinfo {author}
  {\bibfnamefont {K.}~\bibnamefont {Srakaew}}, \bibinfo {author} {\bibfnamefont
  {S.}~\bibnamefont {Agrawal}}, \bibinfo {author} {\bibfnamefont
  {P.}~\bibnamefont {Weckesser}}, \bibinfo {author} {\bibfnamefont
  {R.}~\bibnamefont {Moessner}}, \bibinfo {author} {\bibfnamefont
  {F.}~\bibnamefont {Pollmann}}, \bibinfo {author} {\bibfnamefont
  {I.}~\bibnamefont {Bloch}},\ and\ \bibinfo {author} {\bibfnamefont
  {J.}~\bibnamefont {Zeiher}},\ }\href@noop {} {\bibfield  {journal} {\bibinfo
  {journal} {Nature}\ ,\ \bibinfo {pages} {1}} (\bibinfo {year}
  {2024})}\BibitemShut {NoStop}%
\bibitem [{\citenamefont {Fromholz}\ and\ \citenamefont
  {Lecheminant}(2020)}]{FromholzPRB2020}%
  \BibitemOpen
  \bibfield  {author} {\bibinfo {author} {\bibfnamefont {P.}~\bibnamefont
  {Fromholz}}\ and\ \bibinfo {author} {\bibfnamefont {P.}~\bibnamefont
  {Lecheminant}},\ }\href {https://doi.org/10.1103/PhysRevB.102.094410}
  {\bibfield  {journal} {\bibinfo  {journal} {Phys. Rev. B}\ }\textbf {\bibinfo
  {volume} {102}},\ \bibinfo {pages} {094410} (\bibinfo {year}
  {2020})}\BibitemShut {NoStop}%
\bibitem [{\citenamefont {Gergs}\ \emph {et~al.}(2016)\citenamefont {Gergs},
  \citenamefont {Fritz},\ and\ \citenamefont {Schuricht}}]{GergsPRB2016}%
  \BibitemOpen
  \bibfield  {author} {\bibinfo {author} {\bibfnamefont {N.~M.}\ \bibnamefont
  {Gergs}}, \bibinfo {author} {\bibfnamefont {L.}~\bibnamefont {Fritz}},\ and\
  \bibinfo {author} {\bibfnamefont {D.}~\bibnamefont {Schuricht}},\ }\href@noop
  {} {\bibfield  {journal} {\bibinfo  {journal} {Physical Review B}\ }\textbf
  {\bibinfo {volume} {93}},\ \bibinfo {pages} {075129} (\bibinfo {year}
  {2016})}\BibitemShut {NoStop}%
\bibitem [{\citenamefont {Rachel}(2018)}]{rachel2018IN}%
  \BibitemOpen
  \bibfield  {author} {\bibinfo {author} {\bibfnamefont {S.}~\bibnamefont
  {Rachel}},\ }\href@noop {} {\bibfield  {journal} {\bibinfo  {journal}
  {Reports on Progress in Physics}\ }\textbf {\bibinfo {volume} {81}},\
  \bibinfo {pages} {116501} (\bibinfo {year} {2018})}\BibitemShut {NoStop}%
\bibitem [{\citenamefont {Wang}\ and\ \citenamefont
  {Senthil}(2015)}]{WangPRB2015}%
  \BibitemOpen
  \bibfield  {author} {\bibinfo {author} {\bibfnamefont {C.}~\bibnamefont
  {Wang}}\ and\ \bibinfo {author} {\bibfnamefont {T.}~\bibnamefont {Senthil}},\
  }\href {https://doi.org/10.1103/PhysRevB.91.239902} {\bibfield  {journal}
  {\bibinfo  {journal} {Phys. Rev. B}\ }\textbf {\bibinfo {volume} {91}},\
  \bibinfo {pages} {239902} (\bibinfo {year} {2015})}\BibitemShut {NoStop}%
\bibitem [{\citenamefont {Ding}\ \emph {et~al.}(2022)\citenamefont {Ding},
  \citenamefont {Liu}, \citenamefont {Shi}, \citenamefont {Guo}, \citenamefont
  {M{\o}lmer},\ and\ \citenamefont {Adams}}]{ding2022EN}%
  \BibitemOpen
  \bibfield  {author} {\bibinfo {author} {\bibfnamefont {D.-S.}\ \bibnamefont
  {Ding}}, \bibinfo {author} {\bibfnamefont {Z.-K.}\ \bibnamefont {Liu}},
  \bibinfo {author} {\bibfnamefont {B.-S.}\ \bibnamefont {Shi}}, \bibinfo
  {author} {\bibfnamefont {G.-C.}\ \bibnamefont {Guo}}, \bibinfo {author}
  {\bibfnamefont {K.}~\bibnamefont {M{\o}lmer}},\ and\ \bibinfo {author}
  {\bibfnamefont {C.~S.}\ \bibnamefont {Adams}},\ }\href@noop {} {\bibfield
  {journal} {\bibinfo  {journal} {Nature Physics}\ }\textbf {\bibinfo {volume}
  {18}},\ \bibinfo {pages} {1447} (\bibinfo {year} {2022})}\BibitemShut
  {NoStop}%
\bibitem [{Sup()}]{SuppMat}%
  \BibitemOpen
  \href@noop {} {\bibinfo  {journal} {Supplemental Materials that include
  Refs.~\cite{QinYi2024PRL,kim2024collective}}\ }\BibitemShut {NoStop}%
\bibitem [{\citenamefont {Kim}\ \emph {et~al.}(2024)\citenamefont {Kim},
  \citenamefont {Han},\ and\ \citenamefont {Park}}]{kim2024collective}%
  \BibitemOpen
\bibfield  {journal} {  }\bibfield  {author} {\bibinfo {author} {\bibfnamefont
  {B.~H.}\ \bibnamefont {Kim}}, \bibinfo {author} {\bibfnamefont {J.-H.}\
  \bibnamefont {Han}},\ and\ \bibinfo {author} {\bibfnamefont {M.~J.}\
  \bibnamefont {Park}},\ }\href@noop {} {\bibfield  {journal} {\bibinfo
  {journal} {Communications Physics}\ }\textbf {\bibinfo {volume} {7}},\
  \bibinfo {pages} {73} (\bibinfo {year} {2024})}\BibitemShut {NoStop}%
\bibitem [{\citenamefont {Hatano}\ and\ \citenamefont
  {Nelson}(1996)}]{HN1996PRL}%
  \BibitemOpen
  \bibfield  {author} {\bibinfo {author} {\bibfnamefont {N.}~\bibnamefont
  {Hatano}}\ and\ \bibinfo {author} {\bibfnamefont {D.~R.}\ \bibnamefont
  {Nelson}},\ }\href@noop {} {\bibfield  {journal} {\bibinfo  {journal}
  {Physical review letters}\ }\textbf {\bibinfo {volume} {77}},\ \bibinfo
  {pages} {570} (\bibinfo {year} {1996})}\BibitemShut {NoStop}%
\bibitem [{\citenamefont {Hatano}\ and\ \citenamefont
  {Nelson}(1997)}]{HN1997PRB}%
  \BibitemOpen
  \bibfield  {author} {\bibinfo {author} {\bibfnamefont {N.}~\bibnamefont
  {Hatano}}\ and\ \bibinfo {author} {\bibfnamefont {D.~R.}\ \bibnamefont
  {Nelson}},\ }\href@noop {} {\bibfield  {journal} {\bibinfo  {journal}
  {Physical Review B}\ }\textbf {\bibinfo {volume} {56}},\ \bibinfo {pages}
  {8651} (\bibinfo {year} {1997})}\BibitemShut {NoStop}%
\bibitem [{\citenamefont {Liu}\ \emph {et~al.}(2018)\citenamefont {Liu},
  \citenamefont {Garrison}, \citenamefont {Deng}, \citenamefont {Gong},\ and\
  \citenamefont {Gorshkov}}]{Liu2018PRL}%
  \BibitemOpen
  \bibfield  {author} {\bibinfo {author} {\bibfnamefont {F.}~\bibnamefont
  {Liu}}, \bibinfo {author} {\bibfnamefont {J.~R.}\ \bibnamefont {Garrison}},
  \bibinfo {author} {\bibfnamefont {D.-L.}\ \bibnamefont {Deng}}, \bibinfo
  {author} {\bibfnamefont {Z.-X.}\ \bibnamefont {Gong}},\ and\ \bibinfo
  {author} {\bibfnamefont {A.~V.}\ \bibnamefont {Gorshkov}},\ }\href
  {https://doi.org/10.1103/PhysRevLett.121.250404} {\bibfield  {journal}
  {\bibinfo  {journal} {Phys. Rev. Lett.}\ }\textbf {\bibinfo {volume} {121}},\
  \bibinfo {pages} {250404} (\bibinfo {year} {2018})}\BibitemShut {NoStop}%
\bibitem [{\citenamefont {Zhang}\ \emph
  {et~al.}(2023{\natexlab{a}})\citenamefont {Zhang}, \citenamefont {Qian},
  \citenamefont {Sun},\ and\ \citenamefont {Zhang}}]{Zhang2023CP}%
  \BibitemOpen
  \bibfield  {author} {\bibinfo {author} {\bibfnamefont {W.}~\bibnamefont
  {Zhang}}, \bibinfo {author} {\bibfnamefont {L.}~\bibnamefont {Qian}},
  \bibinfo {author} {\bibfnamefont {H.}~\bibnamefont {Sun}},\ and\ \bibinfo
  {author} {\bibfnamefont {X.}~\bibnamefont {Zhang}},\ }\bibfield  {journal}
  {\bibinfo  {journal} {Communications Physics}\ }\textbf {\bibinfo {volume}
  {6}},\ \href {https://doi.org/10.1038/s42005-023-01245-6}
  {10.1038/s42005-023-01245-6} (\bibinfo {year}
  {2023}{\natexlab{a}})\BibitemShut {NoStop}%
\bibitem [{\citenamefont {Kwan}\ \emph {et~al.}(2024)\citenamefont {Kwan},
  \citenamefont {Segura}, \citenamefont {Li}, \citenamefont {Kim},
  \citenamefont {Gorshkov}, \citenamefont {Eckardt}, \citenamefont
  {Bakkali-Hassani},\ and\ \citenamefont {Greiner}}]{Joyce2023arxiv}%
  \BibitemOpen
  \bibfield  {author} {\bibinfo {author} {\bibfnamefont {J.}~\bibnamefont
  {Kwan}}, \bibinfo {author} {\bibfnamefont {P.}~\bibnamefont {Segura}},
  \bibinfo {author} {\bibfnamefont {Y.}~\bibnamefont {Li}}, \bibinfo {author}
  {\bibfnamefont {S.}~\bibnamefont {Kim}}, \bibinfo {author} {\bibfnamefont
  {A.~V.}\ \bibnamefont {Gorshkov}}, \bibinfo {author} {\bibfnamefont
  {A.}~\bibnamefont {Eckardt}}, \bibinfo {author} {\bibfnamefont
  {B.}~\bibnamefont {Bakkali-Hassani}},\ and\ \bibinfo {author} {\bibfnamefont
  {M.}~\bibnamefont {Greiner}},\ }\href
  {https://doi.org/10.1126/science.adi3252} {\bibfield  {journal} {\bibinfo
  {journal} {Science}\ }\textbf {\bibinfo {volume} {386}},\ \bibinfo {pages}
  {1055} (\bibinfo {year} {2024})},\ \Eprint
  {https://arxiv.org/abs/https://www.science.org/doi/pdf/10.1126/science.adi3252}
  {https://www.science.org/doi/pdf/10.1126/science.adi3252} \BibitemShut
  {NoStop}%
\bibitem [{\citenamefont {Qin}\ \emph {et~al.}(2025)\citenamefont {Qin},
  \citenamefont {Lee},\ and\ \citenamefont {Li}}]{qin2025CP}%
  \BibitemOpen
  \bibfield  {author} {\bibinfo {author} {\bibfnamefont {Y.}~\bibnamefont
  {Qin}}, \bibinfo {author} {\bibfnamefont {C.~H.}\ \bibnamefont {Lee}},\ and\
  \bibinfo {author} {\bibfnamefont {L.}~\bibnamefont {Li}},\ }\href@noop {}
  {\bibfield  {journal} {\bibinfo  {journal} {Communications Physics}\ }\textbf
  {\bibinfo {volume} {8}},\ \bibinfo {pages} {18} (\bibinfo {year}
  {2025})}\BibitemShut {NoStop}%
\bibitem [{\citenamefont {Koch}\ and\ \citenamefont
  {Budich}(2022)}]{koch2022quantum}%
  \BibitemOpen
  \bibfield  {author} {\bibinfo {author} {\bibfnamefont {F.}~\bibnamefont
  {Koch}}\ and\ \bibinfo {author} {\bibfnamefont {J.~C.}\ \bibnamefont
  {Budich}},\ }\href@noop {} {\bibfield  {journal} {\bibinfo  {journal}
  {Physical Review Research}\ }\textbf {\bibinfo {volume} {4}},\ \bibinfo
  {pages} {013113} (\bibinfo {year} {2022})}\BibitemShut {NoStop}%
\bibitem [{\citenamefont {Sarkar}\ \emph {et~al.}(2022)\citenamefont {Sarkar},
  \citenamefont {Mukhopadhyay}, \citenamefont {Alase},\ and\ \citenamefont
  {Bayat}}]{sarkar2022free}%
  \BibitemOpen
  \bibfield  {author} {\bibinfo {author} {\bibfnamefont {S.}~\bibnamefont
  {Sarkar}}, \bibinfo {author} {\bibfnamefont {C.}~\bibnamefont
  {Mukhopadhyay}}, \bibinfo {author} {\bibfnamefont {A.}~\bibnamefont
  {Alase}},\ and\ \bibinfo {author} {\bibfnamefont {A.}~\bibnamefont {Bayat}},\
  }\href@noop {} {\bibfield  {journal} {\bibinfo  {journal} {Physical Review
  Letters}\ }\textbf {\bibinfo {volume} {129}},\ \bibinfo {pages} {090503}
  (\bibinfo {year} {2022})}\BibitemShut {NoStop}%
\bibitem [{\citenamefont {Mukhopadhyay}\ and\ \citenamefont
  {Bayat}(2024)}]{MukhopadhyayPRL2024}%
  \BibitemOpen
  \bibfield  {author} {\bibinfo {author} {\bibfnamefont {C.}~\bibnamefont
  {Mukhopadhyay}}\ and\ \bibinfo {author} {\bibfnamefont {A.}~\bibnamefont
  {Bayat}},\ }\href {https://doi.org/10.1103/PhysRevLett.133.120601} {\bibfield
   {journal} {\bibinfo  {journal} {Phys. Rev. Lett.}\ }\textbf {\bibinfo
  {volume} {133}},\ \bibinfo {pages} {120601} (\bibinfo {year}
  {2024})}\BibitemShut {NoStop}%
\bibitem [{\citenamefont {Montenegro}\ \emph {et~al.}(2024)\citenamefont
  {Montenegro}, \citenamefont {Mukhopadhyay}, \citenamefont {Yousefjani},
  \citenamefont {Sarkar}, \citenamefont {Mishra}, \citenamefont {Paris},\ and\
  \citenamefont {Bayat}}]{montenegro2024R}%
  \BibitemOpen
  \bibfield  {author} {\bibinfo {author} {\bibfnamefont {V.}~\bibnamefont
  {Montenegro}}, \bibinfo {author} {\bibfnamefont {C.}~\bibnamefont
  {Mukhopadhyay}}, \bibinfo {author} {\bibfnamefont {R.}~\bibnamefont
  {Yousefjani}}, \bibinfo {author} {\bibfnamefont {S.}~\bibnamefont {Sarkar}},
  \bibinfo {author} {\bibfnamefont {U.}~\bibnamefont {Mishra}}, \bibinfo
  {author} {\bibfnamefont {M.~G.}\ \bibnamefont {Paris}},\ and\ \bibinfo
  {author} {\bibfnamefont {A.}~\bibnamefont {Bayat}},\ }\href@noop {}
  {\bibfield  {journal} {\bibinfo  {journal} {arXiv preprint arXiv:2408.15323}\
  } (\bibinfo {year} {2024})}\BibitemShut {NoStop}%
\bibitem [{\citenamefont {Liang}\ \emph {et~al.}(2022)\citenamefont {Liang},
  \citenamefont {Xie}, \citenamefont {Dong}, \citenamefont {Li}, \citenamefont
  {Li}, \citenamefont {Gadway}, \citenamefont {Yi},\ and\ \citenamefont
  {Yan}}]{liang2022dynamic}%
  \BibitemOpen
  \bibfield  {author} {\bibinfo {author} {\bibfnamefont {Q.}~\bibnamefont
  {Liang}}, \bibinfo {author} {\bibfnamefont {D.}~\bibnamefont {Xie}}, \bibinfo
  {author} {\bibfnamefont {Z.}~\bibnamefont {Dong}}, \bibinfo {author}
  {\bibfnamefont {H.}~\bibnamefont {Li}}, \bibinfo {author} {\bibfnamefont
  {H.}~\bibnamefont {Li}}, \bibinfo {author} {\bibfnamefont {B.}~\bibnamefont
  {Gadway}}, \bibinfo {author} {\bibfnamefont {W.}~\bibnamefont {Yi}},\ and\
  \bibinfo {author} {\bibfnamefont {B.}~\bibnamefont {Yan}},\ }\href
  {https://doi.org/10.1103/PhysRevLett.129.070401} {\bibfield  {journal}
  {\bibinfo  {journal} {Phys. Rev. Lett.}\ }\textbf {\bibinfo {volume} {129}},\
  \bibinfo {pages} {070401} (\bibinfo {year} {2022})}\BibitemShut {NoStop}%
\bibitem [{\citenamefont {Zhao}\ \emph {et~al.}(2025)\citenamefont {Zhao},
  \citenamefont {Wang}, \citenamefont {He}, \citenamefont {Poon}, \citenamefont
  {Pak}, \citenamefont {Liu}, \citenamefont {Ren}, \citenamefont {Liu},\ and\
  \citenamefont {Jo}}]{zhao2025two}%
  \BibitemOpen
  \bibfield  {author} {\bibinfo {author} {\bibfnamefont {E.}~\bibnamefont
  {Zhao}}, \bibinfo {author} {\bibfnamefont {Z.}~\bibnamefont {Wang}}, \bibinfo
  {author} {\bibfnamefont {C.}~\bibnamefont {He}}, \bibinfo {author}
  {\bibfnamefont {T.~F.~J.}\ \bibnamefont {Poon}}, \bibinfo {author}
  {\bibfnamefont {K.~K.}\ \bibnamefont {Pak}}, \bibinfo {author} {\bibfnamefont
  {Y.-J.}\ \bibnamefont {Liu}}, \bibinfo {author} {\bibfnamefont
  {P.}~\bibnamefont {Ren}}, \bibinfo {author} {\bibfnamefont {X.-J.}\
  \bibnamefont {Liu}},\ and\ \bibinfo {author} {\bibfnamefont {G.-B.}\
  \bibnamefont {Jo}},\ }\href@noop {} {\bibfield  {journal} {\bibinfo
  {journal} {Nature}\ }\textbf {\bibinfo {volume} {637}},\ \bibinfo {pages}
  {565} (\bibinfo {year} {2025})}\BibitemShut {NoStop}%
\bibitem [{\citenamefont {Sahin}\ \emph {et~al.}(2025)\citenamefont {Sahin},
  \citenamefont {Jalil},\ and\ \citenamefont {Lee}}]{sahin2025topolectrical}%
  \BibitemOpen
  \bibfield  {author} {\bibinfo {author} {\bibfnamefont {H.}~\bibnamefont
  {Sahin}}, \bibinfo {author} {\bibfnamefont {M.}~\bibnamefont {Jalil}},\ and\
  \bibinfo {author} {\bibfnamefont {C.~H.}\ \bibnamefont {Lee}},\ }\href@noop
  {} {\bibfield  {journal} {\bibinfo  {journal} {APL Electronic Devices}\
  }\textbf {\bibinfo {volume} {1}} (\bibinfo {year} {2025})}\BibitemShut
  {NoStop}%
\bibitem [{\citenamefont {Yang}\ \emph
  {et~al.}(2024{\natexlab{b}})\citenamefont {Yang}, \citenamefont {Song},
  \citenamefont {Cao},\ and\ \citenamefont {Yan}}]{yang2024circuit}%
  \BibitemOpen
  \bibfield  {author} {\bibinfo {author} {\bibfnamefont {H.}~\bibnamefont
  {Yang}}, \bibinfo {author} {\bibfnamefont {L.}~\bibnamefont {Song}}, \bibinfo
  {author} {\bibfnamefont {Y.}~\bibnamefont {Cao}},\ and\ \bibinfo {author}
  {\bibfnamefont {P.}~\bibnamefont {Yan}},\ }\href@noop {} {\bibfield
  {journal} {\bibinfo  {journal} {Physics Reports}\ }\textbf {\bibinfo {volume}
  {1093}},\ \bibinfo {pages} {1} (\bibinfo {year}
  {2024}{\natexlab{b}})}\BibitemShut {NoStop}%
\bibitem [{\citenamefont {Zheng}\ \emph {et~al.}(2022)\citenamefont {Zheng},
  \citenamefont {Chen}, \citenamefont {Zhang}, \citenamefont {Sun},\ and\
  \citenamefont {Zhang}}]{zheng2022exploring}%
  \BibitemOpen
  \bibfield  {author} {\bibinfo {author} {\bibfnamefont {X.}~\bibnamefont
  {Zheng}}, \bibinfo {author} {\bibfnamefont {T.}~\bibnamefont {Chen}},
  \bibinfo {author} {\bibfnamefont {W.}~\bibnamefont {Zhang}}, \bibinfo
  {author} {\bibfnamefont {H.}~\bibnamefont {Sun}},\ and\ \bibinfo {author}
  {\bibfnamefont {X.}~\bibnamefont {Zhang}},\ }\href@noop {} {\bibfield
  {journal} {\bibinfo  {journal} {Physical Review Research}\ }\textbf {\bibinfo
  {volume} {4}},\ \bibinfo {pages} {033203} (\bibinfo {year}
  {2022})}\BibitemShut {NoStop}%
\bibitem [{\citenamefont {Zhang}\ \emph
  {et~al.}(2023{\natexlab{b}})\citenamefont {Zhang}, \citenamefont {Chen},
  \citenamefont {Li}, \citenamefont {Lee},\ and\ \citenamefont
  {Zhang}}]{zhang2023electrical}%
  \BibitemOpen
  \bibfield  {author} {\bibinfo {author} {\bibfnamefont {H.}~\bibnamefont
  {Zhang}}, \bibinfo {author} {\bibfnamefont {T.}~\bibnamefont {Chen}},
  \bibinfo {author} {\bibfnamefont {L.}~\bibnamefont {Li}}, \bibinfo {author}
  {\bibfnamefont {C.~H.}\ \bibnamefont {Lee}},\ and\ \bibinfo {author}
  {\bibfnamefont {X.}~\bibnamefont {Zhang}},\ }\href@noop {} {\bibfield
  {journal} {\bibinfo  {journal} {Physical Review B}\ }\textbf {\bibinfo
  {volume} {107}},\ \bibinfo {pages} {085426} (\bibinfo {year}
  {2023}{\natexlab{b}})}\BibitemShut {NoStop}%
\bibitem [{\citenamefont {Shang}\ \emph {et~al.}(2024)\citenamefont {Shang},
  \citenamefont {Liu}, \citenamefont {Jiang}, \citenamefont {Shao},
  \citenamefont {Zang}, \citenamefont {Lee}, \citenamefont {Thomale},
  \citenamefont {Manchon}, \citenamefont {Cui},\ and\ \citenamefont
  {Schwingenschl{\"o}gl}}]{shang2024observation}%
  \BibitemOpen
  \bibfield  {author} {\bibinfo {author} {\bibfnamefont {C.}~\bibnamefont
  {Shang}}, \bibinfo {author} {\bibfnamefont {S.}~\bibnamefont {Liu}}, \bibinfo
  {author} {\bibfnamefont {C.}~\bibnamefont {Jiang}}, \bibinfo {author}
  {\bibfnamefont {R.}~\bibnamefont {Shao}}, \bibinfo {author} {\bibfnamefont
  {X.}~\bibnamefont {Zang}}, \bibinfo {author} {\bibfnamefont {C.~H.}\
  \bibnamefont {Lee}}, \bibinfo {author} {\bibfnamefont {R.}~\bibnamefont
  {Thomale}}, \bibinfo {author} {\bibfnamefont {A.}~\bibnamefont {Manchon}},
  \bibinfo {author} {\bibfnamefont {T.~J.}\ \bibnamefont {Cui}},\ and\ \bibinfo
  {author} {\bibfnamefont {U.}~\bibnamefont {Schwingenschl{\"o}gl}},\
  }\href@noop {} {\bibfield  {journal} {\bibinfo  {journal} {Advanced Science}\
  }\textbf {\bibinfo {volume} {11}},\ \bibinfo {pages} {2303222} (\bibinfo
  {year} {2024})}\BibitemShut {NoStop}%
\bibitem [{\citenamefont {Zhang}\ \emph
  {et~al.}(2025{\natexlab{a}})\citenamefont {Zhang}, \citenamefont {Mei},
  \citenamefont {Li}, \citenamefont {Lee}, \citenamefont {Ma}, \citenamefont
  {Xiao},\ and\ \citenamefont {Jia}}]{zhang2025observation}%
  \BibitemOpen
  \bibfield  {author} {\bibinfo {author} {\bibfnamefont {J.-H.}\ \bibnamefont
  {Zhang}}, \bibinfo {author} {\bibfnamefont {F.}~\bibnamefont {Mei}}, \bibinfo
  {author} {\bibfnamefont {Y.}~\bibnamefont {Li}}, \bibinfo {author}
  {\bibfnamefont {C.~H.}\ \bibnamefont {Lee}}, \bibinfo {author} {\bibfnamefont
  {J.}~\bibnamefont {Ma}}, \bibinfo {author} {\bibfnamefont {L.}~\bibnamefont
  {Xiao}},\ and\ \bibinfo {author} {\bibfnamefont {S.}~\bibnamefont {Jia}},\
  }\href@noop {} {\bibfield  {journal} {\bibinfo  {journal} {Nature
  Communications}\ }\textbf {\bibinfo {volume} {16}},\ \bibinfo {pages} {2050}
  (\bibinfo {year} {2025}{\natexlab{a}})}\BibitemShut {NoStop}%
\bibitem [{\citenamefont {Zou}\ \emph {et~al.}(2024)\citenamefont {Zou},
  \citenamefont {Chen}, \citenamefont {Meng}, \citenamefont {Ang},
  \citenamefont {Zhang},\ and\ \citenamefont {Lee}}]{zou2024experimental}%
  \BibitemOpen
  \bibfield  {author} {\bibinfo {author} {\bibfnamefont {D.}~\bibnamefont
  {Zou}}, \bibinfo {author} {\bibfnamefont {T.}~\bibnamefont {Chen}}, \bibinfo
  {author} {\bibfnamefont {H.}~\bibnamefont {Meng}}, \bibinfo {author}
  {\bibfnamefont {Y.~S.}\ \bibnamefont {Ang}}, \bibinfo {author} {\bibfnamefont
  {X.}~\bibnamefont {Zhang}},\ and\ \bibinfo {author} {\bibfnamefont {C.~H.}\
  \bibnamefont {Lee}},\ }\href@noop {} {\bibfield  {journal} {\bibinfo
  {journal} {Science Bulletin}\ }\textbf {\bibinfo {volume} {69}},\ \bibinfo
  {pages} {2194} (\bibinfo {year} {2024})}\BibitemShut {NoStop}%
\bibitem [{\citenamefont {Stegmaier}\ \emph {et~al.}(2024)\citenamefont
  {Stegmaier}, \citenamefont {Fritzsche}, \citenamefont {Sorbello},
  \citenamefont {Greiter}, \citenamefont {Brand}, \citenamefont {Barko},
  \citenamefont {Hofer}, \citenamefont {Schwingenschl{\"o}gl}, \citenamefont
  {Moessner}, \citenamefont {Lee} \emph {et~al.}}]{stegmaier2024topological}%
  \BibitemOpen
  \bibfield  {author} {\bibinfo {author} {\bibfnamefont {A.}~\bibnamefont
  {Stegmaier}}, \bibinfo {author} {\bibfnamefont {A.}~\bibnamefont
  {Fritzsche}}, \bibinfo {author} {\bibfnamefont {R.}~\bibnamefont {Sorbello}},
  \bibinfo {author} {\bibfnamefont {M.}~\bibnamefont {Greiter}}, \bibinfo
  {author} {\bibfnamefont {H.}~\bibnamefont {Brand}}, \bibinfo {author}
  {\bibfnamefont {C.}~\bibnamefont {Barko}}, \bibinfo {author} {\bibfnamefont
  {M.}~\bibnamefont {Hofer}}, \bibinfo {author} {\bibfnamefont
  {U.}~\bibnamefont {Schwingenschl{\"o}gl}}, \bibinfo {author} {\bibfnamefont
  {R.}~\bibnamefont {Moessner}}, \bibinfo {author} {\bibfnamefont {C.~H.}\
  \bibnamefont {Lee}}, \emph {et~al.},\ }\href@noop {} {\bibfield  {journal}
  {\bibinfo  {journal} {arXiv preprint arXiv:2407.10191}\ } (\bibinfo {year}
  {2024})}\BibitemShut {NoStop}%
\bibitem [{\citenamefont {Nagulu}\ \emph {et~al.}(2022)\citenamefont {Nagulu},
  \citenamefont {Ni}, \citenamefont {Kord}, \citenamefont {Tymchenko},
  \citenamefont {Garikapati}, \citenamefont {Al{\`u}},\ and\ \citenamefont
  {Krishnaswamy}}]{nagulu2022chip}%
  \BibitemOpen
  \bibfield  {author} {\bibinfo {author} {\bibfnamefont {A.}~\bibnamefont
  {Nagulu}}, \bibinfo {author} {\bibfnamefont {X.}~\bibnamefont {Ni}}, \bibinfo
  {author} {\bibfnamefont {A.}~\bibnamefont {Kord}}, \bibinfo {author}
  {\bibfnamefont {M.}~\bibnamefont {Tymchenko}}, \bibinfo {author}
  {\bibfnamefont {S.}~\bibnamefont {Garikapati}}, \bibinfo {author}
  {\bibfnamefont {A.}~\bibnamefont {Al{\`u}}},\ and\ \bibinfo {author}
  {\bibfnamefont {H.}~\bibnamefont {Krishnaswamy}},\ }\href@noop {} {\bibfield
  {journal} {\bibinfo  {journal} {Nature Electronics}\ }\textbf {\bibinfo
  {volume} {5}},\ \bibinfo {pages} {300} (\bibinfo {year} {2022})}\BibitemShut
  {NoStop}%
\bibitem [{\citenamefont {Koh}\ \emph {et~al.}(2024)\citenamefont {Koh},
  \citenamefont {Tai},\ and\ \citenamefont {Lee}}]{koh2024realization}%
  \BibitemOpen
  \bibfield  {author} {\bibinfo {author} {\bibfnamefont {J.~M.}\ \bibnamefont
  {Koh}}, \bibinfo {author} {\bibfnamefont {T.}~\bibnamefont {Tai}},\ and\
  \bibinfo {author} {\bibfnamefont {C.~H.}\ \bibnamefont {Lee}},\ }\href@noop
  {} {\bibfield  {journal} {\bibinfo  {journal} {Nature Communications}\
  }\textbf {\bibinfo {volume} {15}},\ \bibinfo {pages} {5807} (\bibinfo {year}
  {2024})}\BibitemShut {NoStop}%
\bibitem [{\citenamefont {Shen}\ \emph
  {et~al.}(2025{\natexlab{b}})\citenamefont {Shen}, \citenamefont {Chen},
  \citenamefont {Yang},\ and\ \citenamefont {Lee}}]{shen2025observation}%
  \BibitemOpen
  \bibfield  {author} {\bibinfo {author} {\bibfnamefont {R.}~\bibnamefont
  {Shen}}, \bibinfo {author} {\bibfnamefont {T.}~\bibnamefont {Chen}}, \bibinfo
  {author} {\bibfnamefont {B.}~\bibnamefont {Yang}},\ and\ \bibinfo {author}
  {\bibfnamefont {C.~H.}\ \bibnamefont {Lee}},\ }\href@noop {} {\bibfield
  {journal} {\bibinfo  {journal} {Nature Communications}\ }\textbf {\bibinfo
  {volume} {16}},\ \bibinfo {pages} {1340} (\bibinfo {year}
  {2025}{\natexlab{b}})}\BibitemShut {NoStop}%
\bibitem [{\citenamefont {Koh}\ \emph {et~al.}(2025)\citenamefont {Koh},
  \citenamefont {Xue}, \citenamefont {Tai}, \citenamefont {Koh},\ and\
  \citenamefont {Lee}}]{koh2025interacting}%
  \BibitemOpen
  \bibfield  {author} {\bibinfo {author} {\bibfnamefont {J.~M.}\ \bibnamefont
  {Koh}}, \bibinfo {author} {\bibfnamefont {W.-T.}\ \bibnamefont {Xue}},
  \bibinfo {author} {\bibfnamefont {T.}~\bibnamefont {Tai}}, \bibinfo {author}
  {\bibfnamefont {D.~E.}\ \bibnamefont {Koh}},\ and\ \bibinfo {author}
  {\bibfnamefont {C.~H.}\ \bibnamefont {Lee}},\ }\href@noop {} {\bibfield
  {journal} {\bibinfo  {journal} {arXiv preprint arXiv:2503.14595}\ } (\bibinfo
  {year} {2025})}\BibitemShut {NoStop}%
\bibitem [{\citenamefont {Shen}\ \emph
  {et~al.}(2025{\natexlab{c}})\citenamefont {Shen}, \citenamefont {Chen},
  \citenamefont {Yang}, \citenamefont {Zhong},\ and\ \citenamefont
  {Lee}}]{shen2025robust}%
  \BibitemOpen
  \bibfield  {author} {\bibinfo {author} {\bibfnamefont {R.}~\bibnamefont
  {Shen}}, \bibinfo {author} {\bibfnamefont {T.}~\bibnamefont {Chen}}, \bibinfo
  {author} {\bibfnamefont {B.}~\bibnamefont {Yang}}, \bibinfo {author}
  {\bibfnamefont {Y.}~\bibnamefont {Zhong}},\ and\ \bibinfo {author}
  {\bibfnamefont {C.~H.}\ \bibnamefont {Lee}},\ }\href@noop {} {\bibfield
  {journal} {\bibinfo  {journal} {arXiv preprint arXiv:2503.08776}\ } (\bibinfo
  {year} {2025}{\natexlab{c}})}\BibitemShut {NoStop}%
\bibitem [{\citenamefont {Smith}\ \emph {et~al.}(2022)\citenamefont {Smith},
  \citenamefont {Jobst}, \citenamefont {Green},\ and\ \citenamefont
  {Pollmann}}]{smith2022crossing}%
  \BibitemOpen
  \bibfield  {author} {\bibinfo {author} {\bibfnamefont {A.}~\bibnamefont
  {Smith}}, \bibinfo {author} {\bibfnamefont {B.}~\bibnamefont {Jobst}},
  \bibinfo {author} {\bibfnamefont {A.~G.}\ \bibnamefont {Green}},\ and\
  \bibinfo {author} {\bibfnamefont {F.}~\bibnamefont {Pollmann}},\ }\href@noop
  {} {\bibfield  {journal} {\bibinfo  {journal} {Physical Review Research}\
  }\textbf {\bibinfo {volume} {4}},\ \bibinfo {pages} {L022020} (\bibinfo
  {year} {2022})}\BibitemShut {NoStop}%
\bibitem [{\citenamefont {Zhang}\ \emph
  {et~al.}(2025{\natexlab{b}})\citenamefont {Zhang}, \citenamefont
  {Carrasquilla},\ and\ \citenamefont {Kim}}]{zhang2025observation2}%
  \BibitemOpen
  \bibfield  {author} {\bibinfo {author} {\bibfnamefont {Y.}~\bibnamefont
  {Zhang}}, \bibinfo {author} {\bibfnamefont {J.}~\bibnamefont
  {Carrasquilla}},\ and\ \bibinfo {author} {\bibfnamefont {Y.~B.}\ \bibnamefont
  {Kim}},\ }\href@noop {} {\bibfield  {journal} {\bibinfo  {journal} {Nature
  Communications}\ }\textbf {\bibinfo {volume} {16}},\ \bibinfo {pages} {3286}
  (\bibinfo {year} {2025}{\natexlab{b}})}\BibitemShut {NoStop}%
\end{thebibliography}

\begin{thebibliography}{3}%
\makeatletter
\providecommand \@ifxundefined [1]{%
 \@ifx{#1\undefined}
}%
\providecommand \@ifnum [1]{%
 \ifnum #1\expandafter \@firstoftwo
 \else \expandafter \@secondoftwo
 \fi
}%
\providecommand \@ifx [1]{%
 \ifx #1\expandafter \@firstoftwo
 \else \expandafter \@secondoftwo
 \fi
}%
\providecommand \natexlab [1]{#1}%
\providecommand \enquote  [1]{``#1''}%
\providecommand \bibnamefont  [1]{#1}%
\providecommand \bibfnamefont [1]{#1}%
\providecommand \citenamefont [1]{#1}%
\providecommand \href@noop [0]{\@secondoftwo}%
\providecommand \href [0]{\begingroup \@sanitize@url \@href}%
\providecommand \@href[1]{\@@startlink{#1}\@@href}%
\providecommand \@@href[1]{\endgroup#1\@@endlink}%
\providecommand \@sanitize@url [0]{\catcode `\\12\catcode `\$12\catcode
  `\&12\catcode `\#12\catcode `\^12\catcode `\_12\catcode `\%12\relax}%
\providecommand \@@startlink[1]{}%
\providecommand \@@endlink[0]{}%
\providecommand \url  [0]{\begingroup\@sanitize@url \@url }%
\providecommand \@url [1]{\endgroup\@href {#1}{\urlprefix }}%
\providecommand \urlprefix  [0]{URL }%
\providecommand \Eprint [0]{\href }%
\providecommand \doibase [0]{https://doi.org/}%
\providecommand \selectlanguage [0]{\@gobble}%
\providecommand \bibinfo  [0]{\@secondoftwo}%
\providecommand \bibfield  [0]{\@secondoftwo}%
\providecommand \translation [1]{[#1]}%
\providecommand \BibitemOpen [0]{}%
\providecommand \bibitemStop [0]{}%
\providecommand \bibitemNoStop [0]{.\EOS\space}%
\providecommand \EOS [0]{\spacefactor3000\relax}%
\providecommand \BibitemShut  [1]{\csname bibitem#1\endcsname}%
\let\auto@bib@innerbib\@empty
\bibitem [{\citenamefont {Qin}\ and\ \citenamefont {Li}(2024)}]{SQinYi2024PRL}%
  \BibitemOpen
  \bibfield  {author} {\bibinfo {author} {\bibfnamefont {Y.}~\bibnamefont
  {Qin}}\ and\ \bibinfo {author} {\bibfnamefont {L.}~\bibnamefont {Li}},\
  }\href {https://doi.org/10.1103/PhysRevLett.132.096501} {\bibfield  {journal}
  {\bibinfo  {journal} {Phys. Rev. Lett.}\ }\textbf {\bibinfo {volume} {132}},\
  \bibinfo {pages} {096501} (\bibinfo {year} {2024})}\BibitemShut {NoStop}%
\bibitem [{\citenamefont {Kim}\ \emph {et~al.}(2024)\citenamefont {Kim},
  \citenamefont {Han},\ and\ \citenamefont {Park}}]{Skim2024collective}%
  \BibitemOpen
  \bibfield  {author} {\bibinfo {author} {\bibfnamefont {B.~H.}\ \bibnamefont
  {Kim}}, \bibinfo {author} {\bibfnamefont {J.-H.}\ \bibnamefont {Han}},\ and\
  \bibinfo {author} {\bibfnamefont {M.~J.}\ \bibnamefont {Park}},\ }\href@noop
  {} {\bibfield  {journal} {\bibinfo  {journal} {Communications Physics}\
  }\textbf {\bibinfo {volume} {7}},\ \bibinfo {pages} {73} (\bibinfo {year}
  {2024})}\BibitemShut {NoStop}%
\bibitem [{\citenamefont {Orito}\ and\ \citenamefont
  {Imura}(2022)}]{Orito2022PRB}%
  \BibitemOpen
  \bibfield  {author} {\bibinfo {author} {\bibfnamefont {T.}~\bibnamefont
  {Orito}}\ and\ \bibinfo {author} {\bibfnamefont {K.-I.}\ \bibnamefont
  {Imura}},\ }\href {https://doi.org/10.1103/PhysRevB.105.024303} {\bibfield
  {journal} {\bibinfo  {journal} {Phys. Rev. B}\ }\textbf {\bibinfo {volume}
  {105}},\ \bibinfo {pages} {024303} (\bibinfo {year} {2022})}\BibitemShut
  {NoStop}%
\end{thebibliography}
